\newif\ifcqg\cqgtrue
\ifcqg
\documentclass[]{iopart}
\usepackage{iopams}
\usepackage{setstack}
\eqnobysec
\def\flamoi{\fl}
\def\mybepsilon{\bepsilon}
\else
\documentclass[aps,prd,preprint,tightenlines,eqsecnum,showpacs,footinbib,amsmath,amssymb,amsfonts]{revtex4}
\usepackage{bm}
\def\Or{\mathrm{O}}
\def\flamoi{}
\def\mybepsilon{\bm{\epsilon}}
\fi

\usepackage[dvips,usenames]{color}
\usepackage{dsfont}
\usepackage{hyperref}

\newcommand{\Christoffel}[3]{\Gamma^{#1}_{#2 #3}}

\newcommand{\Lie}{\mathcal{L}}
\newcommand{\be}{\begin{equation}}
\newcommand{\ee}{\end{equation}}
\newcommand{\bea}{\begin{eqnarray}}
\newcommand{\eea}{\end{eqnarray}}
\newcommand{\gr}[1]{\mathbf{#1}}

\newcommand{\demi}{\frac{1}{2}}

\newcommand{\ii}{\mathrm{i}}

\newcommand{\dirac}[1]{\delta_{D}^{#1}}

\newcommand{\frd}[2]{\frac{\dd #1}{\dd #2}}

\newcommand{\eb}{\bar{e}}
\newcommand{\mf}[1]{\mathcal{#1}}
\newcommand{\N}[1]{N_{#1}}
\newcommand{\uline}[1]{\underline{#1}}
\def\st{\sigma_{\mathrm T}}
\def\me{m_{\mathrm{e}}}
\def\gel{g_{\mathrm e}}
\def\dd{{\rm d}}
\def\HH{\mathcal{H}}
\def\Tr{\mathcal{T}}
\definecolor{Myblue}{rgb}{0,0,1}
\definecolor{Myred}{rgb}{1,0,0}
\def\zB{{\color{Myred}O}}
\def\iB{{\color{Myred}I}}
\def\jB{{\color{Myred}J}}
\def\kB{{\color{Myred}K}}
\def\lB{{\color{Myred}L}}

\def\mB{{\color{Myred}M}}
\def\nB{{\color{Myred}N}}
\def\aB{{\color{Myred}A}}
\def\bB{{\color{Myred}B}}
\def\cB{{\color{Myred}C}}
\def\dB{{\color{Myred}D}}

\def\zT{{{\color{Myblue}o}}}
\def\iT{{{\color{Myblue}i}}}
\def\jT{{{\color{Myblue}j}}}
\def\kT{{{\color{Myblue}k}}}
\def\lT{{{\color{Myblue}l}}}
\def\mT{{{\color{Myblue}m}}}

\def\aT{{{\color{Myblue}a}}}
\def\bT{{{\color{Myblue}b}}}
\def\cT{{{\color{Myblue}c}}}
\def\dT{{{\color{Myblue}d}}}

\def\fT{{{\color{Myblue}f}}}
\def\hT{{{\color{Myblue}h}}}
\def\zTt{{\color{Myblue}{\tilde o}}}
\def\iTt{{\color{Myblue}{\tilde \imath}}}
\def\jTt{{\color{Myblue}{\tilde \jmath}}}

\def\aTt{{\color{Myblue}{\tilde a}}}
\def\bTt{{\color{Myblue}{\tilde b}}}
\def\cTt{{\color{Myblue}{\tilde c}}}
\def\dTt{{\color{Myblue}{\tilde d}}}

\def\hTt{{\color{Myblue}{\tilde h}}}

\def\aS{\alpha}
\def\bS{\beta}
\def\cS{\mu}
\def\dS{\nu}
\def\eS{\sigma}
\def\fS{\lambda}
\def\gS{\omega}
\def\hS{\tau}
\def\xS{\kappa}

\def\ib{{\mathrm b}}
\def\ir{{\mathrm r}}
\def\ie{{\mathrm e}}
\def\ic{{\mathrm c}}
\def\ip{{\mathrm p}}
\def\NG{{\rm NG}}

\def\hashamoi{\#}

\newcommand{\dsd}[2]{\frac{\partial {#1}}{\partial {#2}}}
\newcommand{\Yd}[3]{{\cal Y}^{#1 #2}_{#3}}
\newcommand{\Yu}[3]{{\cal Y}_{#1 #2}^{#3}}
\newcommand{\Ysd}[3]{{\cal Y}^{\star\, #1 #2}_{#3}}
\newcommand{\Ysu}[3]{{\cal Y}_{#1 #2}^{\star \,#3}}

\newcommand{\kPlmsn}[4]{{}^{#4}_{#3} {\uparrow \!\!\!K}^{#2}_{#1}}
\newcommand{\kMlmsn}[4]{{}^{#4}_{#3} {\downarrow \!\!\!K}^{#2}_{#1}}

\newcommand{\llmn}[3]{{}^{#3} \lambda^{#2}_{#1}\,}

\begin{document}
\title[The radiative transfer at second order]{The radiative transfer at second order:\\ a full treatment of the Boltzmann equation with polarization}
\author{Cyril Pitrou}
\ifcqg
\ead{cyrilp@astro.uio.no}
\address{Institute of Theoretical Astrophysics,
         University of Oslo, 
         P.O. Box 1029 Blindern, 0315 Oslo, Norway.}
\else
\email{cyrilp@astro.uio.no}
\affiliation{Institute of Theoretical Astrophysics,
         University of Oslo, 
         P.O. Box 1029 Blindern, 0315 Oslo, Norway.}
\fi


\date{\today}

\begin{abstract}
This article investigates the full Boltzmann equation up to
second order in the cosmological perturbations. Describing the distribution of
polarized radiation by a tensor valued distribution function, we study the
gauge dependence of the distribution function and summarize the construction of the gauge-invariant distribution function.
The Liouville operator which describes the free streaming of electrons, and the collision term which describes the
scattering of photons on free electrons are computed up to second order. Finally, the remaining dependence in the direction of the photon momentum is handled by expanding in projected symmetric
trace-free multipoles and also in the more commonly used normal modes components. The
results obtained remain to be used for computing numerically the contribution
in the cosmic microwave background bispectrum which arises from the evolution
of second order perturbations, in order to disentangle the primordial
non-Gaussianity from the one generated by the subsequent non-linear evolution.
\end{abstract}
\ifcqg
\else
\maketitle
\fi
\section*{Introduction}

The cosmic microwave background (CMB) has become in the past twenty years a
central observable of modern cosmology. The properties of the CMB temperature fluctuations depend both on the initial conditions set at the end
of the primordial inflationary era, and on their evolution through time in the
post-inflationary eras. The theory of cosmological perturbations around a
space-time with maximally symmetric spatial sections is a cornerstone of our
understanding of the large scale structure of the universe. The relativistic matter (photons, neutrinos) is
described with a statistical approach~\cite{Ehlers1971,Bernstein1988,Stewart1971}, also
referred to as kinetic theory, in which we use a distribution function whose evolution is given by
a Boltzmann equation. As for non-relativistic matter [baryons, cold dark
matter], a fluid approximation is usually sufficient. Given the typical 
amplitude ($10^{-5}$) of the metric fluctuations, the dynamics of the
cosmological  perturbations was so far mainly studied at linear order around a Friedmann-Lema\^itre (FL) space-time, and
characterized statistically by the power spectrum. This method allows to relate the
CMB angular power spectrum to the initial power spectrum, which opens a window
on the early universe. However, the linear order of perturbations fails to capture the
intrinsic non-linear features of General Relativity which enter both the
initial conditions and the evolution. The CMB measurements, though already of
high precision~\cite{Spergel2007,Komatsu2008}, are soon going to improve with
the new forthcoming missions such as Planck~\cite{PLANCK}, and will be
sensitive to these non-linear effects. It becomes thus necessary to go beyond this
linear perturbation scheme by studying the second-order perturbations. This is
even more crucial if we want to estimate the bispectrum in the cosmic microwave
background since this can only arise from non-Gaussian initial conditions set
by inflation or from non-linear evolution. In order to improve our theory of
the early universe and discriminate between different models of inflation, it is thus necessary to disentangle the primordial
non-Gaussianity from the one induced by non-linear evolution. We thus need to extend the
program followed at first order in perturbations up to second order. \\

Since cosmological perturbations are plagued by the gauge freedom,
we need to build a full set of gauge invariant perturbation variables and derive the perturbed
Boltzmann and Einstein equations in terms of these
variables to obtain their dynamical equations. The first-order gauge invariant
perturbation variables in the fluid approach were built in~\cite{Bardeen1980},
and in the kinetic theory in~\cite{Durrer1988,Durrer1994}. In the inflationary era and  for slow-roll one
field inflation, the quantization in the linear equations of the canonical
degrees of freedom~\cite{Mukhanov1992} which transfer to Gaussian classical
fluctuations enable us to fix the initial conditions for the post-inflationary
dynamics on super-Hubble scales. By using quantities which are conserved for modes larger than the Hubble radius~\cite{Rigopoulos2003,Malik2004,Vernizzi2005}, we can ignore the details between
the end of inflation and the subsequent eras. In this program, the evolution of radiation\footnote{Usually,
  the word radiation is used for relativistic particle. In the context of CMB,
  it is often used just for photons, as it is the case in this paper.} requires a special
care since the distribution function is shaped by Compton scattering by free
electrons and in particular this generates polarization. At first order, the evolution equations describing
polarized radiation were studied intensively in~\cite{Hu1994b,Hu1997,Ma1995},
and codes have been made available~\cite{CMBFAST,CAMB} for integrating numerically these equations and thus analyzing the CMB data. \\

At second-order, the program remains so far incomplete and this paper aims at
filling this gap. The quantization of the canonical degrees of freedom in the
models of slow-roll one field inflation with non-linear couplings has been investigated in the interaction
picture~\cite{Maldacena2003,Bernardeau2003a} up to the loop corrections~\cite{Weinberg2005,Sloth2006,Sloth2006a} aiming at
computing the level of primordial non-Gaussianity. This generically predicts negligible amounts of non-Gaussianity~\cite{Maldacena2003} whereas multi-field
inflation~\cite{Bernardeau2002,Bernardeau2003,Bartolo2004,Seery2005a,Vernizzi2006,Rigopoulos2006,Arroja2008,Langlois2008a,Langlois2008b}
can generate significant levels of non-Gaussianity. Since the use of adapted
estimators~\cite{Komatsu2005,Creminelli2006a,Yadav2007} tends to show that there might be a detectable amount of
non-Gaussianity in the CMB~\cite{Spergel2007,Komatsu2008,Yadav2007b}, the
understanding of non-linear evolution is of first importance in order to
constrain these models of inflation. In particular all non-linear effects in the foreground~\cite{Cooray2000,Goldberg1999,Castro2004,Serra2008}
(see~\cite{Aghanim2007} for a review) have to be understood and estimated. In order to develop the mathematical
tools for this general study, the gauge issue in the
fluid approach was studied in~\cite{Bruni1997} and gauge invariant variables
were built in~\cite{Malik2004,Nakamura2007} up to second order in perturbations. This
fluid approximation has already been used to understand the general form of the
bispectrum generated by evolutionary effects~\cite{PUB2008}. As for the more
general kinetic approach, the gauge issue was studied in our previous
paper~\cite{Pitrou2007}, and the evolution equations through free-streaming
and Compton collision on free electrons were derived in~\cite{Hu1994,Tomita2005,Bartolo2006,Bartolo2007}, but it is so far
restricted to unpolarized radiation, which is inconsistent since Compton
scattering does generate polarization. In this article, we will extend these works
to polarized radiation, leaving open the issue of second-order numerical
integration. It should be mentioned that an alternative program consists in
working directly with covariantly defined quantities in the so-called $1+3$
covariant formalism. The first order canonical degrees of freedom required to fix the
initial conditions in this approach have been identified~\cite{Pitrou2007a}, the conserved
quantities, used to ignore the detail between the inflationary era and the
subsequent eras, have been built~\cite{Langlois2005a,Langlois2005} and the dynamical equations
up to second order~\cite{Challinor1999,Challinor2000,Gebbie2000,Gebbie2000b,Maartens1999,Tsagas2007,Ellis1998}
have also been extensively studied, but so far leaving aside for the second
order the problem of mode decomposition and the treatment of polarized
polarization. The results presented in this paper can be used in a
straightforward manner for the description of collisions in this $1+3$
formalism, and we even correct a small mistake for the unpolarized case. The
two formalisms should of course lead to the same conclusion and they have been
compared in~\cite{Bruni1991,Osano2007,Enqvist2007}.\\

This paper is organized as follows. We first review the description of
polarized radiation with a tensor valued distribution function in
section~\ref{Sec_Genform}. We then recall the gauge dependence and the construction of gauge invariant variables when using
a fluid approximation~\ref{SecGIT}. In section~\ref{SecGIf}, we summarize the
results of our previous paper~\cite{Pitrou2007} which focused on the gauge
dependence of the distribution function and the construction of a gauge
invariant distribution function in the kinetic theory, and we extend
the results to the tensor valued distribution function. This formalism is then used to
compute the second order gauge invariant Liouville operator, for radiation in
section~\ref{SecLiouville} and for matter in section~\ref{SecLiouvilleBar}. The
collision term for both cases is computed in section~\ref{Sec_Collision}.
Finally, in section~\ref{SecSTFGlm}, we develop the necessary tools to express
these results in terms of the normal modes decomposition.

\section{Kinetic theory}\label{Sec_Genform}

\subsection{Momentum and tetrad}

In the kinetic description of radiation, the momentum of photons is
usually decomposed onto an orthonormal basis, that is using a tetrad field.
Though not compulsory, this facilitates the separation between the magnitude
of the momentum and its direction represented by a spacelike unit vector. The tetrad vectors $\gr{e}_\aT$ and their corresponding tetrad forms $\gr{e}^\aT$ satisfy the orthonormality conditions
\be\label{deftetrad}
\gr{e}_\aT.\gr{e}_\bT\equiv e_{\aT}^{\,\,\mu} e_{\bT}^{\,\,\nu} g_{\mu
  \nu}=\eta_{\aT \bT},\qquad\gr{e}^\aT.\gr{e}^\bT\equiv e^{\aT}_{\,\,\mu} e^{\bT}_{\,\,\nu}
g^{\mu \nu}=\eta^{\aT \bT},
\ee
where $g_{\mu\nu}$ is the space-time metric, $g^{\mu\nu}$ its inverse and
$\eta^{\aT \bT}=\eta_{\aT \bT}$ the Minkowski metric.
In the previous expressions and throughout this paper, we use Greek indices ($\mu,\nu,\rho,\sigma\dots$) for abstract indices and the
beginning of the Latin alphabet ($\aT,\bT,\cT,\dT\dots$) for tetrad labels.
Since the tetrad labels run from $0$ to $3$, we also use Latin indices
starting from the letter $\iT$ (that is $\iT,\jT,\kT,\lT\dots$) with values
ranging from $1$ to $3$ to label the spacelike vectors or forms of a tetrad. We
then reserve the label $\zT$ for the timelike vector and form in a
tetrad. A momentum can then be decomposed as
\be
\gr{p}=p^\aT \gr{e}_\aT=p^\zT \gr{e}_\zT +p^\iT \gr{e}_\iT\,,  
\ee
where the components can be extracted as
\be
p^\aT=\gr{p}.\gr{e}^\aT \equiv e^{\aT}_{\,\mu}p^\mu\,.
\ee
It can be decomposed into a magnitude $p^\zT$ and the a direction vector $\gr{n}$ according to
\be
p^\mu=p^\zT (e_{\zT}^{\,\mu} +n^\mu)\,,\quad \gr{n}.\gr{e}_\zT \equiv n_\mu e_{\zT}^{\,\mu}=0\,, \quad n^\mu n_\mu\equiv\gr{n}.\gr{n}=1\,.
\ee
This decomposition can be used to define a screen projector
\be\label{DefScreen}
{S_{\gr{e_\zT}}}_{\mu\nu}(\gr{p})=g_{\mu \nu}+e^\zT_{\,\mu} e^\zT_{\,\nu}-n_\mu n_\nu\,, 
\ee
which projects on a space which is both orthogonal to $\gr{e}_\zT$ and orthogonal to the direction $\gr{n}$, since 
\be
\flamoi{S_{\gr{e}_\zT}}_{\mu \nu} e_\zT^{\,\mu}=0,\qquad {S_{\gr{e}_\zT}}_{\mu \nu}p^\mu=0,\qquad {S_{\gr{e}_\zT}}_{\mu \nu}n^\nu=0,\qquad {S_{\gr{e}_\zT}}_{\mu}^{\,\,\nu}{S_{\gr{e}_\zT}}_{\nu\sigma}={S_{\gr{e}_\zT}}_{\mu\sigma}\,.
\ee
In this paper, a projected tensorial quantity is orthogonal to $\gr{e}_\zT$ only
and a screen projected quantity is orthogonal to both $\gr{n}$ and $\gr{e}_\zT$. 
If there is no risk of confusion about the choice of the observer in this
screen projector, then we will use the notation $S_{\mu \nu}$ instead of
${S_{\gr{e}_\zT}}_{\mu\nu}$. If we also omit the dependence of the screen
projector in the photon momentum, then we abbreviate $S_{\mu \nu}(\gr{p})$ in
$S_{\mu \nu}$, and $S_{\mu \nu}(\gr{p}')$ in $S'_{\mu \nu}$.
The polarization of a photon is represented by its polarization vector
$\mybepsilon(\gr{p})$ which is a complex spacelike unit vector ($\epsilon_\mu^\star
(\gr{p})\epsilon^\mu(\gr{p})=1$) and taken in the Lorentz gauge ($\epsilon_\mu(\gr{p}) p^\mu =0$).
Since there is a residual gauge freedom (of electromagnetism) in the choice of
the polarization vector, we will work with the screen projected polarization vector $S^\mu_{\,\,\nu}\epsilon^\nu(\gr{p})$ which is unique.

\subsection{The (screen-projected) polarization tensor}   

The radiation is represented by a Hermitian tensor valued distribution
function (which is thus complex valued), also called polarization tensor~\cite{Bildhauer1989,Kosowsky1996,Tsagas2007,Stebbins2007} satisfying
\be
F_{\mu \nu}(x^\aB,p^\aT)\qquad p^\mu F_{\mu\nu}(x^\aB,p^\aT)=0\,,
\ee
from which we can form the distribution function of photons in a given state of
polarization $\mybepsilon$ by
\be
F_{\mu \nu}(x^\aB,p^\aT){\epsilon^\star}^\mu(\gr{p}) \epsilon^\nu(\gr{p})\,.
\ee
Here the $x^\aB$ are the coordinates used to label points on the space-time manifold. Throughout this paper, indices
which refer to these coordinates are $\aB,\bB,\cB,\dots$ if they
run from $0$ to $3$. Furthermore, indices which are $\iB,\jB,\kB,\dots$ run from $1$
to $3$, and the time component index is $\zB$.\\
For a given electromagnetic plane wave with potential vector amplitude
$A^\mu=A \epsilon^\mu$ and wave vector $k^\mu$ in the geometric optics limit, this polarization tensor can be defined~\cite{MTW} by
\be\label{Def_buildpolartensor}
\dirac{1}(\gr{p}.\gr{p})F_{\mu \nu}(p^\aT) \equiv \frac12 (2\pi)^3 \dirac{4} (\gr{k} -\gr{p})  A^2 \epsilon_\mu(\gr{k}) \epsilon^\star_\nu(\gr{k}) \,,
\ee 
where $\dirac{n}$ is the Dirac function of dimension $n$. This tensor $F_{\mu\nu}$ should not be confused with the Faraday tensor. Since
the remaining gauge freedom of electromagnetism also affects the polarization tensor, we can
define the screen-projected distribution function by
\be
f_{\mu\nu}(x^\aB,p^\aT)=S_\mu^\rho S_\nu^\sigma F_{\rho \sigma}(x^\aB,p^\aT)\,.
\ee
This tensor is no more dependent on the residual gauge freedom and encodes all
the polarization properties of radiation as seen by an observer having a
velocity $\gr{e}_\zT$. Similarly to the definition of the screen projector, if the context requires it, we will use an index
notation of the type $f_{\gr{e}_\zT}^{\mu\nu}$ to remind with which velocity, and thus with which screen projector, it is defined. 
The screen projected tensor has four degrees of freedom which can be split according to
\be
 f_{\mu \nu}(x^\aB,p^\aT)\equiv\frac{1}{2} I(x^\aB,p^\aT)S_{\mu
  \nu}+P_{\mu \nu}(x^\aB,p^\aT)+\frac{\ii}{2} V(x^\aB,p^\aT) \epsilon_{\mu \nu \sigma}
n^\sigma,
\ee
with $\epsilon_{\mu \nu \sigma}\equiv e_\zT^\rho\epsilon_{\rho \mu \nu
  \sigma}$, and  where $\epsilon_{\rho  \mu \nu \sigma}$ is the completely
antisymmetric tensor.
$P_{\mu \nu}$, which encodes the degree of linear polarization, is real symmetric and trace free, as well as orthogonal to
$\gr{e}_\zT$ and $\gr{n}$, and thus encodes two degrees of freedom, which can be described by the Stokes functions $Q$ and $U$~\cite{Tsagas2007}.
$I(x^\aB,p^\aT)$ and $V(x^\aB,p^\aT)$ are respectively the intensity (or distribution function) for both polarizations and the degree of
circular polarization. We also define the normalized (screen-projected)
polarization tensor by
\be
U_{\mu \nu}\equiv \frac{f_{\mu\nu}}{f^{\alpha}_{\,\,\alpha}}=\frac{f_{\mu\nu}}{I}\,\,.
\ee

\subsection{Stress-energy tensor and energy-integrated functions}

For a distribution of photons, a stress-energy tensor can be defined by
\be
T^{\mu \nu}(x^\aB)\equiv e_\aT^{\,\mu} e_\bT^{\,\nu} \left(\int \dirac{1}(\gr{p}.\gr{p})I(x^\aB,p^\cT) p^\aT p^\bT \frac{\dd p^\zT \dd^3
  p^{\iT}}{(2 \pi)^3}\right).
\ee
Performing the integral over $p^\zT$ (we choose the convention in which we
integrate on the two mass hyperboloides), we eliminate the Dirac function and it leads to
\be
T^{\mu \nu}(x^\aB)\equiv e_\aT^{\,\mu} e_\bT^{\,\nu} \left(\int
I(x^\aB,p^\iT) p^\aT p^\bT \frac{\dd^3 p^{\iT}}{p^\zT (2\pi)^3}\right)\,,
\ee
where now the intensity distribution function has to be considered as a
function of $p^\iT$, and $p^\zT$ is positive and taken on the mass shell, that
is $p^\zT=\sqrt{p_\iT p^\iT}$, and is thus identified with the energy of the photon. Splitting the integral over $\dd^3 p^\iT$ into magnitude and
angular direction leads to
\be\label{Def_Tmunu3}
T^{\mu \nu}(x^\aB)\equiv \frac{1}{(2\pi)^3}e_\aT^{\,\mu} e_\bT^{\,\nu} \left(\int
I(x^\aB,p^\iT) \left(p^\zT\right)^3 N^\aT N^\bT \dd p^\zT \dd^2 \Omega\right)\,,
\ee
where $\dd^2 \Omega$ is the differential solid angle associated with the
unit vector $n^\iT$, and where $N^\iT \equiv n^\iT$ and $N^\zT\equiv 1$.
This motivates the definition of the (energy-) integrated counterparts of $I$,
$V$ and $P_{\mu \nu}$ which are
\bea\label{Brightnessdef}
{\cal I}(x^\aB,n^\iT)&\equiv &\frac{4 \pi}{(2 \pi)^3} \int
I(x^\aB,p^\zT,n^\iT)(p^\zT)^3 \dd p^\zT\,,\\*
{\cal V}(x^\aB,n^\iT)&\equiv& \frac{4 \pi}{(2 \pi)^3} \int
V(x^\aB,p^\zT,n^\iT)(p^\zT)^3 \dd p^\zT\,,\\*
{\cal P}_{\mu \nu}(x^\aB,n^\iT)&\equiv& \frac{4 \pi}{(2 \pi)^3} \int
P_{\mu \nu}(x^\aB,p^\zT,n^\iT)(p^\zT)^3 \dd p^\zT\,.
\eea
$\cal I$ is the brightness, ${\cal V}$ is the degree of linear polarization in
units of ${\cal I}$, and ${\cal P}_{\mu \nu}$ is the tensor of linear
polarization in units of ${\cal I}$.

\subsection{Description of massive particles}

For massive particles such as electrons ($\ie$), protons ($\ip$) or cold dark
matter ($\ic$), we do not need to describe polarization and thus we can rely solely on a
the distribution functions $\gel$, $g_{\mathrm p}$ and $g_{\mathrm c}$ (chosen
to describe the two helicities). Additionally, following common practice in
cosmology, we will refer to electrons and protons together as baryons though
electrons are leptons. This is motivated by the fact that most of the mass is
carried by protons which are baryons, and the Compton interaction between
protons and electrons makes these two components highly interdependent. For a particle with impulsion $q^\aT$, we use the following notation
\be
\flamoi n^\iT\equiv \frac{q^\iT}{q^\zT}\,,\quad \beta=\sqrt{n^\iT n_\iT}\,,\quad\lambda \equiv
\beta q^\zT\,, \quad \hat n^\iT= n^\iT/\beta\,, \quad \gamma= \left(1 -\beta^2 \right)^{-1/2}\,,
\ee  
where now we have to distinguish between the unit vector $\hat n^\iT$ and the
velocity vector $n^\iT$ because of the mass $m$ of the particles.
The stress-energy tensor can be defined in a similar manner to equation~(\ref{Def_Tmunu3}) by
\be
T^{\mu \nu}(x^\aB)\equiv e_\aT^{\,\mu} e_\bT^{\,\nu} \left(\int
  \dirac{1}(\gr{q}.\gr{q}-m^2) g(x^\aB,q^\cT) q^\aT q^\bT \frac{\dd q^\zT \dd^3
  q^{\iT}}{(2 \pi)^3}\right).
\ee
Integrating over $q^\zT$, this leads to
\be\label{Def_Tmunu4}
T^{\mu \nu}(x^\aB)\equiv \frac{1}{(2\pi)^3}e_\aT^{\,\mu} e_\bT^{\,\nu} \left(\int
g(x^\aB,q^\iT) q^\zT N^\aT N^\bT \dd^3 q^\iT\right)\,,
\ee
where we recall that $N^\aT \equiv (1,n^\iT)$, and $q^\zT$ is taken on the
mass shell ($q^\zT=\sqrt{q_\iT q^\iT+m^2}$).\\

\subsection{The fluid limit}\label{Fluidlimit}

The stress-energy tensor of radiation or matter is equivalent to the one of an
imperfect fluid with stress-energy tensor
\begin{equation}
 T_{\mu\nu} = \rho u_\mu u_\nu + P\left( g_{\mu\nu} + u_\mu u_\nu\right)
 +\Pi_{\mu \nu}\,.
\label{defstressenergytensor}
\end{equation}
In this decomposition $\rho$ is the energy density, $P$ is the pressure,
$u^\mu$ is the fluid velocity and $\Pi^{\mu \nu}$, which satisfies
$\Pi^{\mu}_{\,\,\mu}=u^\mu \Pi_{\mu \nu}=0$, is the anisotropic stress. 
For an isotropic distribution of radiation, that is where ${\cal I}(x^\aB,p^\iT)$ depends only on the magnitude of $p^\iT$, it is
straightforward to show by comparison of the expressions~(\ref{Def_Tmunu3})
and (\ref{defstressenergytensor}) that $P=\rho/3$. Similarly, for a set of heavy particles (or non-relativistic particles), that is with $\sqrt{q^\iT q_\iT} \ll m$, the pressure
satisfies $P\ll \rho$ and the anisotropic stress tensor is also similarly small.
For cold dark matter, it is assumed that the mass of particles is large
enough so that we can use this approximation.\\
In the case of electrons and protons, that is baryons, the Coulomb interaction ensures that the
distribution of momenta follows a Fermi-Dirac distribution in the reference
frame where they have no bulk velocity~\cite{Bernstein1988}, at least as long as the baryonic
matter is ionized. This distribution is isotropic in this adapted frame and
depends only on $\lambda$
\be
g_{\mathrm{fd}}(\lambda)=\left(\exp\left[\frac{\sqrt{(\lambda^2+m^2)}-\mu}{T}\right]+1\right)^{-1}\,\,\,.
\ee
As a consequence, the anisotropic stress vanishes, and in this adapted
frame the baryons are then ideally described by the energy density and the pressure
\bea
\rho_\ib&\equiv& \frac{4\pi}{(2\pi)^3}\int g_{\mathrm{fd}}(\lambda)\sqrt{\lambda^2+m^2}\lambda^2 \dd \lambda\,,\\*
P_\ib&\equiv& \frac{4\pi}{3(2\pi)^3}\int g_{\mathrm{fd}}(\lambda)\frac{\lambda^4}{\sqrt{\lambda^2+m^2}} \dd \lambda\,.
\eea
If we can neglect the chemical potential $\mu$, which is the case in the
cosmological context, then for non-relativistic particles we obtain
\bea
\rho_\ib&\simeq&m\left(1+\frac{3T}{2m}\right) n_\ib\,,\\*
P_\ib&\simeq&n_\ib T\,.
\eea
The baryonic matter is ionized roughly until recombination where
the temperature of photons is of order $T_{\mathrm{LSS}} \simeq 0.25\,\, \mathrm{eV}$~\cite{Komatsu2008}. For electrons, 
the thermal correction is of order $T_{\mathrm{LSS}}/\me\simeq 0.25/511000\simeq 0.5\,\, 10^{-6}$,
and this ratio is even $m_{\ip}/m_{\ie}\simeq 1836$ times smaller for protons.
We thus deduce that the baryons either have no anisotropic stress because of
Coulomb interaction, or have a very tiny pressure and anisotropic stress after
recombination has occurred. However, as discussed in
section~\ref{Sec_pertscheme} this thermal correction is still of order of the
metric perturbations and should not {\it in principle} be ignored for second order
computations. Nevertheless, the thermal corrections are not relevant for computing the bispectrum and
it is for this reason that we will, from now on, drop terms in $T/m$ and
describe baryons as cold matter (but not dark since it can interact with radiation). It should be mentionned that for neutrinos
these conclusions are not valid anymore since they are very light~\cite{Lesgourgues2006}. We will
here assume that they are light enough to be treated as collisionless
radiation, and the equations which govern their evolution can be found by
setting $\st=0$ in the equations for photons, where $\st$ is the Thomson cross section. 

\subsection{Multipole expansion for radiation}

Functions of $p^\aT$ can be viewed as functions of $(p^\zT,n^\aT)$ and we can
separate the dependence into the energy and the direction of the momentum. The
dependence in the direction can be further expanded in multipoles using projected symmetric trace-free (PSTF) tensors, where
projected means that they are orthogonal to $\gr{e}_\zT$. For instance, $I$ can be expanded as
\be\label{Def_multipoleS}
I(x^\aB,p^\zT,n^\aT)=\sum_{\ell=0}^\infty I_{\uline{\aT_\ell}}(x^\aB,p^\zT)n^{\uline{\aT_\ell}}\,,
\ee
where the $I_{\uline{\aT_\ell}}\equiv I_{\aT_1\dots\aT_\ell}$ are PSTF. For
the lowest multipole, i.e. the one corresponding to $\ell=0$, we use the
notation $I_{\emptyset}$. Note that we have defined the notation $n^{\uline{\aT_\ell}}\equiv n^{\aT_1}\dots n^{\aT_\ell}$.
We also remind that $n^\aT\equiv \gr{n}.\gr{e}^\aT\equiv n^\mu e^\aT_{\,\mu}$.
 Since $\gr{n}$ is projected, $n^\zT=0$ and thus if any of the indices in
$I_{\uline{\aT_\ell}}$ is $\zT$, then the multipoles is chosen to vanish. These multipoles can be obtained by performing the following integrals
\be
\flamoi I_{\uline{\aT_\ell}}(x^\aB,p^\zT)=\Delta_\ell^{-1}\int  I(x^\aB,p^\zT,n^\aT)
n_{\langle \uline{\aT_\ell}\rangle}\dd^2 \Omega\,,\qquad \Delta_\ell\equiv 4 \pi \frac{\ell!}{(2\ell+1)!!}\,\,\,,
\ee
where $\langle \dots\rangle$ means the symmetric trace-free part. A similar expansion can be performed on $V$, by replacing $I$ by $V$ in the above
expressions, as well as for their energy integrated counterparts. Note that in
particular
\bea
{\cal I}^{\emptyset}&=&T^{\zT \zT}\,,\\*
{\cal I}^\iT&=&4 \pi \Delta_1^{-1} T^{\zT \iT}\,,\\*
{\cal I}^{\iT \jT}&=&4 \pi  \Delta_2^{-1} T^{\langle \iT \jT \rangle}\,.
\eea
As for $P_{\aT \bT}\equiv P_{\mu \nu} e_\aT^{\,\mu}e_\bT^{\,\nu}$, it can be
expanded in electric and magnetic type components according to~\cite{Challinor2000,Tsagas2007,Dautcourt1978,Thorne1980}
\be\label{Def_multipoleP}
\flamoi P_{\aT \bT}(x^\aB,p^\aT)=\sum_{\ell=2}^\infty \left[E_{\aT \bT
  \uline{\cT_{\ell-2}}}(x^\aB,p^\zT)n^{ \uline{\cT_{\ell-2}}} \,- n_{\cT}\epsilon^{\cT
  \dT}_{\,\,\,\,\,(\aT}B_{\bT)\dT \uline{\cT_{\ell-2}}}(x^\aB,p^\zT)n^{\uline{\cT_{\ell-2}}}\right]^{\mathrm{TT}},
\ee
where the notation $\mathrm{TT}$ denotes the transverse (to $\gr{n}$)
symmetric trace-free part, which for a second rank tensor is 
\be
\left[X_{\aT\bT}\right]^{\mathrm{TT}}\equiv S_{(\aT}^{\,\,\cT}
S_{\bT)}^{\,\,\dT}X_{\cT \dT}-\frac{1}{2}S_{\aT\bT} S^{\cT\dT}X_{\cT \dT}\,.
\ee
The electric and magnetic multipoles can be obtained by performing the
integrals
\bea
E_{\uline{\aT_\ell}}(x^\aB,p^\zT)&=&M_\ell^2\Delta_\ell^{-1}\int 
\,n_{\langle
  \uline{\aT_{\ell-2}}}P_{\aT_{\ell-1}\aT_{\ell}\rangle}(x^\aB,p^\zT,n^\aT)\dd^2 \Omega\,\,,\\*
B_{\uline{\aT_\ell}}(x^\aB,p^\zT)&=&M_\ell^2\Delta_\ell^{-1}\int 
\,n_\bT \epsilon^{\bT
  \dT}_{\,\,\,\,\,\langle \aT_\ell}n_{\uline{\aT_{\ell-2}}}P_{\aT_{\ell-1}\rangle \dT}(x^\aB,p^\zT,n^\aT)\dd^2 \Omega\,\,,
\eea
where 
\be
M_\ell=\sqrt{\frac{2 \ell (\ell-1)}{(\ell+1)(\ell+2)}}\,\,.
\ee
\subsection{Transformation rules under a change of frame}\label{Sec_Changeframe}

\subsubsection{The photon momentum}

So far, everything was defined with respect to an observer having a velocity
$\gr{e}_\zT$. How does all this machinery transform when the radiation is
observed by an observer with a different velocity ${\tilde{\gr{e}}}_{\zT}$? Since two velocities can be related by a
Lorentz transformation, there exists a vector $\gr{v}$ such that
\be
{\tilde{\gr{e}}}_{\zT}=\gamma(\gr{e}_\zT+\gr{v}),\qquad\gamma\equiv\frac{1}{\sqrt{1-\gr{v}.\gr{v}}}\,\qquad
v^\zT\equiv\gr{v}.\gr{e}^\zT=0\,\,.
\ee
We deduce immediately that the magnitude and the direction unit vector of the photon momentum transform as
\bea\label{Eq_Tmomentum1}
p^{\zTt}&\equiv& \gr{p}.\tilde{\gr{e}}^{\zT}=\gamma p^\zT\left(1-\gr{n}.\gr{v} \right)\,,\\*
{\tilde{\gr{n}}}&=&\frac{1}{\gamma(1-
  \gr{v}.\gr{n})}(\gr{e}_\zT+\gr{n})-\gamma(\gr{e}_\zT + \gr{v})\,.
\eea
We remind that the direction, as observed by the transformed observer, is given
by the decomposition $
\gr{p}=p^{\zTt} ({\tilde{\gr{e}}}_{\zT} +{\tilde{\gr{n}}})$.
These rules imply the following transformation rule for the screen projector
\be
\tilde S_{\mu \nu}=S_{\mu \nu}+\frac{2
  \gamma}{p^{\zTt}}p_{(\mu}S_{\nu)\rho}v^\rho+\left(\frac{\gamma}{p^{\zTt}}\right)^2
p_\mu p_\nu S_{\alpha \beta}v^\alpha v^\beta\,,
\ee
which implies directly the following useful relations
\bea\label{TruleforS}
\tilde S_{\mu \nu}=\tilde S_\mu^{\,\,\rho} \tilde S_\nu^{\,\,\sigma} S_{\rho
  \sigma}\,,\qquad \tilde n^\mu \tilde \epsilon_{\mu \rho \sigma}=n^\mu
\epsilon_{\mu \alpha \beta} \tilde S^\alpha_{\,\,\rho}\tilde S^\beta_{\,\,\sigma}\,.
\eea  
The last relation can also be demonstrated easily by noting that $n^\mu \epsilon_{\mu \alpha \beta} \tilde S^\alpha_{\,\,\rho}\tilde
S^\beta_{\,\,\sigma}$ is by construction orthogonal to $\tilde{\gr{e}}_\zT$
and ${\tilde{\gr{n}}}$, and since it is also obviously antisymmetric in its two free indices, it has to be proportional to $\tilde
  n^\mu \tilde \epsilon_{\mu \rho \sigma}$. By contracting both expressions with themselves we obtain that they are indeed equal.\\
The rest of the tetrad can be transformed without rotation, that is with a
pure boost, by 
\be
\tilde e^{\aT} = \Lambda^\aT_{\,\,\bT}e^\bT,\qquad\tilde e_\aT = e_\bT ({\Lambda^{-1}})^\bT_{\,\,\aT}=\Lambda_\aT^{\,\,\bT}e_\bT\,,
\ee
and the components of this transformation are given by
\be
\Lambda^\zT_{\,\zT}=\gamma,\qquad \Lambda^\zT_{\,\,\iT}=\Lambda^\iT_{\,\,\zT}=-\gamma v_\iT,\qquad
\Lambda^\iT_{\,\jT}=\delta^\iT_\jT+\frac{\gamma^2}{1+\gamma}v^\iT v_\jT,
\ee
where we remind that $v^\iT$ is the component of $\gr{v}$ along the tetrad
$\gr{e}^\iT$, that is $v^\iT=\gr{v}.\gr{e}^\iT $ .
The transformation rule for the photon direction when expressed along tetrads
is thus
\be
\tilde n^{\iTt}\equiv{\tilde{\gr{n}}}. {\tilde{\gr{e}}}^\iT=\frac{1}{\gamma(1-\gr{n}.\gr{v})}\left[n^\iT+\frac{\gamma^2}{(1+\gamma)}\gr{n}.\gr{v} \,v^\iT-\gamma v^\iT \right].
\ee
\subsubsection{The radiation multipoles}\label{Transfomultipoles}

It can be easily checked that for a vector orthogonal to $\gr{p}$, such as
the polarization vector, $\tilde
S^\mu_{\,\,\nu}S^\nu_{\,\,\sigma}\epsilon^\sigma=\tilde
S^\mu_{\,\,\sigma}\epsilon^\sigma$. As an immediate consequence, we deduce
from equation~(\ref{Def_buildpolartensor}) that the screen-projected polarization tensor transforms according to
\be\label{Eq_Tpropertydebase}
\tilde f_{\mu \nu}(x^\aB,p^{\zTt},\tilde n^\aTt)=\tilde S^\mu_{\,\,\alpha} \tilde S^\nu_{\,\,\beta}
f_{\alpha \beta}(x^\aB,p^{\zT},n^\aT)\,.
\ee
We deduce from equation~(\ref{TruleforS}) that $P_{\mu \nu}$ transforms following the same rule, whereas $I$
and $V$ transform as scalars 
\be
\tilde I(p^{\zTt},\tilde n^\aTt)=I(p^{\zT},n^\aT)\,,\qquad\tilde V(p^{\zTt},\tilde n^\aTt)=V(p^{\zT},n^\aT)\,.
\ee
Here and in the rest of this paper, we omit the dependence in the
position $x^\aB$ to simplify the notation.
We can deduce from equation~(\ref{Eq_Tmomentum1}) that the differential solid angle
transforms according to (this can also be deduced from using the transformation rule of $p^\zT$ and the fact that $\dd^3 p^\iT/p^\zT =p^\zT \dd p^\zT \dd^2 \Omega$ is a scalar)
\be
\dd \tilde \Omega =\left[\frac{1}{\gamma(1-\gr{v}.\gr{n})}\right]^2\dd \Omega\,,
\ee
and this can be used to deduce the transformation rules of the multipoles
\be\label{Eqinttochangeframe}
\flamoi\tilde I_{\uline{\tilde{\aT}_\ell}}(p^{\zTt})=\Delta_\ell^{-1}\int \dd \Omega
\left[\gamma(1-\gr{v}.\gr{n})\right]^{-2}\sum_{\ell'=0}^\infty I_{\uline{\bT_{\ell'}}}[p^{\zTt}\gamma^{-1}(1-n^\cT v_\cT)^{-1}]n^{\uline{\bT_{\ell'}}} \tilde n_{\langle \uline{\tilde{\aT}_\ell}\rangle}\,.
\ee
In the previous integral, $I_{\uline{\bT_{\ell'}}}[p^{\zTt}\gamma^{-1}(1-n^\cT v_\cT)^{-1}]$ has to be considered as a function of the direction $n^\aT$. It
is thus necessary to Taylor expand it as
\be
I_{\uline{\bT_\ell}}[p^{\zTt}\gamma^{-1}(1-n^\cT
v_\cT)^{-1}]=\sum_{n=0}^\infty\frac{1}{n!}\left( \frac{\gamma^{-1}n^\cT
v_\cT}{1-\gamma^{-1}n^\cT v_\cT}\right)^n I^{\{n\}}_{\uline{\bT_\ell}}[p^{\zTt}\gamma^{-1}]\,,
\ee 
where we define
\be 
I^{\{n\}}_{\uline{\bT_\ell}}(p^\zT)\equiv (p^\zT)^n\frac{\partial^n I_{\uline{\bT_\ell}}(p^\zT)}{\partial (p^\zT)^n}\,,
\ee 
with the conventions $I'_{\uline{\bT_\ell}} \equiv
I^{\{1\}}_{\uline{\bT_\ell}}$ and $I''_{\uline{\bT_\ell}} \equiv I^{\{2\}}_{\uline{\bT_\ell}}$. Under this form, it is then possible to perform the integration using the following well known integrals~\cite{Uzan1998}
\ifcqg
\be\label{Intonn}
\flamoi \int n^{\iT_1}\dots n^{\iT_k}\frac{\dd^2 \Omega}{4 \pi} = \cases{0\,, &if\;\;$k=
2p + 1$,\\\frac{1}{k+1}\left[\delta^{(\iT_1
    \iT_2}\dots\delta^{\iT_{(k-1)}\iT_k)}\right]\,, &if\;\;$k= 2p$.}
\ee
\else
\be\label{Intonn}
\flamoi \int n^{\iT_1}\dots n^{\iT_k}\frac{\dd^2 \Omega}{4 \pi} = \begin{cases}0\,, &if\;\;$k=
2p + 1$,\\\frac{1}{k+1}\left[\delta^{(\iT_1
    \iT_2}\dots\delta^{\iT_{(k-1)}\iT_k)}\right]\,, &if\;\;$k= 2p$.\end{cases}
\ee
\fi
Note that we have used the standard notation $(\dots)$ for the symmetrization
of indices which leaves unchanged an symmetric tensor and we will also use the
notation $[\dots]$ for the antisymmetrization of indices which leaves unchanged an antisymmetric tensor.
The integrals~(\ref{Intonn}) used in the transformation
rule~(\ref{Eqinttochangeframe}) are ideally suited for a tensor calculus
package, and we used \emph{xAct}~\cite{xAct} to compute them. \\
Since the result of equation~(\ref{Eqinttochangeframe}) will involve terms like $I^{\{n\}}_{\uline{\bT_\ell}}[p^{\zTt}\gamma^{-1}]=I^{\{n\}}_{\uline{\bT_\ell}}[p^\zT(1-n^\cT
v_\cT)]$, it will require an additional Taylor expansion in order to have the
result expressed only in function of the
$I^{\{n\}}_{\uline{\bT_\ell}}[p^\zTt]$ or $I^{\{n\}}_{\uline{\bT_\ell}}[p^\zT]$. 
Note that the expression of $\tilde I_{\uline{\tilde{\aT}_\ell}}(p^{\zTt})$ in
function of the $I^{\{n\}}_{\uline{\bT_\ell}}[p^\zTt]$, which is the choice
that we make in the expressions that we report below, does not depend on the direction
$n^\iT$, whereas it does when expressed in function of the
$I^{\{n\}}_{\uline{\bT_\ell}}[p^\zT]$ since $p^\zT$ is unambiguously defined
from $p^\zTt$ only once a direction $n^\iT$ is specified.  A similar method, with similar definitions
can be used to determine the transformation rules of the electric and magnetic multipoles~\cite{Tsagas2007}. At first order in the velocity $\gr{v}$, we obtain the
following transformation rules
\bea
\flamoi\tilde I_{ \uline{\aTt_\ell}}(p^{\zTt})&=&I_{\uline{\aT_\ell}}(p^{\zTt})+\frac{(\ell+2)(\ell+1)}{(2\ell+3)}v^\bT I_{\bT \uline{\aT_\ell}}(p^{\zTt})+\frac{(\ell+1)}{(2\ell+3)}v^\bT I'_{\bT \uline{\aT_\ell}}(p^{\zTt})\nonumber\\*
\flamoi&&-(\ell-1)v_{\langle \aT_\ell}I_{\uline{\aT_{\ell-1}}\rangle}(p^{\zTt}) +v_{\langle
  \aT_\ell}I'_{\uline{\aT_{\ell-1}}\rangle}(p^{\zTt})\,,
\eea

\bea
\flamoi\tilde E_{ \uline{\aTt_\ell}}(p^{\zTt})&=&E_{\uline{\aT_\ell}}(p^{\zTt})+\frac{(\ell-1)(\ell+2)(\ell+3)}{(\ell+1)(2\ell+3)}v^\bT E_{\bT \uline{\aT_\ell}}(p^{\zTt})+\frac{(\ell-1)(\ell+3)}{(\ell+1)(2\ell+3)}v^\bT E'_{\bT \uline{\aT_\ell}}(p^{\zTt})\nonumber\\*
\flamoi&&-(\ell-1)v_{\langle \aT_\ell}E_{\uline{\aT_{\ell-1}}\rangle}(p^{\zTt}) +v_{\langle
  \aT_\ell}E'_{\uline{\aT_{\ell-1}}\rangle}(p^{\zTt})\nonumber\\*
\flamoi&&-\frac{2}{(\ell+1)}v_\bT \epsilon^{\bT \cT}_{\,\,\,\,\,\langle
  \aT_\ell}B_{\uline{\aT_{\ell-1}}\rangle \cT}(p^{\zTt})-\frac{2}{(\ell+1)}v_\bT \epsilon^{\bT
  \cT}_{\,\,\,\,\,\langle \aT_\ell}B'_{\uline{\aT_{\ell-1}}\rangle \cT}(p^{\zTt})\,,
\eea

\bea
\flamoi\tilde B_{\uline{\aTt_\ell}}(p^{\zTt})&=&B_{\uline{\aT_\ell}}(p^{\zTt})+\frac{(\ell-1)(\ell+2)(\ell+3)}{(\ell+1)(2\ell+3)}v^\bT B_{\bT \uline{\aT_\ell}}(p^{\zTt})+\frac{(\ell-1)(\ell+3)}{(\ell+1)(2\ell+3)}v^\bT B'_{\bT \uline{\aT_\ell}}(p^{\zTt})\nonumber\\*
\flamoi&&-(\ell-1)v_{\langle \aT_\ell}B_{\uline{\aT_{\ell-1}}\rangle}(p^{\zTt}) +v_{\langle  \aT_\ell}B'_{\uline{\aT_{\ell-1}}\rangle}(p^{\zTt})\nonumber\\*
\flamoi&&+\frac{2}{(\ell+1)}v_\bT \epsilon^{\bT \cT}_{\,\,\,\,\,\langle \aT_\ell}E_{\uline{\aT_{\ell-1}}\rangle \cT}(p^{\zTt})+\frac{2}{(\ell+1)}v_\bT \epsilon^{\bT \cT}_{\,\,\,\,\,\langle \aT_\ell}E'_{\uline{\aT_{\ell-1}}\rangle \cT}(p^{\zTt})\,.
\eea
As for the energy-integrated counterparts, they transform at first order in $\gr{v}$
as
\be
\flamoi\tilde {\cal I}_{\uline{\aTt_\ell}}={\cal I}_{\uline{\aT_\ell}}+\frac{(\ell-2)(\ell+1)}{(2\ell+3)}v^\bT {\cal
  I}_{\bT \uline{\aT_\ell}}-(\ell+3)v_{\langle \aT_\ell}{\cal I}_{\uline{\aT_{\ell-1}}\rangle}\,,
\ee

\be
\flamoi\tilde {\cal E}_{ \uline{\aTt_\ell}}={\cal
  E}_{\uline{\aT_\ell}}-\frac{(\ell-1)(\ell-2)(\ell+3)}{(\ell+1)(2\ell+3)}v^\bT {\cal
  E}_{\bT \uline{\aT_\ell}}-(\ell+3)v_{\langle \aT_\ell}{\cal E}_{\uline{\aT_{\ell-1}}\rangle}-\frac{6}{(\ell+1)}v_\bT \epsilon^{\bT \cT}_{\,\,\,\,\,\langle
  \aT_\ell}{\cal B}_{\uline{\aT_{\ell-1}}\rangle \cT}\,,
\ee
and
\be
\flamoi\tilde {\cal B}_{\uline{\aTt_\ell}}={\cal
  B}_{\uline{\aT_\ell}}-\frac{(\ell-1)(\ell-2)(\ell+3)}{(\ell+1)(2\ell+3)}v^\bT {\cal
  B}_{\bT \uline{\aT_\ell}}-(\ell+3)v_{\langle \aT_\ell}{\cal B}_{\uline{\aT_{\ell-1}}\rangle}+\frac{6}{(\ell+1)}v_\bT \epsilon^{\bT \cT}_{\,\,\,\,\,\langle
  \aT_\ell}{\cal E}_{\uline{\aT_{\ell-1}}\rangle \cT}\,.
\ee
We report in~\ref{AppTrule2} the transformation rules up to second
order in $\gr{v}$ for multipoles of further interest.

\subsection{The Liouville equation}

In this section, we present the equation which governs the free-streaming of
photons directly in the tetrad basis though nothing prevents it from being
expressed with formal indices. The evolution of the polarization tensor is dictated by the Boltzmann equation
\be\label{EqBoltzmann}
L[f_{\aT \bT}(x^\aB,p^\zT,n^\iT)]=C_{\aT \bT}(x^\aB,p^\zT,n^\iT)\,.
\ee
$L$ is the Liouville operator whose action on TT tensors like the
screen-projected polarization tensor $f_{\aT \bT}$ is given by~\cite{Tsagas2007}
\be
\flamoi L[f_{\aT \bT}(x^\aB,p^\aT)]\equiv S_\aT^{\,\cT}
S_\bT^{\,\dT}\left[p^\hT \nabla_\hT f_{\cT
    \dT}(x^\aB,p^\aT)+ \frac{\partial f_{\cT
      \dT}(x^\aB,p^\aT)}{\partial p^\hT}\frac{\dd p^\hT}{\dd s}
\right]\,,
\ee
where $s$ is the affine parameter along the particle geodesic. $\nabla$ is the covariant derivative which in the tetrad basis is related to the partial derivative by
\be
\flamoi p^\hT \nabla_\hT f_{\cT \dT}(x^\aB,p^\aT)\equiv p^\hT \partial_\hT f_{\cT
  \dT}(x^\aB,p^\aT)+p^\hT\omega_{\hT\cT}^{\,\,\,\,\,\,\bT}f_{\bT
  \dT}(x^\aB,p^\aT)+p^\hT\omega_{\hT\dT}^{\,\,\,\,\,\,\bT}f_{\cT \bT}(x^\aB,p^\aT)\,,
\ee
where the $\omega_{\aT\bT\cT}$ are the Ricci rotation coefficients (see \ref{app_TRS}
and \cite{Wald1984} for details).
$C_{\aT \bT}$ is the collision tensor whose expression will be detailed in section~\ref{Sec_Collision}.\\
In the case where the collision tensor can be ignored, that is when the collision of photons with electrons or protons can be
neglected [the latter type of collision can be always ignored compared to the
former since its cross section is reduced by a factor $(m_\ie/m_\ip)^2$], this reduces to the Liouville
equation. The Liouville equation arises from the fact that the polarization
vector $\mybepsilon$ of a photon is parallel transported and thus satisfies
$p^\mu \nabla_\mu \epsilon^\nu=0$, and then it follows from the construction~(\ref{Def_buildpolartensor}) of the
(not screen projected) polarization tensor $F_{\mu \nu}$ that 
\be
p^\hT \nabla_\hT F_{\cT \dT}(x^\aB,p^\aT)+ \frac{\partial F_{\cT
      \dT}(x^\aB,p^\aT)}{\partial p^\hT}\frac{\dd p^\hT}{\dd s}=0\,.
\ee
By using the expression~(\ref{DefScreen}) for $S^{\mu \nu}$ and the property $F_{\mu
  \nu}(x^\aB,p^\aT)p^\mu=0$, we obtain directly from the previous equation
that the screen-projected tensor satisfies $L[f_{\aT \bT}(x^\aB,p^\aT)]=0$.
It can also be shown that the Liouville operator preserves the decomposition
of $f_{\mu \nu}$ in an antisymmetric part ($V$), a trace ($I$) and a symmetric
traceless part ($P_{\mu \nu}$), that is
\be
\flamoi L[f_{\aT \bT}(x^\aB,p^\aT)]=\frac{1}{2}L[I(x^\aB,p^\dT)]S_{\aT \bT}+L[P_{\aT
  \bT}(x^\aB,p^\aT)]+\frac{\ii}{2}L[V(x^\aB,p^\dT)]n^\cT \epsilon_{\cT \aT \bT}\,,
\ee 
with the Liouville operator acting on a scalar valued function like $I$ or $V$
according to
\be
L[I(x^\aB,p^\aT)]\equiv p^\hT \partial_\hT I(x^\aB,p^\aT)+ \frac{\partial I(x^\aB,p^\aT)}{\partial p^\hT}\frac{\dd p^\hT}{\dd s}\,\,.
\ee 
To see this, we need only to use the
property $\omega_{\aT [\bT \cT]}=\omega_{\aT \bT \cT}$ and remark that 
\be
S^\cT_{(\aT}S_{\,\,\bT)}^{\dT} p^\hT \omega_{\hT\cT \fT}S^\fT_{\,\dT}= p^\hT \omega_{\hT\cT \fT}S^\cT_{(\aT}S^\fT_{\,\bT)}=0\,, 
\ee 
\be
S^\cT_{[\aT}S_{\,\,\bT]}^{\dT} p^\hT \omega_{\hT\cT \fT} \epsilon^{\fT}_{\,\,\dT}\propto S^\cT_{[\aT}S_{\,\,\bT]}^{\dT}
\epsilon_{\cT \fT} \epsilon^{\fT}_{\,\,\dT}=\epsilon_{\fT[\aT} \epsilon^{\,\,\fT}_{\bT]}=0\,,
\ee 
where we have used the definition $\epsilon_{\mu \nu} \equiv n^\alpha
\epsilon_{\alpha \mu\nu}$.\\
Additionally, since the affine parameter $s$ is a scalar, the transformation
properties of the Liouville operator and the collision tensor under a local change of frame is the same as the transformation
property of $f_{\mu \nu}$ given in equation~(\ref{Eq_Tpropertydebase})~\cite{Tsagas2007}.

\section{Gauge transformations and gauge invariance for tensors}\label{SecGIT}

The formalism presented in the previous section is very general and can be
applied to the description of matter and radiation in any type of space-time.
In the cosmological context it proves useful to use the high symmetries of the
large scale space-time to find the solutions of the Einstein and Boltzmann
equations using a perturbation scheme. We will review the standard
perturbation theory for tensors in this section as well as for a scalar
valued distribution function in section~\ref{SecGIf} and extend it to the TV
distribution function.

\subsection{First- and second-order perturbations}\label{sec1part1}

We assume that, at lowest order, the universe is well described by a
Friedmann-Lema\^{\i}tre space-time (FL) with Euclidian spatial sections. The most
general form of the metric for an almost FL universe is
\begin{eqnarray}\label{metric}
\flamoi \dd s^2 &=& g_{\mu\nu}\dd x^\mu\dd x^\nu \\*
\flamoi &=& a(\eta)^2 \big\{-(1 + 2\Phi )\dd\eta^2 + 2
 \omega_{\iB} \dd x^{\iB}\dd\eta + [(1-2 \Psi)\delta_{\iB\jB} + h_{\iB\jB}]\dd x^{\iB}\dd x^{\jB}\big\},\nonumber
\end{eqnarray}
where $\eta$ is the conformal time for which the corresponding index is $\zB$,
and $a(\eta)$ is the scale factor. We perform a scalar-vector-tensor (SVT) decomposition as
\begin{equation}\label{eq:I-2}
 \omega_{\iB}=\partial_{\iB} B + B_{\iB}\,,
\end{equation}
\begin{equation}
 h_{\iB \jB}=2 H_{\iB\jB} + \partial_{\iB}E_\jB + \partial_\jB E_\iB + 2 \partial_{\iB} \partial_{\jB} E,
\end{equation}\label{eq:I-3}
where $B_\iB$, $E_\iB$ and $H_{\iB \jB}$ are transverse ($\partial^\iB
E_\iB= \partial^\iB B_\iB=\partial^\iB H_{\iB \jB}=0$), and $H_{\iB \jB}$
is traceless ($H^\iB_{\,\iB}=0$). There are four scalar degrees of freedom
($\Phi,\,\Psi,\,B,\,E$), four vector degrees of freedom ($B_\iB,\,E_\iB$) and two
tensor degrees of freedom ($H_{\iB\jB}$). As we shall see in
section~\ref{Sec_GaugetransfoT}, the perturbation variables live on the background space-time and
thus their indices are lowered and raised by the background (conformally
transformed) spatial metric and its inverse, that is with $\delta^{\iB \jB}$ and
$\delta_{\iB \jB}$.
Each of these perturbation variables can be split in first and second-order parts as
\begin{equation}\label{decomposition-ordre2}
 W=W^{(1)}+\frac{1}{2}W^{(2)}\,.
\end{equation}
This expansion scheme will refer, as we shall see, to the way gauge transformations and gauge-invariant (GI) variables are defined. First-order variables are solutions of first-order
equations which have been extensively studied (see~\cite{PeterUzan} for a
review). Second-order equations will involve purely second-order terms, e.g.
$W^{(2)}$ and terms quadratic in the first-order variables, e.g.
$[W^{(1)}]^2$. There will thus never be any ambiguity about the order of perturbation variables involved as long as the
order of the equation considered is known. Consequently, we will often omit to specify the order
superscript when there is no risk of confusion.

As we shall see in section~\ref{Sec_GaugetransfoT}, 4 of the 10 metric perturbations are gauge degrees of freedom and
the 6 remaining degrees of freedom reduce to 2 scalars, 2 vectors and 2 tensors. The three types of
perturbations decouple at first order and can thus be treated separately. As long as no vector
source terms are present, which is generally the case when no magnetic
field or topological defect is taken into account, the first order vector modes decay as
$a^{-2}$. Thus, we can safely discard them and set $E^{(1)}_\iB=B^{(1)}_\iB=0$. In
the following of this work, we shall not include first-order vector modes for the sake of
clarity. We checked that our arguments and derivation can trivially (but
at the expense of much lengthy expressions) take them into account. 

In the fluid description, the four-velocity of each fluid is decomposed as
\begin{equation}
 u^{\aB}=\frac{1}{a}\left(\delta^{\aB}_\zB+V^\aB\right)\,.
\label{defvelocity}
\end{equation}
The indices of $V^\aB$ are raised and lowered with the (conformally
transformed) background metric, that is with $\eta_{\aB \bB}$ and $\eta^{\aB \bB}$.
The perturbation $V^\aB$ has only three independent degrees of freedom since 
$\gr{u}$ must satisfy $u_{\mu}u^{\mu}=-1$. The spatial components can be decomposed as
\begin{equation}\label{decv}
 V^{\iB}=\partial^{\iB}V + {\tilde V}^{\iB}\,,
\end{equation}
$\tilde V^{\iB}$ being the vector degree of freedom ($\partial_\iB \tilde V^\iB=0$), and $V$ the scalar degree of
freedom. As for the energy density and pressure, similarly to any quantity which
does not vanish in the background space-time, they are decomposed according
to
\begin{equation}
 \rho = \bar\rho + \rho^{(1)}+\frac{1}{2}\rho^{(2)}+\dots,\qquad
 P=\bar P + P^{(1)}+\frac{1}{2}P^{(2)}+\dots\,.\label{defdeltamatter}
\end{equation}
It is also common practice to define a conformally transformed anisotropic
stress by
\be
\pi^{\cB\dB}=\frac{1}{a^2}\Pi^{\cB\dB}\,,\qquad\pi^{\cB}_{\,\dB}=\Pi^{\cB}_{\,\dB}\,,\qquad\pi_{\cB\dB}=a^2\Pi_{\cB\dB}\,.
\ee
It should be mentionned that due to the symmetries of the background
space-time, the anisotropic stress is already a perturbed quantity. 
At first order, the formalism developed by the seminal work of~\cite{Bardeen1980} provides a full set of gauge-invariant variables. Thanks to
the general covariance of the equations at hand (Einstein equations,
conservation equations, Boltzmann equation), it was shown that it was possible to get first-order equations involving only these
gauge-invariant variables. In addition, if these gauge invariant
variables reduce, in a particular gauge, to the perturbation variables that we
use in this particular gauge, then the computation of the equation can be
simplified. Actually, we only need to derive the equations in this particular
gauge, as long as it is completely fixed, and then to promote by identification our perturbation variables to the
gauge-invariant variables. Thus, provided we know this full set of gauge
invariant variables, the apparent loss of generality by fixing the
gauge in a calculation, is in fact just a way to simplify computations.
Eventually we will reinterpret the equations as being satisfied by gauge
invariant variables. The full set of first-order gauge-invariant variables is
well known and is reviewed in~\cite{PeterUzan} and~\cite{KodamaSasaki}. As gauge
transformations up to any order were developed, it remained uncertain
\cite{Bruni1997}, whether or not a full set of gauge-invariant variables
could be built for second and higher orders. This has been recently clarified
\cite{Nakamura2007}, and the autosimilarity of the transformation rules for
different orders can be used as a guide to build the gauge-invariant variables
at any order. We present a summary of the ideas presented in~\cite{Bruni1997} about gauge transformations and the construction of
gauge-invariant variables \cite{Nakamura2007} in a shorter version than in
our previous paper~\cite{Pitrou2007}. A summary emphasizing the differences
between the active and passive point of views can also  be found in~\cite{Malik2008}.

\subsection{Points identification on manifolds}

When working with perturbations, we consider two manifolds: a background
manifold, $\mf{M}_0$, with associated metric $\bar{g}$, which in our case is
the FL space-time, and the physical space-time $\mf{M}_1$ with the metric $g$.
Considering the variation of metric boils down to a comparison between tensor
fields on distinct manifolds. Thus, in order to give a sense to ``$g^{(1)}(P)= g(P) - \bar{g}(\bar{P})$'', we need to identify the points $P$ and $\bar{P}$ between these two
manifolds and also to set up a procedure for comparing tensors. This will also
be necessary for the comparison of any tensor field.

One solution to this problem \cite{Bruni1997} is to consider an embedding
$4+1$ dimensional manifold $\mf{N} = \mf{M} \times [0,1]$, endowed with the
trivial differential structure induced, and the projections
$\mathcal{P_{\lambda}}$ on submanifolds  with $\mathcal{P}_0(\mf{N})=\mf{M}
\times \{0\}={\cal M}_0$ and $\mathcal{P}_1(\mf{N})=\mf{M} \times \{1\}={\cal M}_1$. The collection of ${\cal M}_{\lambda}\equiv{\cal
  P}_{\lambda}({\cal N})$ is a foliation of ${\cal N}$, and each element is
diffeomorphic to the physical space-time ${\cal M}_1$ and the background space-time ${\cal M}_0$. The gauge choice on this stack of space-times is defined as a vector field $X$ on $\mf{N}$ which satisfies $X^4=1$ (the component along the space-time slicing $\mathds{R}$). 
A vector field defines integral curves that are always tangent to the vector field
itself, hence inducing a one parameter group of diffeomorphisms
$\phi(\lambda,.)$, also noted $\phi_{\lambda}(.)$, a flow, leading in our case from $\phi(0,p \in
\mathcal{P}_0(\mf{N}))= p \in \mathcal{P}_0(\mf{N})$ along the integral curves
to  $\phi\left(1,p \in \mathcal{P}_0(\mf{N})\right)= q \in
\mathcal{P}_1(\mf{N})$. Due to the never vanishing last component of $X$,
the integral curves will always be transverse to the stack of space-times and
the points lying on the same integral curve, belonging to distinct
space-times, will be identified. Additionally  the property $X^4=1$ ensures
that $\phi_{\lambda,X}(\mathcal{P}_0(\mf{N}))= \mathcal{P}_{\lambda}(\mf{N})$,
i.e. the flow carries a space-time slice to another. This points
identification is necessary when comparing tensors, but we already see that
the arbitrariness in the choice of a gauge vector field $X$ should not have
physical meaning, and this is the well known {\it gauge freedom}.   

\subsection{Gauge transformations and gauge invariance}\label{Sec_GaugetransfoT}

The induced transport, along the flow, of tensors living on the tangent
bundle, is determined by the push-forward $\phi_{\star \lambda}$ and the
pull-back $\phi^{\star}_{\lambda}$ \cite{Wald1984,Malik2008} associated with an element
$\phi_{\lambda}$ of the group of diffeomorphisms. These two functions encapsulate the transformation properties of the tangent and co-tangent spaces at each point and its image. Indeed, the pull-back can be linked to the local differential properties of the vector field embedded by the Lie derivatives along the vector field in a Taylor-like fashion (see~\cite{Wald1984} or~\cite{Bruni1997})
\begin{equation}
\phi^{\star}_{X,\lambda}(T)= \sum_{k=0}^{k=\infty} \frac{\lambda^k}{k!}\Lie_X^k T,
\label{expansion-pullback}
\end{equation}  
for any tensor $T$.
The expansion of equation~(\ref{expansion-pullback}) on $\mathcal{P}_0(\mf{N})$ provides a way to compare a tensor field on $\mathcal{P}_{\lambda}(\mf{N})$ to the corresponding one on the background space-time $\mathcal{P}_{0}(\mf{N})$. The background value being $T_0 \equiv \Lie_X^0 T|_{\mathcal{P}_0(\mf{N})}$, we obtain a natural definition for the tensor perturbation 
\begin{equation}
\Delta_X T_\lambda \equiv \sum_{k=1}^{k=\infty} \frac{\lambda^k}{k!}\Lie_X^k T  \Big|_{\mathcal{P}_0(\mf{N})}= \phi^{\star}_{X,\lambda}(T) - T_0.
\label{perturbation-tenseur}
\end{equation} 
The subscript $X$ reminds the gauge dependence. We can read the $n$-th order
perturbation as
\begin{equation}
{}_X \! T^{(n)} \equiv \Lie_X^n T \Big|_{\mathcal{P}_0(\mf{N})}\,,
\end{equation}
which is consistent with the expansion of perturbation variables of the physical metric in equation~(\ref{decomposition-ordre2}), since the physical space-time is labeled by $\lambda=1$. However, the fact that the intermediate space-time slices ${\cal P}_{\lambda}({\cal N})$ are labeled by $\lambda$ removes the absolute meaning of order by order perturbations, as it can be seen from equation~(\ref{perturbation-tenseur}). The entire structure embedded by $\mf{N}$ is more than just a convenient construction and this will have important consequences in gauge changes as we will now detail.

If we consider two gauge choices $X$ and $Y$, a gauge transformation from $X$ to $Y$ is defined as the diffeomorphism 
\begin{equation}
\phi_{X \rightarrow Y, \lambda} = (\phi_{X,\lambda})^{-1}(\phi_{Y,\lambda}),
\end{equation}
and it induces a pull-back which carries the tensor $\Delta_X T_{\lambda}$, which is the perturbation in the gauge $X$, to $\Delta_Y T_{\lambda}$, which is the perturbation in gauge $Y$.
As demonstrated in~\cite{Bruni1997,Enqvist2007} this family (indexed by $\lambda$)
of gauge transformations fails to be a one parameter group due to the lack of the composition rule. It should be Taylor expanded using the so called knight-diffeormorphism along a sequence of vector fields $\xi_n$. For the two first orders, the expression of this knight-diffeomorphism is 
\begin{eqnarray}
\phi^{\star}_{Y,\lambda}(T) &=& \phi^{\star}_{X \rightarrow Y, \lambda} \phi^{\star}_{X,\lambda}(T)  \\*\label{knight}
&=&   \phi^{\star}_{X,\lambda}(T) + \lambda \Lie_{\xi_1}  \phi^{\star}_{X,\lambda}(T)  + \frac{\lambda^2}{2!} (\Lie_{\xi_2} +\Lie_{\xi_1}^2) \phi^{\star}_{X,\lambda}(T)+\dots \nonumber
\end{eqnarray}
The vector fields $\xi_1$, $\xi_2$ are related to the gauge vector fields $X$
and $Y$ by $\xi_1= Y-X$ and $\xi_2=[X,Y]$. By substitution of the perturbation by its expression in equation~(\ref{perturbation-tenseur}), we identify order by order in $\lambda$, and obtain the transformation rules for perturbations order by order. The first and second order transformation rules, on which we will focus our attention, are
\begin{eqnarray}
{}_Y \!T^{(1)} -{}_X \! T^{(1)} &=& \Lie_{\xi_1} T_0, \nonumber\\*
{}_Y\! T^{(2)} -{}_X \! T^{(2)} &=& 2 \Lie_{\xi_1} \,{}_X \!T^{(1)} + ( \Lie_{\xi_2}+ \Lie_{\xi_1}^2)T_0\,.\label{transforule}
\end{eqnarray}

General covariance, i.e. the fact that physics should not depend on a
particular choice of coordinates is an incentive to work with gauge-invariant quantities. As we notice from equation~(\ref{transforule}), a tensor $T$ is gauge-invariant up to $n$-th order if it satisfies $\Lie_{\xi}\,{}_X \! T^{(r)}=0$ for any vector field $\xi$ and any $r \leq n$, as can be deduced by recursion. A consequence of this strong condition is that a tensor is gauge-invariant up to order $n$ if and only if $T_0$ and all its perturbations of order lower than $n$ either vanish, or are constant scalars, or are combinations of Kronecker deltas with constant coefficients. Einstein equation is of the form $G-T=0$, and for this reason is totally gauge invariant. However, we cannot find non-trivial tensorial quantities (that is, different from $G-T$) gauge-invariant up to the order we intend to study perturbations, with which we could express the perturbed set of Einstein equations.

Consequently, we will build, by combinations of perturbed tensorial quantities, gauge-invariant variables. These combinations will not be the perturbation of an underlying tensor. This method will prove to be very conclusive since a general procedure exists for perturbations around FL. Eventually we shall identify observables among these gauge-invariant variables and the fact that they are not the perturbation of a tensor will not matter. It has to be emphasized that the transformation rules of these combinations are not intrinsic and cannot be deduced directly from the knight-diffeomorphism since they are not tensorial quantities. Instead, we have to form the combination before and after the gauge change in order to deduce their transformation rules.

We now summarize the standard way to build gauge-invariant variables up to
second order. For simplicity we will not consider the vector part of the gauge transformations at
first order, since we will not consider first order vector modes in the metric and fluid perturbation
variables (again, this could be done, but would just obfuscate the
explanations). In the following, we split $\xi^{\mu}_1$ and $\xi^{\mu}_2$ as
\begin{equation}
{\xi_1}^\zB=T^{(1)},\,\,\,\,\,\xi^{\iB}_1=\partial^{\iB}
L^{(1)},\quad{\xi_2}^\zB=T^{(2)},\,\,\,\,\,\xi^{\iB}_2=\partial^{\iB}
L^{(2)}+L^{\iB(2)}\,,
\end{equation}
with $\partial_\iB L^{\iB(2)}=0$. We conventionally choose to lower and raise
the indices of $\partial^\iB L + L^\iB$ with the (conformally transformed)
background metric, which implies that
\be
{\xi_1}_{\iB}=a^2 \partial_{\iB}
L^{(1)}\,,\qquad{\xi_2}_{\iB}=a^2 \partial_{\iB}
L^{(2)}+a^2 L_{\iB}^{(2)}\,.
\ee

\subsection{First-order gauge-invariant variables}

In the subsequent work we present the transformation rules of perturbed
quantities in a simplified notation. Instead of writing
$_Y\!W^{(r)}=\,_X\!W^{(r)}+f\left(\xi_{1},..,\xi_{r}\right)$, in order to state that
the difference between the expression of the $r$-th order pertubed variable
$W$ in gauge $Y$ and in gauge $X$ is a function $f$ of the
knight-diffeomorphism fields $\xi_{1},...,\xi_{r}$, we prefer to write
$W^{(r)} \rightarrow W^{(r)} + f\left(\xi_{1},..,\xi_{r}\right)$. We remind
that the expressions of the fields $\left(\xi_{n}\right)_{1\leq n\leq r}$
necessary for the knight-diffeomorphism are expressed in function of the gauge
fields $X$ and $Y$ [see below equation~(\ref{knight})]. From the transformation rules~(\ref{transforule}) we deduce that the first-order perturbations of the metric tensor (\ref{metric}) transform as
\begin{eqnarray}
\Phi^{(1)} & \rightarrow & \Phi^{(1)} + T'^{(1)} + \HH T^{(1)}\\*
B^{(1)} & \rightarrow & B^{(1)} -T^{(1)} + L'^{(1)}\\*
\Psi^{(1)} & \rightarrow & \Psi^{(1)} -\HH T^{(1)}\\*
E^{(1)} & \rightarrow & E^{(1)} + L^{(1)}\\*
H^{(1)}_{\iB \jB} & \rightarrow & H^{(1)}_{\iB \jB}\,,
\end{eqnarray}
while the quantities related to matter transform as 
\begin{eqnarray}\label{Tfluide1}
\rho^{(1)} & \rightarrow & \rho^{(1)} + \bar{\rho}'T^{(1)} \nonumber\\*
P^{(1)} & \rightarrow & P^{(1)} + \bar{P}'T^{(1)} \nonumber\\*
V^{(1)} & \rightarrow & V^{(1)} - L'^{(1)}\\*
\pi^{\iB \jB(1)} & \rightarrow& \pi^{\iB \jB(1)}\,,
\end{eqnarray}
where a prime denotes a derivative w.r.t. to conformal time $\eta$, and where $\HH \equiv a'/a$.
From now on, we shall refer to these first-order transformation rules defined by $\xi_1$ as $\Tr_{\xi_1}(\Phi^{(1)}),\Tr_{\xi_1}(B^{(1)}),...$ or simply $\Tr(\Phi^{(1)}),\Tr(B^{(1)}),...$ For instance $\Tr(\Phi^{(1)})= \Phi^{(1)} +  T'^{(1)}+ \HH T^{(1)}$.\\

We first note that the first-order tensorial modes and the first-order
anisotropic stress are automatically gauge
invariant. We can define gauge invariant variables, by transforming $\Phi^{(1)}$ and $\Psi^{(1)}$ towards the Newtonian gauge
(NG) \cite{Mukhanov1992}. This transformation is defined by the vector field
$\xi^{(1)}_{\rightarrow \NG}$ decomposed in $T^{(1)}_{\rightarrow \NG}=
B^{(1)}-E'^{(1)} ,\,\,\, L^{(1)}_{\rightarrow \NG}=-E^{(1)} $, and it transforms the perturbation variables as
\begin{eqnarray}\label{transfo_order_1}
\flamoi B^{(1)} & \rightarrow & 0 \\*
\flamoi  E^{(1)} & \rightarrow & 0 \\*
\flamoi \Phi^{(1)} & \rightarrow & \hat{\Phi}^{(1)} \equiv \,{}_{_{\NG}}\!\Phi^{(1)}=\Phi^{(1)} + \HH\left(B^{(1)}-E'^{(1)}\right)+\left(B^{(1)}-E'^{(1)}\right)' \\*
\flamoi \Psi^{(1)} & \rightarrow & \hat{\Psi}^{(1)} \equiv \,{}_{_{\NG}}\!\Psi^{(1)}=\Psi^{(1)}- \HH\left(B^{(1)}-E'^{(1)}\right).
\end{eqnarray}
Similarly the gauge-invariant variables that would reduce to $\delta \rho$, $\delta P$  and $v$ are
\begin{eqnarray}\label{deffluideGI1}
\hat{\rho}^{(1)}& \equiv &\,_{_{\NG}} \rho^{(1)}= \rho^{(1)} + \bar{\rho}' \left(B^{(1)}-E'^{(1)}\right)\nonumber\\*
\delta^{(1)} \hat{P}& \equiv &\,_{_{\NG}} P^{(1)}= P^{(1)} + \bar{P}' \left(B^{(1)}-E'^{(1)}\right)\nonumber\\*
\hat{V}^{(1)}& \equiv &\,_{_{\NG}}V^{(1)}= V^{(1)} + E'^{(1)}\nonumber\\*
\hat{\pi}^{\iB \jB(1)} & \equiv & \,_{_{\NG}} \pi^{\iB \jB(1)} = \pi^{\iB \jB(1)}.
\end{eqnarray}

Since we have ignored the vector gauge degrees of freedom, $B^{(1)}$ and
$E^{(1)}$ are the two gauge variant variables of the metric perturbation while
$\hat{\Phi}^{(1)}$ and $\hat{\Psi}^{(1)}$ are the gauge-invariant part. As
mentionned before, we then force the gauge-invariant variables in the
perturbed metric by replacing $\Phi^{(1)}$ with $\hat{\Phi}^{(1)} -
\HH\left(B^{(1)}-E'^{(1)}\right)+\left(B^{(1)}-E'^{(1)}\right)'$ and applying
similar procedures for $\Psi^{(1)}$, $\rho^{(1)}$ and $P^{(1)}$. 
When developping Einstein equations, we know that general covariance will
eventually keep only gauge-invariant terms. 
Thus, we can either do a full calculation and witness the terms involving the degrees of freedom $B^{(1)}$ and $E^{(1)}$ disappear, or perform the calculations with $B^{(1)}$ and $E^{(1)}$ set to zero and obtain the perturbed Einstein equations only in function of gauge-invariant variables. The latter simplifies the computation, which is useful when going to higher orders. 
This procedure means that we decompose the perturbed metric in a gauge-invariant part and a gauge variant part as

\begin{equation}
g^{(1)} \equiv \hat{g}^{(1)} + \Lie_{-\xi^{(1)}_{\rightarrow \NG}} \bar{g},
\end{equation}
as it can be seen from the transformation rules under a gauge change characterised by $\xi_1$
\begin{eqnarray}
\hat{g}^{(1)} &\rightarrow&  \hat{g}^{(1)},\nonumber\\*
-\xi^{(1)}_{\rightarrow \NG} &\rightarrow& -\xi^{(1)}_{\rightarrow \NG} + \xi_1\,,
\end{eqnarray}
and that eventually, only $\hat{g}^{(1)}$ will appear in the equations.
This property which is not general but happens to hold in the case of cosmological perturbation (i.e. around  FL metric) is the key to extend this construction to second order.

It should be noted that this procedure, although achieved by defining gauge
invariant variables which reduce to the perturbation variables in the
Newtonian gauge, can be extended to other types of gauge-invariant variables
which reduce to perturbation variables in another gauge.

\subsection{Second-order gauge-invariant variables}\label{defGIV}

For second-order perturbations, equation~(\ref{transforule}) gives the following transformation rules
\begin{eqnarray}\label{transfo_order_2}
\Phi^{(2)} & \rightarrow & \Phi^{(2)} + T'^{(2)} + \HH T^{(2)} + S_{\Phi}\nonumber\\*
B^{(2)} & \rightarrow & B^{(2)} -T^{(2)} + L'^{(2)} + S_{B}\nonumber\\*
\Psi^{(2)} & \rightarrow & \Psi^{(2)} -\HH T^{(2)} + S_{\Psi}\nonumber\\*
E^{(2)} & \rightarrow & E^{(2)} + L^{(2)} + S_{E}\nonumber\\*
B_{\iB}^{(2)} &\rightarrow& B_{\iB}^{(2)}+L_{\iB}^{'(2)}+ S_{B_\iB}\,\nonumber\\*
E_{\iB}^{(2)} &\rightarrow& E_{\iB}^{(2)}+L_{\iB}^{(2)}+ S_{E_\iB}\,\nonumber\\*
H^{(2)}_{\iB \jB} & \rightarrow & H^{(2)}_{\iB \jB}+ {S_{H_{\iB \jB}}}\nonumber\\*
\rho^{(2)} & \rightarrow & \rho^{(2)} + \bar{\rho}'T^{(2)} + S_{\rho}\nonumber\\*
P^{(2)} & \rightarrow & P^{(2)} + \bar{P}'T^{(2)} + S_{P}\nonumber\\*
V^{(2)} & \rightarrow & V^{(2)} -L'^{(2)} + S_{V}\nonumber\\*
\tilde{V}^{\iB(2)} & \rightarrow & \tilde{V}^{\iB(2)} -L'^{\iB(2)} + S_{\tilde{V}^\iB}\nonumber\\*
\pi^{\iB \jB(2)} & \rightarrow& \pi^{\iB \jB(2)} + S_{\pi^{\iB \jB}},
\end{eqnarray}
where the source terms are quadratic in the first-order gauge transformation variables
$T^{(1)}$,$L^{(1)}$, and the first order metric perturbations $\Phi^{(1)}$,$\Psi^{(1)}$,$B^{(1)}$,$E^{(1)}$ and $E_{\iB\jB}^{(1)}$. We collect the expressions of these terms in~\ref{app_sources}.
In the rest of this paper, we shall refer to these second-order transformation rules associated with $(\xi)\equiv (\xi_1,\xi_2)$ as $\Tr_{(\xi)}(\Phi^{(2)}),\Tr_{(\xi)}(B^{(2)}),...$ or simply $\Tr(\Phi^{(2)}),\Tr(B^{(2)}),...$.
These transformation rules are much more complicated than their first-order
counterparts. However, the combination defined by $F \equiv g^{(2)} + 2
\Lie_{\xi^{(1)}_{\rightarrow \NG}} g^{(1)} + \Lie_{\xi^{(1)}_{\rightarrow
    \NG}}^2 \bar{g} $ enjoys the simple transformation rule $F \rightarrow  F +
\Lie_{\xi_2 + [\xi^{(1)}_{\rightarrow \NG}, \xi_1 ]} \bar{g}$ under a gauge change
defined by $\xi_2$ and $\xi_1$ (see~\cite{Nakamura2007}). As a result,
its transformation rule mimics the one of first-order pertubations under a
gauge change. This means that if we decompose $F$ in the same way as we did
for the metric with 
\begin{eqnarray}
\Phi_F &\equiv&\Phi^{(2)}+S_{\Phi}(\xi^{(1)}_{\rightarrow \NG})\nonumber\\* 
\Psi_F &\equiv& \Psi^{(2)}+S_{\Psi}(\xi^{(1)}_{\rightarrow \NG}) \nonumber\\* 
B_F &\equiv& B^{(2)}+S_{B}(\xi^{(1)}_{\rightarrow \NG}) \nonumber\\* 
E_F &\equiv& E^{(2)}+S_{E}(\xi^{(1)}_{\rightarrow \NG})\nonumber\\*  
B_{F_\iB} &\equiv& B^{(2)}_{\iB}+S_{B_\iB}(\xi^{(1)}_{\rightarrow \NG})\nonumber\\*
E_{F_\iB} &\equiv& E^{(2)}_{\iB}+S_{E_\iB}(\xi^{(1)}_{\rightarrow \NG})\nonumber\\*
H_{F_{\iB \jB}} &\equiv& H^{(2)}_{\iB \jB}+S_{H_{\iB \jB}}(\xi^{(1)}_{\rightarrow \NG}),  
\end{eqnarray}
then the transformation rules for these quantities will be similar to those of
equation~(\ref{transfo_order_1}), but with the vector $\xi_2 + [\xi_{\rightarrow
  \NG}, \xi_1 ]$ instead of $\xi_1$. Consequently, we shall use the same
combinations (taking into account the vector contribution at second order since it is does
not vanish at this order) in order to construct gauge-invariant variables which are
\begin{eqnarray}
\hat{\Phi}^{(2)} &\equiv& \Phi_F + \left(B_F-E_F'\right)' + \HH\left(B_F-E_F'\right)\nonumber\\*
\hat{\Psi}^{(2)} &\equiv& \Psi_F -\HH\left(B_F-E_F'\right)\nonumber\\*
\hat{\Phi}_{\iB}^{(2)} & \equiv & B_{F_\iB}-E'_{F_\iB}\nonumber\\*
\hat{H}_{\iB \jB}^{(2)} & \equiv & H_{F_{\iB \jB}}.
\end{eqnarray}
This procedure is equivalent to transforming quantities in the Newtonian
gauge since it transforms $B$, $E$ and $E^\iB$ into a null value up to second order. This transformation is defined by
$\xi^{(2)}_{\rightarrow \NG}$ that we decompose in 
\begin{eqnarray}
T^{(2)}_{\rightarrow  \NG}&=&B^{(2)}-E^{'(2)} +
S_{B}\left(\xi^{(1)}_{\rightarrow
    \NG}\right)-S_{E}^{'}\left(\xi^{(1)}_{\rightarrow  \NG}\right)\nonumber\\*
L^{(2)}_{\rightarrow \NG}&=&-E^{(2)} -S_{E}\left(\xi^{(1)}_{\rightarrow
    \NG}\right)\nonumber\\*
L^{\iB(2)}_{\rightarrow \NG}&=&-E^{\iB (2)} -S_{E^\iB}\left(\xi^{(1)}_{\rightarrow  \NG}\right)\,.
\end{eqnarray}
The second-order gauge-invariant variables can thus also be defined by
\begin{eqnarray}
\hat{\Phi}^{(2)} &\equiv& \,_{_{\NG}} \Phi^{(2)}\nonumber\\*
\hat{\Psi}^{(2)} &\equiv& \,_{_{\NG}}\Psi^{(2)}\nonumber\\*
\hat{\Phi}^{(2)}_{\iB} & \equiv & \,_{_{\NG}} E^{(2)}_{\iB}\nonumber\\*
\hat{H}^{(2)}_{\iB \jB} & \equiv & \,_{_{\NG}} H^{(2)}_{\iB \jB}
\end{eqnarray} 
\begin{eqnarray}\label{deffluideGI2}
\hat{\rho}^{(2)}&\equiv&\,_{_{\NG}} \rho^{(2)}\nonumber\\*
\hat{P}^{(2)}&\equiv& \,_{_{\NG}}P^{(2)}\nonumber\\*
\hat{V}^{(2)}& \equiv& \,_{_{\NG}}V^{(2)}\nonumber\\*
\hat{\tilde{V}}^{\iB(2)}& \equiv& \,_{_{\NG}}\tilde{V}^{\iB(2)}\nonumber\\*
\hat{\pi}^{\iB \jB(2)} & \equiv & \,_{_{\NG}} \pi^{\iB \jB(2)}.
\end{eqnarray}
where the index $\NG$ indicates that we transformed the quantity with the
formula (\ref{transforule}), with the vector fields $\xi^{(1)}_{\rightarrow
  \NG}$ and $\xi^{(2)}_{\rightarrow \NG}$ defined above. This procedure means that we have split the second-order metric according to
\begin{equation}
g^{(2)} = \tilde{g}^{(2)} + \Lie_{-\xi^{(2)}_{\rightarrow \NG}} \bar{g}+2\Lie_{-\xi^{(1)}_{\rightarrow \NG}} g^{(1)}-\Lie^{2}_{-\xi^{(1)}_{\rightarrow \NG}} \bar{g}\,,
\end{equation}
where $\tilde{g}^{(2)}$  is the gauge-invariant part and
$-\xi^{(2)}_{\rightarrow \NG}$ the gauge variant part, as it can be seen from the transformation rules under a gauge change characterised by $(\xi_1,\xi_2)$
\begin{eqnarray}
\tilde{g}^{(2)} &\rightarrow&  \tilde{g}^{(2)},\nonumber\\*
-\xi^{(2)}_{\rightarrow \NG} &\rightarrow& -\xi^{(2)}_{\rightarrow \NG} + \xi_2 + [\xi^{(1)}_{\rightarrow \NG}, \xi_1 ].
\end{eqnarray}

\section{Gauge transformations and gauge invariance for a distribution function}\label{SecGIf}

\subsection{pre-Riemannian distribution function in the unpolarized case}

So far, we have set up the mathematical framework to identify points between
the background space-time and the perturbed space-times through a gauge
field $X$. This enabled us to define the perturbation of tensors and to
calculate their transformation properties under a gauge transformation.
However this allows only to perform a fluid treatment of the radiation. In
the statistical description for a set of particles, we assume that each
particle has a given impulsion $p^{\mu}$ and is located at a given position
\cite{Hillery1983}. The equations then have to describe the phase space distribution of the particles. If the number of particles is high enough, we can define a
probability density, the one-particle distribution function, of finding a particle in an
infinitesimal volume of the phase space. Now, let us focus our attention on
this distribution function. The distribution function is a function of the
point considered (i.e. its coordinates $x^\aB$), and also a function of the
tangent space at this point whose coordinate we label by
$p^{\nu}\partial_{\nu}$. There is no special reason for this function to be
linear in $p^{\nu}\partial_{\nu}$, but we can expand it, without any loss of generality, in power series of tensors according to 

\begin{equation}\label{Multipoleabstractindices}
f\left(x^{\aB},p^{\nu} \right) = \sum_k F_{\mu_1..\mu_k}(x^{\aB})p^{\mu_1}...p^{\mu_k}.
\end{equation}
Here $f$ stands for either the intensity $I$, the degree of circular
polarization $V$ or the matter distribution functions $\gel$, $g_{\mathrm b}$
and $g_{\mathrm c}$. From the previous section we know the transformation rules for the tensorial
 quantities $F_{\mu_1..\mu_k}$, thus $f$ transforms according to

\begin{equation}
\Tr_{(\xi)}\left[ f\left(x^{\aB},p^{\nu}  \right)\right]  \equiv \sum_k \Tr_{(\xi)}\left[F_{\mu_1..\mu_k}(x^{\aB})\right]p^{\mu_1}...p^{\mu_k},
\end{equation}
where $\Tr_{(\xi)}$ refers to the knight-diffeomorphism with the set of vectors
$(\xi_1,\xi_2,...)$.

As we do not necessarily want to refer explicitly to the decomposition in multipoles, we
use the fact that for any vector $\xi = \xi^{\mu}\partial_{\mu}$, which defines a flow on the background space-time $\mathcal{P}_{0}(\mf{N})$, we can define an induced flow (a natural lift) on the vector tangent bundle $T\mathcal{P}_{0}(\mf{N})$ directed by the vector field $T\xi =
 \left[\xi^{\mu}\partial_{\mu},p^{\nu}(\partial_{\nu}\xi^{\mu})\frac{\partial}{\partial p^{\mu}}\right] $. This implies the useful property
 
\begin{equation}
\Lie_{\xi}\left(F_{\mu_1..\mu_p}\right)p^{\mu_1}..p^{\mu_p} = \Lie_{T\xi}\left(F_{\mu_1..\mu_p}p^{\mu_1}..p^{\mu_p}\right).
\end{equation}
With this definition, we can rewrite the transformation rule for $f$ as

\begin{equation}\label{Txi}
\Tr_{(\xi)}\left[ f\left(x^{\aB},p^{\nu} \right)\right]  = \Tr_{(T\xi)}\left[ f\left(x^{\aB},p^{\nu} \right)\right],
\end{equation} 
where now $\Tr_{(T \xi)}$ refers to the knight-diffeomorphism with the set of
vectors $(T\xi_1,T\xi_2,...)$.
The evolution of the distribution function is dictated by the
Boltzmann equation, and its collision term can be expressed in the local Minkowskian frame defined by a tetrad fields $\gr{e}_a$, from known particles physics. For this reason, the framework developed to define gauge
transformations for a general manifold has to be extended to the case of
Riemannian manifold. Instead of using the coordinates basis $\partial_{\aB}$ to
express a vector of tangent space as $\gr{p}=p^{\aB}\partial_{\aB}$, we prefer to use the tetrads basis $\gr{e}_a$ and write
$\gr{p}=p^\aT \gr{e}_\aT$. In terms of coordinates, this means that the distribution function is a function
of $x^{\aB}$ and $p^\aT$, and that we favor the
expansion~(\ref{Def_multipoleS}) instead of the expansion~(\ref{Multipoleabstractindices}). When expressing the physics with the tetrad fields, the metric is not just one of the many tensors of the theory whose properties under a gauge transformation we need to know, but
rather a central feature of the manifold, since it determinates the tetrads
(up to a Lorentz tranformation) required to express the distribution function. 

\subsection{Perturbation of tetrads}

\subsubsection{Pulling back the tetrad}

With the formalism developed for tensors, we carry this tetrad field
onto the background space-time using a gauge field $X$ with
\begin{eqnarray}
&&_X\!\gr{e}_{\aT} \equiv \phi^{\star}_{\lambda,X}(\gr{e}_\aT)= \sum_{k=0}^{k=\infty}\frac{\lambda^k}{k!}\Lie_X^k \gr{e}_\aT\nonumber\\*
&&\,_{X}\! \gr{e}^{(n)}_\aT \equiv \Lie_X^n \gr{e}_\aT \Big|_{\mathcal{P}_0(\mf{N})}\,\qquad\bar{\gr{e}}_\aT \equiv \,_X \gr{e}^{(0)}_\aT,
\end{eqnarray}  
and similar formulas for $\gr{e}^\aT$.

As $\bar{\gr{e}}_\aT$ is a basis of the tangent space on the background space-time
(and $\bar{\gr{e}}^\aT$ a basis of its dual space),
$_X \gr{e}_{\aT}$ and $_X \gr{e}^{\aT}$ which also lie on the background
space-time can be expressed in the
generic form
\begin{equation}\label{defRS}
\flamoi _X \!\gr{e}_{\aT} =\, _X\!R^{\,\,\bT}_{\aT}\gr{\eb}_\bT\,,\qquad _X\!\gr{e}^\bT =
\gr{\eb}^\aT\, _X\!S^{\,\,\bT}_{\aT}\,,\qquad _X\!R^{\,\,\cT}_{\aT}\, _X\!S^{\,\,\bT}_{\cT}= \,_X\!S^{\,\,\cT}_{\aT}\,_X\!R^{\,\,\bT}_{\cT}=\delta^{\bT}_{\,\aT},
\end{equation}
where
\be\label{Devsimplifies}
_X\!R_{\aT \bT}  \equiv \sum_k\frac{\lambda^k}{k!}\,_X\!R^{(k)}_{\aT \bT}\,,\qquad_X\!S_{\aT \bT} \equiv \sum_k \frac{\lambda^k}{k!}\,_X\!S^{(k)}_{\aT \bT}\,.
\ee
Order by order, this reads
\begin{equation}
_X\! \gr{e}^{(n)}_\aT =\,_X\!R^{(n)\bT}_{\aT} \gr{\eb}_\bT,\qquad\,_X \!\gr{e}^{\bT(n)} = \gr{\eb}^\aT
\,_X\!S^{(n)\bT}_{\aT}. 
\end{equation}

\subsubsection{Normalization condition}\label{sec:choicetetrad}

Tetrads are four vector fields which satisfy equation~(\ref{deftetrad})
and are thus related to the metric. Consequently, the perturbations of the
tetrad defined above are partly related to the perturbations of the metric.
When pulled back to the background space-time, equation~(\ref{deftetrad}) implies
\begin{eqnarray}\label{Eqnormalisationtetradepulledback}
\phi^{\star}_{\lambda,X}(\eta_{\aT \bT})=\eta_{\aT\bT} &=&\phi^{\star}_{\lambda,X}(e^{\mu}_\aT e^{\nu}_\bT g_{\mu\nu})\nonumber\\*
&=&\phi^{\star}_{\lambda,X}(e^{\mu}_\aT) \phi^{\star}_{\lambda,X}(e^{\nu}_\bT) \phi^{\star}_{\lambda,X}(g_{\mu\nu}).
\end{eqnarray}
Identifying order by order we get in particular for
the first and second orders
\begin{eqnarray}\label{Constraints}
&&\gr{\eb}_\bT.\,_X\! \gr{e}^{(1)}_\aT + \gr{\eb}_\aT.\,_X\! \gr{e}^{(1)}_\bT + \,_{_X}\!g^{(1)}(\gr{\eb}_\aT,\gr{\eb}_\bT)=0\,,\nonumber\\*
&&\gr{\eb}_\bT.\,_X\! \gr{e}^{(2)}_\aT + \gr{\eb}_\aT.\,_X \!\gr{e}^{(2)}_\bT + \,_{_X}\!g^{(2)}(\gr{\eb}_\aT,\gr{\eb}_\bT)+ \,_X \!\gr{e}^{(1)}_\bT .\, _X \!\gr{e}^{(1)}_\aT\\*
&&  + \,_{_X} \!g^{(1)}\left(\,_X \!\gr{e}^{(1)}_\aT,\gr{\eb}_\bT\right)+ \,_{_X}\!
g^{(1)}\left(\gr{\eb}_\aT,\,_X \!\gr{e}^{(1)}_\bT\right)=0\,,\nonumber
\end{eqnarray}
where a dot product here stands for $\bar{g}\left(\,\,\,,\,\,\right)$.
From the constraints (\ref{Constraints}), we can determine the symmetric part
of $R^{(n)}_{\aT\bT}$ as
\begin{eqnarray}\label{Rfdeg}
_X\!R^{(1)}_{(\aT\bT)}&=&-\demi \,_{_X}\! g^{(1)}(\gr{\eb}_\aT,\gr{\eb}_\bT)\,,\\*
_X\!R^{(2)}_{(\aT\bT)}&=&-\demi \,_{_X}\! g^{(2)}(\gr{\eb}_\aT,\gr{\eb}_\bT) - \,_{_X}\!
g^{(1)}\left(\,_X\!R^{(1)}_{\aT\cT}\gr{\eb}^\cT,\gr{\eb}_\bT\right) \nonumber\\*
&& -\,_{_X} \!g^{(1)}\left(\gr{\eb}_\aT,\,_X\!R^{(1)}_{\bT\cT} \gr{\eb}^\cT\right)-\,_X
\!R^{(1)\cT}_{\aT}\,_X\!R^{(1)}_{\bT\cT}\,, 
\end{eqnarray}
which are related to the components of the inverse by
\begin{eqnarray}
_X S^{(1)}_{\aT\bT} &=& -\, _XR^{(1)}_{\aT\bT}\,,\\*
_X S^{(2)}_{\aT\bT} &=& -\,_XR^{(2)}_{\aT\bT}+ 2\,_XR^{(1)\cT}_{\aT}\,_XR^{(1)}_{\cT\bT}\,.
\end{eqnarray}
The antisymmetric part, $_X\! R_{[\aT \bT]}$, still remains to be chosen as it corresponds
to the Lorentz transformation freedom (boost and rotation), which is allowed by the
definition~(\ref{deftetrad}).
A first and easy choice would be $_X\!R^{(n)}_{[\aT \bT]}=0$ for any $n$. However, as
mentioned above, we eventually want to decompose a vector $p^{\aB}\partial_{\aB}$ on tangent space as
\begin{equation}\label{ptopi}
p^{\bB}\partial_{\bB}= p^\aT \gr{e}_\aT =p^\aT e_\aT^{\,\bB}\partial_{\bB},
\end{equation}
and identify $p^\zT$ with the energy and $p^\iT$ with the momentum (although
conserved quantities are generally ill-defined in general relativity, energy
and momentum can be defined when performing perturbations around a maximally symmetric
background \cite{Deruelle1996} as it is the case here). When working with coordinates, we
want to express physical quantities, as measured by a set of fundamental observers that
we need to choose. If we choose the velocity of these fundamental observers as orthonormal to the
constant time coordinate hypersurfaces, then we require~\cite{Gourgoulhon2007}
$\gr{e}^\zT \sim \gr{\dd\eta} $, which is equivalent to choose $_X
\!S^{(n)}_{\iT \zT}=0$ for any $n$. This choice allows us to fix the
boost in $S^{(n)}$ by imposing the condition $_X\!S^{(n)}_{[\iT \zT]}=
-\,_X\!S^{(n)}_{[\zT \iT]}=-\,_X\!S^{(n)}_{(\iT \zT)}$, and it can be checked by recursion that this implies 
\begin{equation}
_X \!R^{(n)}_{[\iT \zT]}= -\,_X\!R^{(n)}_{[\zT  \iT]}=-\,_X\!R^{(n)}_{(\iT \zT)}\,.
\end{equation}
We also fix the rotation by requiring $_X\! S^{(n)}_{[\iT \jT]}=0$, and it can be
checked similarly, that this implies $_X\! R^{(n)}_{[\iT \jT]}=0$.

\subsection{Gauge transformation of tetrads}\label{Sec_GToftetrads}

Under a gauge transformation, we can deduce the transformation
properties of the tetrad from those of the perturbed metric. In the FL case,
we use a natural background tetrad associated to Cartesian coordinates
\be
\bar{\gr{e}}_\zT \equiv
\left(\partial_{\zB}\right)/a\,,\qquad\bar{\gr{e}}_{\iT} \equiv
\left(\partial_{\iB}\right)/a \,,
\ee
in order to evaluate equation~(\ref{Rfdeg}). 
We report the detailed expressions for the transformation of the tetrads for
the first and second orders in \ref{app_TRS}. Since the tetrads are
constrained by equation~(\ref{Eqnormalisationtetradepulledback}), the
transformation is necessarily of the form
\be
\Tr(\gr{e}_\aT)=\Lambda_\aT^{\,\,\bT}\Tr_{\mathrm{knight}}(\gr{e}_\bT)\,,\qquad\Tr(\gr{e}^\aT)=\Lambda_\bT^{\,\,\aT}\Tr_{\mathrm{knight}}(\gr{e}^\bT)\,,
\ee
where $\Tr_{\mathrm{knight}}$ refers to the transformation that one would
obtain by considering each of the four vectors fields in the tetrad as
independent vector fields, whose transformation rule is calculated using
the knight diffeomorphism of equations~(\ref{transforule}). As a result of the
choice $R_{[\iT \jT]}=S_{[\iT \jT]}=0$, we get that
$\Lambda_{\iT}^{\,\,\jT}=\delta_{\iT}^{\jT}$, and the three vectors
$\gr{e}_\iT $ in a tetrad transform as independent vector fields. Similarly, if a
vector field $\gr{W}$ satisfies $\gr{W}.\gr{e}_\zT=0$, then $W^\iT$ transform
as a scalar field under gauge transformations (recall that $W^\iT \equiv
\gr{e}^\iT.\gr{W} $). This property extends of course to higher rank tensors,
which means that the tetrad components of a projected tensor are to be
considered as scalars under gauge transformations.

\subsection{Gauge transformation of a distribution function}

\subsubsection{The perturbed distribution function with a tetrad}

Now that the transformation properties of the tetrads are known, we turn to the general transformation of a distribution function $f(x^\aB,p^\aT)$.
First, by using the tetrad we can deduce a multipole decomposition of the type~(\ref{Def_multipoleS}) from
the decomposition~(\ref{Multipoleabstractindices}) by
\begin{eqnarray}\label{decmultipoles}
f(x^\aB,p^{\mu}) & \equiv & \sum_\ell F_{\mu_1\dots\mu_\ell}(x^{\aB})p^{\mu_1}\dots p^{\mu_\ell}\nonumber\\*
& = &\sum_\ell
\left[F_{\mu_1\dots\mu_p}(x^{\aB})e_{\aT_1}^{\,\mu_1}\dots e_{\aT_\ell}^{\,\mu_\ell}\right]p^{\aT_1}\dots
p^{\aT_\ell}\nonumber\\*
&\equiv& \sum_\ell F_{\uline{\aT_\ell}}(x^{\aB})p^{\uline{\aT_\ell}}\nonumber\\*
&\equiv& f(x^\aB,p^{\aT}).
\end{eqnarray}
Since the momentum is constrained to be on the mass shell, only three of the
four components are independent, and we can thus choose the multipoles $F_{\uline{\aT_\ell}}(x^{\aB})$ to be
projected, that is to say if any of their indices is $\zT$, then they vanish.
Note that in the case of radiation, that is for $f\equiv I$, then $F_{\uline{\aT_\ell}}=(p^\zT)^\ell I_{\uline{\aT_\ell}}$.\\
We do not need any additional identification procedure for the tangent spaces
through a gauge field, in order to identify points of the tangent space of the slices $T{\mathcal P}_{\lambda}({\mathcal
    N})$. Indeed, once the metric and a gauge field $X$ are chosen, there exists a natural
  identification with the tetrad fields. First, and as mentioned before, we identify the points of ${\cal
    N}$ which lie on the same integral curves of $X$, that is, we identify a point $P \in{\cal P}_{0}({\cal N})$ and $\Phi_{\lambda,X}(P) \in {\cal
    P}_{\lambda}({\cal N})$. Then, we identify vectors of their respective tangent
  spaces, if the coordinates of these vectors in their respective local tetrad
  frames $\bar{\gr{e}}_\aT$ and $\gr{e}_\aT$, are the same. To be short, we identify $p^\aT
  \gr{e}_\aT$ and $p^\aT \bar{\gr{e}}_\aT$. The multipoles $F_{\uline{\aT_\ell}}(x^{\aB})$ are scalar fields, and they can be pulled back on the background space-time using the gauge field $X$, and we define in this way perturbations
\begin{equation}
\Phi^{\star}_{\lambda,X}\left[F_{\uline{\aT_\ell}}(x^{\aB})\right]\equiv \,_X\!F_{\uline{\aT_\ell}}(x^{\aB})\equiv\sum_{\lambda}\frac{\lambda^n}{n!} \,_X\! F^{(n)}_{\uline{\aT_\ell}}(x^{\aB})\,.
\end{equation}
This perturbation scheme induces a perturbation procedure for the distribution
function $f$ as
\begin{eqnarray}\label{fXdev}
_X\!f(x^{\aB},p^\aT) &\equiv & \sum_{n}
\frac{\lambda^n}{n!} \,_X \!f^{(n)}(x^{\aB},p^\aT),\nonumber\\*
_X \!f^{(n)}(x^\aB,p^{\aT}) &\equiv& \sum_\ell  \,_X \!F^{(n)}_{\uline{\aT_\ell}}(x^{\aB})p^{\uline{\aT_\ell}}.
\end{eqnarray}
It is essential to stress that $p^\aT$ is not a perturbed quantity, it is a
coordinate of the locally Minkowskian tangent space. However, the tetrad
field allows us to see $\gr{p}$ as a perturbed vector since $\gr{p}(p^\aT)
= \gr{e}_\aT p^\aT$. In other words, for a given $p^\aT$, there is an
associated vector whose order by order perturbation in a given gauge $X$ is given by $_{_X}\!
\gr{p}^{(n)} \equiv \,_X\!\gr{e}_{\aT}^{(n)} p^\aT$.
 
\subsubsection{Gauge transformation of a distribution function}

We can deduce the transformation rule under a gauge change directly on the form~(\ref{decmultipoles}), pulled back to the background space-time,  
\begin{equation}\label{Tbrute}
\flamoi \Tr\left[\,_X\!f(x^{\aB},p^{\aT})\right]\equiv \sum_\ell \Tr\left[\,_X\!{\mathcal
    F}_{\mu_1\dots\mu_\ell}(x^{\aB})\right]\Tr\left(\,_X\!e_{\aT_1}^{\mu_1}\right)\dots
\Tr\left(\,_X\!e_{\aT_\ell}^{\mu_\ell}\right)p^{\aT_1}\dots
p^{\aT_\ell}\,.
\end{equation}
The first factor in this expression is tensorial. Exactly as for the pre-Riemannian case, its transformation rule
is dictated by the knight-diffeomorphism, whereas we get the transformation
rules of the tetrads from equations~(\ref{Ttetrads1}) and equations~(\ref{Ttetrads2}). As we
do not necessarily want to refer explicitly to the multipole expansion, the first factor
is rewritten by considering $f$ as a function of $p^{\mu}$ using $p^\aT =\,_X
\! e^\aT_{\,\nu} p^{\nu}$, and applying equation~(\ref{Txi}). We then have to consider
the resulting distribution function as a function of $p^\aT$, knowing that the inversion is now given by
 $\gr{p}(p^\aT)=\Tr(\gr{e}_\aT)p^\aT$. This will account for $\Tr\left(\,_{_X}\! e_{\aT_1}^{\mu_1}\right)\dots \Tr\left(\,_{_X}\! e_{\aT_\ell}^{\mu_\ell}\right)$ in equation~(\ref{Tbrute}). In a compact form it reads

\begin{equation}\label{Tcompacte}
\Tr\left[\,_{_X}\!f(x^{\aB},p^\aT)\right]=
     \Tr_{(T\xi)} \left\{\,_{_X}\!f\left[x^{\aB},e^\aT_{\,\mu}p^{\mu}\right]\right\}\Big|_{p^{\mu}=\Tr(e_\bT^{\,\mu})p^\bT}.
\end{equation}

To obtain an order by order formula, we explicit these three steps using a Taylor expansion. First, we use that
\begin{equation}
\flamoi  _{_X}\! f (x^{\aB},p^\aT)=\left[\exp\left(\eb^\bT_{\,\mu}p^{\mu}\,_X\!S_{\bT}^{\,\,\,\aT}\dsd{}{p^\aT}\right){}_{_X}\!f\right] (x^{\aB},\eb^\bT_{\mu}p^{\mu})\equiv \,_{_X}\!g(x^{\aB},p^{\mu}),
\end{equation}
in order to consider $f$ as a function of $p^{\mu}$. We then Taylor expand
again the result of the knight-diffeomorphism in order to read the result as a function of $p^\aT$,
\begin{equation}
\flamoi \Tr\left[\,_{_X}\!f (x^{\aB},p^\aT)\right]=\left[\exp\left(\eb_\bT^{\mu}p^\aT \Tr
    \left(\,_X\!R_{\aT}^{\,\,\,\bT}\right)\dsd{}{p^{\mu}}\right)\Tr_{(T\xi)}
  \left({}_{_X}g\right)\right](x^{\aB},\eb^{\mu}_\aT p^\aT).
\end{equation}

The derivatives in the previous expressions have to be ordered on the right
in each term of the expansion in power series of the exponential. When identifying order by
order, we need to take into account the expansion in $R_{ab}$ and $S_{ab}$, in
the exponentials and also in the knight-diffeomorphism. 
The detail of the transformation properties of the first and second-order
distribution function can be found in~\cite{Pitrou2007}. We report here
the final result of the transformation rule for the first order and the second
order distribution function for radiation
\be\label{Tf1}
\flamoi\Tr\left[ {}_{_X} \!f^{(1)}\right]={}_{_X}\! f^{(1)}+\frac{\partial \bar{f}}{\partial
  p^\zT} p^\zT n^\iT \partial_\iB T + T\dsd{\bar{f}}{\eta}\,,
\ee

\begin{eqnarray}\label{Tf2}
\flamoi \Tr\left( {}_{_X}\! f^{(2)}\right) &=&{}_{_X}\! f^{(2)}+\dsd{\bar{f}}{\eta}(T^{(2)}+T T'+ \partial_\iB T \partial^\iB L) \nonumber\\*
\flamoi &&+\dsd{\bar{f}}{p^\zT}\left\{ p^\iT \partial_\iB T^{(2)} -2 p^\jT \left[
  \left( \partial_\iB \partial_\jB  E + E_{\iB \jB} + \partial_\iB \partial_\jB L\right) \partial^\iB T - \Psi \partial_\jB T\right]  \right.\nonumber\\*
\flamoi && \left.\qquad \qquad  + p^\zT \partial_\iB T \partial^\iB T+ p^\iT (T \partial_\iB  T)'+ p^\iT\partial_\iB \left( \partial^\jB L \partial_\jB T\right) + 2 \Phi p^\iT \partial_\iB T\right\}\nonumber\\*
\flamoi && + \frac{\partial^2 \bar{f}}{\partial
  \left(p^\zT\right)^2} \left( p^\iT \partial_\iB T p^\jT
  \partial_\jB T\right) +2\frac{\partial^2 \bar{f}}{\partial \eta  \partial p^\zT}T p^\iT \partial_\iB T  + \frac{\partial^2 \bar{f}}{\partial
  \eta^2}T^2 \nonumber\\*
\flamoi && + 2 \dsd{{}_{_X}\!f^{(1)}}{p^\zT} p^\jT \partial_\jB T+ 2 \dsd{{}_{_X}\! f^{(1)}}{p^\iT} p^\zT \partial^\iB T+ 2 \partial^\iB L \partial_\iB{}_{_X} \!f^{(1)} + 2 T \dsd{{}_{_X} \!f^{(1)}}{\eta}\,. 
\end{eqnarray}
In order to derive this formula, no use has been made of the normalization of
the momentum. Consequently in the above transformation rule, $f$ stands either
for $I$, $V$ or for $\gel$, $g_{\mathrm b}$ (where we conventionally prefer to
use the notation $\gr{q}$ instead of $\gr{p}$ for the momentum of particles).
Note here that the Einstein implicit summation rule still applies on index of tetrad type
($\iT,\jT,\kT\dots$) when contracted with indices of coordinate type
($\iB,\jB,\kB\dots$). This type of contraction arises from the fact that we
have chosen a background tetrad field adapted to the coordinate system, that is with
$\bar{e}_{\aT}^{\,\bB} \sim \delta_{\aT}^{\,\bB}$ and $\bar{e}^{\aT}_{\,\bB} \sim
\delta^{\aT}_{\,\bB}$.\\

\subsubsection{Gauge transformation of the linear polarization tensor} 

Now if we want to describe polarized radiation, it is straightforward to
generalize this calculation. Indeed it is projected, that is $f_{\zT \iT}=f_{\iT \zT}=f_{\zT \zT}=0$,
and thanks to the remark at the end of section~\ref{Sec_GToftetrads}, each of
the nine components $f_{\iT \jT}$ can be treated like a scalar valued
distribution function, and the same property is valid for $P_{\iT \jT}$.
Furthermore, since for symmetry reasons the radiation is not polarized at the
background level, that is $\bar P_{\iT \jT}=0$, its transformation rule
reduces to 
\be\label{Tf1P}
\flamoi \Tr\left[ {}_{_X}\! P^{(1)}_{\kT \lT}\right]={}_{_X}\! P^{(1)}_{\kT \lT}\,,
\ee

\begin{eqnarray}\label{Tf2P}
\flamoi\Tr\left( {}_{_X}\! P^{(2)}_{\kT \lT}\right) &=&{}_{_X} \!P^{(2)}_{\kT \lT}  + 2
\dsd{{}_{_X}\! P^{(1)}_{\kT \lT}}{p^\zT} p^\jT \partial_\jB T\\*
\flamoi&&+ 2 \dsd{{}_{_X} \!P^{(1)}_{\kT \lT}}{p^\iT} p^\zT \partial^\iB T+ 2 \partial^\iB
L \partial_\iB{}_{_X} \!P^{(1)}_{\kT \lT} + 2 T \dsd{{}_{_X} \!P^{(1)}_{\kT \lT}}{\eta}\,.\nonumber 
\end{eqnarray}
We can then form first-order and second-order gauge invariant intensity function, degree of circular
polarization and tensor of linear polarization, by transforming these
quantities in the Newtonian gauge

\begin{eqnarray}
\hat{I}^{(1)}  &\equiv& _{_{\NG}}\! I^{(1)} \equiv \Tr_{\xi^{(1)}_{\rightarrow \NG}} \left(\,_X\! I^{(1)}\right) \\*
\hat{V}^{(1)}  &\equiv& _{_{\NG}} \!V^{(1)} \equiv \Tr_{\xi^{(1)}_{\rightarrow \NG}} \left(\,_X \!V^{(1)}\right) \\*
\hat{P}^{(1)}_{\iT \jT}  &\equiv& _{_{\NG}} \!P_{\iT \jT}\equiv \Tr_{\xi^{(1)}_{\rightarrow \NG}} \left(\,_{X}\!
  P^{(1)}_{\iT \jT}\right)  \,,\\*
\end{eqnarray}
\begin{eqnarray}
\hat{I}^{(2)}  &\equiv& _{_{\NG}}\! I^{(2)} \equiv \Tr_{\left(\xi^{(1)}_{\rightarrow \NG},\,\xi^{(2)}_{\rightarrow \NG}\right)} \left(\,_X\! I^{(2)}\right) \\*
\hat{V}^{(2)}  &\equiv& _{_{\NG}} \!V^{(2)} \equiv \Tr_{\left(\xi^{(1)}_{\rightarrow \NG},\,\xi^{(2)}_{\rightarrow \NG}\right)} \left(\,_X\! V^{(2)}\right) \\*
\hat{P}^{(2)}_{\iT \jT}  &\equiv& _{_{\NG}}\! P^{(2)}_{\iT \jT} \equiv \Tr_{\left(\xi^{(1)}_{\rightarrow \NG},\,\xi^{(2)}_{\rightarrow \NG}\right)} \left(\,_X\!  P^{(2)}_{\iT \jT}\right) \,. 
\end{eqnarray}
The energy integrated counterparts $\hat{\cal I}^{(n)}$, $\hat{\cal V}^{(n)}$
and $\hat{\cal P}_{\iT\jT}^{(n)}$ with $n=1,2$, follow from a definition similar to equations~(\ref{Brightnessdef}). Since the transformation rule~(\ref{Tf2}) is valid also for the electrons and
protons (or for baryons which is the sum of these two components) distribution functions, the corresponding gauge invariant
distribution functions can be defined in a similar manner and we name them
$\hat g^{(1)}$ and $\hat g^{(2)}$ for the first and second order respectively. 

\section{The gauge invariant Liouville equation for radiation}\label{SecLiouville}

Now that we have a complete formalism to handle the perturbations of the fluid
quantities and of the distribution functions, we will use it to express the
perturbations of the dynamical equations satisfied by these quantities. In this
section we focus on the perturbative expansion of the Liouville operator and in section~\ref{Sec_Collision} we detail the
perturbative expansion of the collision term. In both cases we will recover
the standard first order result and the novelty lies in the second order perturbation.\\

In practice, we want to express the Boltzmann (or the Liouville
equation if relevant) in function of $\eta$ rather than the affine parameter
$s$, since we want to perform an integration on coordinates. We should thus multiply
equation~(\ref{EqBoltzmann}) by $\frac{\dd s}{\dd \eta}=1/p^\zB$. However, there is no point multiplying with the full expression of $\frac{\dd
  s}{\dd \eta}$ since when focusing on the $n$-th order of the Boltzmann
equation, the perturbations  $\left(\frac{\dd  s}{\dd \eta}\right)^{(k)}$ with
$0<k<n$ will multiply the $(n-k)$-th order of the perturbed Boltzmann equation
and these terms will thus vanish. This is the reason why we will instead
multiply only by $\left(\frac{\dd s}{\dd \eta}\right)^{(0)} =
\left(1/p^\zB\right)^{(0)}=a/p^\zT$, and we will use the notation
\be\label{Defstar}
L^\hashamoi[] \equiv L[ ]\left(\frac{\dd s}{\dd \eta}\right)^{(0)}\,,\quad C^\hashamoi_{\aT \bT}[]\equiv C_{\aT \bT}\left(\frac{\dd s}{\dd \eta}\right)^{(0)}\,\,.
\ee
In order to compute the Liouville operator, we need to calculate the perturbed expression of $\frac{\dd p^\zT}{\dd \eta}$ and
$\frac{\dd n^\iT}{\dd \eta}$. They can be obtained from the geodesic equation
\begin{equation}
\frd{p_\aT}{s}+ \omega_{\bT\aT\cT}p^\cT p^\bT=0,\quad\mathrm{with}\quad
\frd{p_\aT}{s}\equiv p^\bT\partial_{\bT}p_\aT\equiv p^\bT
e_\bT^{\,\,\cB}\partial_\cB (p_\aT)\,,
\end{equation}
that we pull back to the background space-time in order to extract order by
order equations
\be\label{Geodesiqueentetrade}
\left(\frd{p_\aT}{s}\right)^{(n)}=- \omega^{(n)}_{\bT\aT\cT}p^\cT
p^\bT\,,\quad n=0,1,2\,,
\ee
and from the perturbative expansion of $p^\zB$
\be
\left(\frac{\dd \eta}{\dd s}\right)^{(n)}=\left(p^\zB\right)^{(n)}=p^\aT e^{(n)\zB}_\aT\,,\quad n=0,1,2\,.
\ee
It should be noted that the transformation properties of $L^\hashamoi[]$ under a
local change of frame is not similar to that of $L[]$ due to the factor
$1/p^\zT$ which transforms according to equation~(\ref{Eq_Tmomentum1})~\cite{Tsagas2007}.

The result of the Liouville operator up to second order is of the form
\be\label{DecL}
\flamoi L[\bar X,X^{(1)},X^{(2)}]=\bar L[\bar X]+L^{(1)}[\bar
X,X^{(1)}]+\frac{1}{2}\left(L^{(2)}[\bar X,X^{(2)}]+L^{(1)(1)}[\bar X,X^{(1)}] \right)\,,
\ee
where $X$ stands for either $I$, $V$ or $P_{\aT \bT}$. Similarly to
Einstein equations, the second order Boltzmann equation will involve either
terms which are linear in purely second order quantities that we gather in
$L^{(2)}[]$, or terms quadratic in first order
quantities that we gather in $L^{(1)(1)}[]$. The linear dependence in these
purely second order quantities is, as usual, the same as for the linear dependence obtained for the first order equation
with respect to the first order quantities, and this means $L^{(2)}[]=L^{(1)}[]$. In principle, once presented the expression of
$L^{(1)}[\bar X,X^{(1)}]$, we only need to report the expression of $L^{(1)(1)}[\bar X,X^{(1)}]$ to fully express the second order
equations. In practice however we will report both since we chose to neglect
the first order vector degrees of freedom but we cannot neglect them at second
order and they will thus contribute to $L^{(2)}[]$. Additionally, we also choose to neglect the first order tensorial
perturbations. Though they are generated in standard models of inflation,
their amplitude is expected to be much smaller than first order scalar
perturbations, and furthermore they decay once they enter the Hubble radius.
Similarly to vector modes, we will not neglect the tensor modes at second order and their
contribution will be also reported in $L^{(2)}[]$.\\
Since the expressions in the decomposition~(\ref{DecL}) have a dependence in
$(p^\zT,n^\iT)$, we will also perform a multipolar expansion according to
equation~(\ref{Def_multipoleS}) for the equations on $I$ and $V$ (with multipoles
$L[I]_{\uline{\aT_\ell}}$ and $L[V]_{\uline{\aT_\ell}}$) and according to equation~(\ref{Def_multipoleP})
for the equation involving $P_{\aT \bT}$ (with electric and magnetic type
multipoles being respectively $L[E]_{\uline{\aT_\ell}}$ and $L[B]_{\uline{\aT_\ell}}$).\\
We will also define the energy integrated Liouville operator by
\be\label{EqLcurly}
 {\cal L}[]\equiv \frac{4 \pi}{(2 \pi)^3} \int L[](p^\zT)^3 \dd p^\zT\,,
\ee
and use similar definitions for $L^\hashamoi[]$, $C_{\aT \bT}$ and $C_{\aT
  \bT}^\hashamoi$, and this will lead to define multipoles for these quantities, with an obvious
choice of notation. We recall that since the polarization tensor is a
function of the position coordinates $x^\aB$, that we do not write explicitly, and of the momentum components
$p^\aT$, then the Liouville operator and the collision tensor have a
dependence in the same variables. However, in order to simplify the notation, we will not write the
dependence in $p^\aT$ throughout this section, which is dedicated to the Liouville
operator for radiation, and throughout section~\ref{SecLiouvilleBar}, which is dedicated to the
Liouville operator for baryons, but we will restore it in section~\ref{Sec_Collision} dedicated to the
collision tensor. 

\subsection{The background Liouville operator}

At the background level, space is homogeneous and isotropic. Consequently, the
radiation is not polarized and its distribution function depends neither on the direction $n^\iT$ of the photon nor on the position in space $x^\iB$. It only depends on $p^\zT$ and $\eta$, which
implies that $\dsd{\bar{I}}{n^\iT}=\dsd{\bar{I}}{x^\iB}=\bar{P}_{\aT\bT}=0$. Since the
background geodesic equation implies $\left(\dsd{p^\zT}{\eta}\right)^{(0)}=-\HH p^\zT$,
the Liouville operator reads at the background level
\begin{equation}\label{Boltzmann0}
\bar L^{\hashamoi}[\bar I]=\dsd{\bar{I}}{\eta}-\HH p^\zT \dsd{\bar{I}}{p^\zT}\,,
\end{equation}
where we recall that the derivative w.r.t. $\eta$ is to be taken at $p^\aT$
fixed. Its energy integrated counterpart reads
\begin{equation}\label{Boltzmann0int}
\bar {\cal L}^{\hashamoi}[\bar {\cal I}]=\dsd{\bar{{\cal I}}}{\eta}+4\HH \bar{{\cal I}}\,.
\end{equation}

\subsection{The Liouville operator at first order}

The equation required to express the Liouville operator is the first order of
equation~\ref{Geodesiqueentetrade} which in the Newtonian gauge leads to
\be\label{Eq_evolution_pi0_ordre_1}
\left(\frd{p^\zT}{\eta}\right)^{(1)}=p^\zT\left[-n^\iT \partial_\iB \Phi^{(1)}
  +\Psi^{(1)'}\right]\,.
\ee
and we obtain that the first order perturbed Liouville operator can be expressed in
function of gauge invariant quantities by
\be
\flamoi L^{\hashamoi(1)}[\bar I,\hat I^{(1)}]=\frac{\partial \hat I^{(1)}}{\partial \eta}+ n^\jT \partial_\jB \hat I^{(1)} -\HH p^\zT \frac{\partial \hat I^{(1)}}{\partial
  p^\zT}+\left[- n^\jT \partial_\jB \hat \Phi^{(1)} + \hat
  \Psi^{(1)'} \right]p^\zT \frac{\partial \bar I}{\partial p^\zT}\,.
\ee
Its energy integrated counterpart is thus given by
\be
\flamoi {\cal L}^{\hashamoi(1)}[\bar {\cal I},\hat {\cal I}^{(1)}]=\frac{\partial \hat
  {\cal I}^{(1)}}{\partial \eta}+ n^\jT \partial_\jB \hat {\cal I}^{(1)} +4
\HH \hat {\cal I}^{(1)}+4\left[ n^\jT \partial_\jB \hat \Phi^{(1)} -\hat \Psi^{(1)'} \right]\bar {\cal I}\,.
\ee
It will prove more convenient to decompose it in multipoles which are given by
\bea
{\cal L}^{\hashamoi(1)}[\bar {\cal I},\hat {\cal I}^{(1)}]_{\uline{\iT_\ell}}&=&\hat {\cal
  I}^{'(1)}_{\uline{\iT_\ell}}+4 \HH \hat {\cal  I}^{(1)}_{\uline{\iT_\ell}}+\frac{\ell+1}{(2 \ell +3)}\partial^\jB \hat {\cal I}^{(1)}_{\jT
  \uline{\iT_\ell}} + \partial_{\langle \iB_\ell} \hat {\cal
  I}^{(1)}_{\uline{\iT_{\ell-1}}\rangle}\\* 
&& +4\delta_{\ell}^{1}\bar {\cal I}\partial_{\iB_1} \hat \Phi^{(1)}-4\delta_{\ell}^{0}\bar {\cal I} \hat \Psi^{'(1)}\,.\nonumber
\eea
As for $V$, it follows the same equation as $I$ but we will see in
section~\ref{Sec_Collision} that it is not excited by the collisions and remains null.\\

Following the same method for the linear polarization tensor leads to 
\bea
L^{\hashamoi(1)}[\hat P^{(1)}_{\cT \dT}]&=&S_\cT^\aT S_\dT^\bT \left[\frac{\partial \hat  P^{(1)}_{\aT\bT}}{\partial \eta}+ n^\jT \partial_\jB \hat P^{(1)}_{\aT\bT} -\HH p^\zT \frac{\partial \hat P^{(1)}_{\aT\bT}}{\partial  p^\zT}\right]\,,\\*
{\cal L}^{\hashamoi(1)}[\hat {\cal P}^{(1)}_{\cT \dT}]&=&S_\cT^\aT S_\dT^\bT \left[\frac{\partial \hat {\cal
    P}^{(1)}_{\aT\bT}}{\partial \eta}+ n^\jT \partial_\jB \hat {\cal
  P}^{(1)}_{\aT\bT} +4 \HH \hat {\cal P}^{(1)}_{\aT\bT}\right]\,.
\eea
We can extract the electric and magnetic type
multipoles of the energy integrated Liouville operator, and they read
\bea
{\cal L}^{\hashamoi(1)}[\hat {\cal E}^{(1)}]_{\uline{\iT_\ell}}&=& \hat {\cal
  E}^{'(1)}_{\uline{\iT_\ell}}+4\HH\hat {\cal
  E}^{(1)}_{\uline{\iT_\ell}}+\frac{(\ell-1)(\ell+3)}{(\ell+1)(2 \ell +3)}\partial^\jB \hat {\cal E}^{(1)}_{\jT
  \uline{\iT_\ell}} + \partial_{\langle \iB_\ell} \hat {\cal
  E}^{(1)}_{\uline{\iT_{\ell-1}}\rangle}\nonumber\\* 
&&-\frac{2}{(\ell+1)} (\mathrm{curl} \,\hat {\cal B}^{(1)})_{\uline{\iT_\ell}}\,,
\eea

\bea\label{LiouvilleB1}
{\cal L}^{\hashamoi(1)}[\hat {\cal B}^{(1)}]_{\uline{\iT_\ell}}&=& \hat {\cal
  B}^{'(1)}_{\uline{\iT_\ell}}+4\HH\hat {\cal  B}^{(1)}_{\uline{\iT_\ell}}+\frac{(\ell-1)(\ell+3)}{(\ell+1)(2 \ell +3)}\partial^\jB \hat {\cal B}^{(1)}_{\jT
  \uline{\iT_\ell}} + \partial_{\langle \iB_\ell} \hat {\cal
  B}^{(1)}_{\uline{\iT_{\ell-1}}\rangle}\nonumber\\* 
&&+\frac{2}{(\ell+1)} (\mathrm{curl} \,\hat {\cal E}^{(1)})_{\uline{\iT_\ell}}\,,
\eea
where 
\be
(\mathrm{curl}\, X)_{\uline{\iT_\ell}}\equiv\epsilon_{\jT \kT \langle
  \iT_\ell}\partial^\jB X_{\uline{\iT_{\ell-1}}\rangle}^{\,\,\,\,\,\,\,\,\,\,\,\,\,\kT} \,\,.
\ee
However, it can be shown that when the first order tensor and vector modes are
neglected, which is our case in this paper, the first order magnetic type
multipoles are not excited and remain null~\cite{Hu1997}.
We will thus discard them from the computation in the following of this paper
in order to simplify the intricate expressions of second order perturbations.
In order to be consistent with equation~(\ref{LiouvilleB1}) we will also have
to neglect $(\mathrm{curl} \,\hat {\cal E}^{(1)})_{\uline{\iT_\ell}}$.

\subsection{The Liouville operator at second order}

Following the same method as for the first order case, we obtain the second
order evolution of the energy $p^\zT$ which in the Newtonian gauge reads
\bea\label{Eq_evolution_pi0_ordre_2}
\left(\frd{p^\zT}{\eta}\right)^{(2)}&=&p^\zT\left[-n^\iT \partial_\iB \Phi^{(2)}
  +\Psi^{(2)'}+\left(\partial_\iB B_{\jB}^{(2)}-H_{\iB\jB}^{(2)'}\right) n^\iT n^\jT\right.\nonumber\\*
&&\qquad \left.+2 (\Phi-\Psi)n^\iT \partial_\iB \Phi + 4\Psi \Psi'  \right].
\eea
We also need the first order photon trajectory which in the Newtonian gauge reads
\be
\left(\frd{x^\iB}{\eta}\right)^{(1)}=\left(\frac{p^\iB}{p^\zB}\right)^{(1)}=n^\iT (\Phi+\Psi)\,,
\ee 
and the first order lensing equation which gives the evolution of the
direction $n^\iT$
\be
\left(\frd{n^\iT}{\eta}\right)^{(1)}=- S^{\iT\jT}\partial_\jB \left(\Psi+\Phi \right)\,.
\ee
We then obtain the gauge invariant second order Liouville operator 
\bea\label{EqLiouville2I}
\flamoi L^{\hashamoi(2)}[\bar I,\hat I^{(2)}]&=&\frac{\partial \hat I^{(2)}}{\partial \eta}+ n^\jT \partial_\jB \hat I^{(2)} -\HH p^\zT \frac{\partial \hat I^{(2)}}{\partial
  p^\zT}\\*
\flamoi &&+\left[- n^\jT \partial_\jB \hat \Phi^{(2)} + \hat  \Psi^{(2)'}
  +\left(\partial_\iB \hat \Phi_{\jB}^{(2)}-\hat H_{\iB\jB}^{(2)'}\right) n^\iT n^\jT\right]p^\zT \frac{\partial \bar I}{\partial p^\zT}\,,\nonumber
\eea
\bea\label{L11I}
\flamoi L^{\hashamoi(1)(1)}[\bar I,\hat I^{(1)}]&=&2 (\hat \Phi+\hat \Psi)n^\iT \partial_\iB\hat
I^{(1)} -2S^{\iT\jT}\left( \partial_\jB \hat \Psi+  \partial_\jB \hat \Phi \right)\frac{\partial \hat I^{(1)}}{\partial n^\iT}\nonumber\\*
\flamoi &&+ 2 \frac{\partial \hat I^{(1)}}{\partial p^\zT}p^\zT
\left[-n^\iT \partial_\iB \hat \Phi+\hat \Psi' \right] + \left[2 (\hat
  \Phi-\hat \Psi)n^\iT \partial_\iB \hat \Phi + 4 \hat \Psi \hat \Psi'
\right]p^\zT \frac{\partial \bar I}{\partial p^\zT}\nonumber\\*
\flamoi &&-2 \hat \Phi L^{\hashamoi(1)}[\bar I,\hat I^{(1)}]\,.
\eea
Its energy integrated version is 
\bea
\flamoi {\cal L}^{\hashamoi(2)}[\bar {\cal I},\hat {\cal I}^{(2)}]&=&\frac{\partial \hat
  {\cal I}^{(2)}}{\partial \eta}+ n^\jT \partial_\jB \hat {\cal I}^{(2)} +4\HH \hat {\cal I}^{(2)}\\*
\flamoi &&-4\bar {\cal I}\left[- n^\jT \partial_\jB \hat \Phi^{(2)} + \hat  \Psi^{(2)'}
  +\left(\partial_\iB \hat \Phi_{\jB}^{(2)}-\hat H_{\iB\jB}^{(2)'}\right) n^\iT
  n^\jT\right]\,,\nonumber
\eea
\bea\label{curlL11I}
\flamoi {\cal L}^{\hashamoi(1)(1)}[\bar {\cal I},\hat {\cal I}^{(1)}]&=&2 (\hat \Phi+\hat \Psi)n^\iT \partial_\iB\hat
{\cal I}^{(1)} -2 S^{\iT\jT}\left( \partial_\jB \hat \Psi+  \partial_\jB \hat \Phi
\right)\frac{\partial \hat {\cal I}^{(1)}}{\partial n^\iT}\nonumber\\*
\flamoi &&-8 \hat {\cal I}^{(1)}\left[-n^\iT \partial_\iB \hat \Phi+\hat \Psi' \right] -4\bar
 {\cal I} \left[2 (\hat \Phi-\hat \Psi)  n^\iT \partial_\iB \hat \Phi + 4 \hat
   \Psi \hat \Psi' \right]\nonumber\\*
\flamoi &&-2 \hat \Phi {\cal L}^{\hashamoi(1)}[\bar{\cal I},\hat{\cal I}^{(1)}]\,.
\eea
We recall again that we have omitted some exponents stating that some quantities are
first order perturbations variables, since at second order, the terms involved
are either involving purely second order perturbations or terms quadratic in
first order quantities, and there is thus never any ambiguity. Note also that,
if $L^\hashamoi[]$ and ${\cal L}^\hashamoi[]$ had been defined with $(\dd
    s/\dd \eta)$ instead of $\left(\dd s/\dd
    \eta\right)^{(0)}$ in the definition~(\ref{Defstar}), the last line of
equations~(\ref{L11I}) and (\ref{curlL11I}) would not be there, and this enables
us to compare with the results of~\cite{Bartolo2006,Bartolo2007} where such
different choice has been made (though it is then inconsistent with the collision term
of~\cite{Bartolo2006} which has been calculated with $\left(\dd s/\dd
    \eta\right)^{(0)}$, but not with the collision term reported
  in~\cite{Bartolo2007} which seems to be defined with $\left(\dd s/\dd
    \eta\right)$ though it is stated that it is the same expression as in~\cite{Bartolo2006}).
We can then extract the multipoles of these equations and we obtain
\bea\label{LiouvilleI2}
\flamoi {\cal L}^{\hashamoi(2)}[\bar {\cal I},\hat {\cal I}^{(2)}]_{\uline{\iT_\ell}}&=&\hat {\cal
  I}^{'(2)}_{\uline{\iT_\ell}}+4 \HH \hat {\cal  I}^{(2)}_{\uline{\iT_\ell}}+\frac{\ell+1}{(2 \ell +3)}\partial^\jB \hat {\cal I}^{(2)}_{\jT
  \uline{\iT_\ell}} + \partial_{\langle \iB_\ell} \hat {\cal
  I}^{(2)}_{\uline{\iT_{\ell-1}}\rangle}\\* 
\flamoi && -4\delta_{\ell}^{2}\bar {\cal I}\left(\partial_{(\iB_1} \hat \Phi_{\iB_2)}^{(2)}-\hat H_{\iB_1\iB_2}^{(2)'}\right)+4\delta_{\ell}^{1}\bar {\cal I}\partial_{\iB_1} \hat \Phi^{(2)}-4\delta_{\ell}^{0}\bar {\cal I} \hat \Psi^{'(2)}\,,\nonumber
\eea
\bea\label{LiouvilleI11}
\flamoi {\cal L}^{\hashamoi(1)(1)}[\hat {\cal I}^{(1)}]_{\uline{\iT_\ell}}&=&\frac{(\ell+1)}{(2 \ell
  +3)}\left[8\partial^\jB \Phi-2(\ell+2)\partial^\jB(\Phi+\Psi)+2 \Psi\partial^\jB
\right]\hat {\cal I}^{(1)}_{\jT \uline{\iT_\ell}}\nonumber\\*
\flamoi &&+\left[8\partial_{\langle \iB_1} \Phi+2(\ell-1)\partial_{\langle
    \iB_1}(\Phi+\Psi)+2\Psi\partial_{\langle \iB_1}\right]\hat {\cal I}^{(1)}_{\uline{\iT_{\ell-1}}\rangle}\nonumber\\*
\flamoi &&-8 \Psi' \hat {\cal I}^{(1)}_{\uline{\iT_\ell}}-2 \Phi\hat {\cal  I}^{(1)'}_{\uline{\iT_\ell}}-8 \HH \Phi\hat {\cal
  I}^{(1)}_{\uline{\iT_\ell}}\nonumber\\*
\flamoi &&+4 \delta_{\ell}^1 \bar {\cal
  I}\left[2(\Psi-2\Phi)\partial_\iB\Phi\right]+4 \delta_{\ell}^0\bar {\cal I}\left[(2\Phi-4\Psi)\Psi'\right]\,.
\eea

Following the same method as for the intensity part of the Liouville operator,
we obtain the Liouville operator for the linear polarization tensor
\be
\flamoi L^{\hashamoi(2)}[\hat P_{\cT\dT}^{(2)}]=S_\cT^\aT S_\dT^\bT \left[\frac{\partial \hat P_{\aT\bT}^{(2)}}{\partial \eta}+ n^\jT \partial_\jB \hat P_{\aT\bT}^{(2)} -\HH p^\zT \frac{\partial \hat P_{\aT\bT}^{(2)}}{\partial  p^\zT}\right]\,,
\ee
\bea
\flamoi L^{\hashamoi(1)(1)}[\hat P_{\cT\dT}^{(1)}]&=&S_\cT^\aT S_\dT^\bT \left[2 (\hat \Phi+\hat \Psi)n^\iT \partial_\iB\hat
P_{\aT\bT}^{(1)} -2 S^{\iT\jT}\left( \partial_\jB \hat \Psi+  \partial_\jB \hat \Phi \right)\frac{\partial \hat P_{\aT\bT}^{(1)}}{\partial n^\iT}\right.\nonumber\\*
\flamoi&&\qquad \left.+ 2 \frac{\partial \hat P_{\aT\bT}^{(1)}}{\partial p^\zT}p^\zT
\left(-n^\iT \partial_\iB \hat \Phi+\hat \Psi' \right) -2 \hat \Phi L^{\hashamoi(1)}[\hat P^{(1)}_{\aT\bT}]\right]\,,
\eea
and its energy integrated counterpart
\be
\flamoi {\cal L}^{\hashamoi(2)}[\hat {\cal P}_{\cT\dT}^{(2)}]=S_\cT^\aT S_\dT^\bT \left[\frac{\partial \hat
  {\cal P}_{\aT\bT}^{(2)}}{\partial \eta}+ n^\jT \partial_\jB \hat {\cal P}_{\aT\bT}^{(2)} +4\HH \hat {\cal P}_{\aT\bT}^{(2)}\right]\,,
\ee
\bea\label{EqPab11}
\flamoi {\cal L}^{\hashamoi(1)(1)}[\hat {\cal P}_{\cT\dT}^{(1)}]&=&S_\cT^\aT S_\dT^\bT \left[2 (\hat \Phi+\hat \Psi)n^\iT \partial_\iB\hat
{\cal P}_{\aT\bT}^{(1)} -2 S^{\iT\jT}\left( \partial_\jB \hat \Psi+  \partial_\jB \hat \Phi
\right)\frac{\partial \hat {\cal P}_{\aT\bT}^{(1)}}{\partial n^\iT}\right.\nonumber\\*
\flamoi&&\qquad \left.-8 \hat {\cal P}_{\aT\bT}^{(1)}\left(-n^\iT \partial_\iB \hat \Phi+\hat
  \Psi' \right)-2 \hat \Phi {\cal L}^{\hashamoi(1)}[\hat{\cal P}^{(1)}_{\aT\bT}]\right]\,.
\eea
We extract the electric and magnetic type multipoles

\bea\label{LiouvilleE2}
\flamoi {\cal L}^{\hashamoi(2)}[\hat {\cal E}^{(2)}]_{\uline{\iT_\ell}}&=&\hat {\cal
  E}^{'(2)}_{\uline{\iT_\ell}}+4\HH\hat {\cal
  E}^{(2)}_{\uline{\iT_\ell}}+\frac{(\ell-1)(\ell+3)}{(\ell+1)(2 \ell +3)}\partial^\jB \hat {\cal E}^{(2)}_{\jT
  \uline{\iT_\ell}} + \partial_{\langle \iB_\ell} \hat {\cal
  E}^{(2)}_{\uline{\iT_{\ell-1}}\rangle}\nonumber\\* 
\flamoi &&-\frac{2}{(\ell+1)} (\mathrm{curl} \,\hat {\cal B}^{(2)})_{\uline{\iT_\ell}}\,,
\eea

\bea
\flamoi {\cal L}^{\hashamoi(1)(1)}[\hat{\cal E}^{(1)}]_{\uline{\iT_\ell}}&=&\frac{(\ell-1)(\ell+3)}{(2 \ell
  +3)(\ell+1)}\left[8\partial^\jB \hat \Phi-2(\ell+2)\partial^\jB(\hat\Phi+\hat\Psi)+2 \hat\Psi\partial^\jB
\right] \hat{\cal E}_{\jT \uline{\iT_\ell}}\\*
\flamoi &&+\left[8\partial_{\langle \iB_\ell} \hat\Phi+2(\ell-1)\partial_{\langle
    \iB_\ell}(\hat\Phi+\hat\Psi)+2\hat\Psi\partial_{\langle
    \iB_\ell}\right]\hat{\cal E}_{\uline{\iT_{\ell-1}}\rangle}\nonumber\\*
\flamoi &&-8 \hat\Psi' \hat{\cal E}_{\uline{\iT_\ell}}-2 \hat\Phi \hat{\cal
  E}'_{\uline{\iT_\ell}}-8\HH \hat\Phi \hat{\cal
  E}_{\uline{\iT_\ell}}\nonumber\,,
\eea

\bea
\flamoi {\cal L}^{\hashamoi(2)}[\hat {\cal B}^{(2)}]_{\uline{\iT_\ell}}&=&\hat {\cal
  B}^{'(2)}_{\uline{\iT_\ell}}+4\HH\hat {\cal
  B}^{(2)}_{\uline{\iT_\ell}}+\frac{(\ell-1)(\ell+3)}{(\ell+1)(2 \ell +3)}\partial^\jB \hat {\cal B}^{(2)}_{\jT
  \uline{\iT_\ell}} + \partial_{\langle \iB_\ell} \hat {\cal
  B}^{(2)}_{\uline{\iT_{\ell-1}}\rangle}\nonumber\\* 
\flamoi &&+\frac{2}{(\ell+1)} (\mathrm{curl} \,\hat {\cal E}^{(2)})_{\uline{\iT_\ell}}\,,
\eea

\bea\label{LiouvilleB11}
\flamoi {\cal L}^{\hashamoi(1)(1)}[\hat{\cal B}^{(1)}]_{\uline{\iT_\ell}}&=&\frac{2}{\ell+1}\epsilon_{\jT \kT\langle
  \iT_\ell}\left[-2 \partial^\jB(\hat\Phi+\hat\Psi)\hat {\cal E}^\kT_{\,\,\,\uline{\iT_{\ell-1}}\rangle}+2(\hat\Phi+\hat\Psi) \partial^\jB\hat {\cal
  E}^\kT_{\,\,\,\uline{\iT_{\ell-1}}\rangle}\right.\nonumber\\*
\flamoi&&\left.\qquad \qquad +8\partial^\jB \hat\Phi \hat {\cal E}^\kT_{\,\,\,\uline{\iT_{\ell-1}}\rangle}\right]\,.
\eea

\subsubsection{Dependence on the choice of the tetrad}

We chose in section~\ref{sec:choicetetrad} to align the tetrad $\gr{e}^\zT$
with the form $(\dd \eta)$,
which means that the corresponding observers follow worldlines always orthogonal
to the constant time hypersurfaces. Alternatively, we could have chosen to
align $\gr{e}_\zT$ with the vector $(\partial_\eta)$, which would then in turn correspond
to observers of constant spatial coordinates~\cite{Durrer1994}. These two choices are related by
a boost parameterized by the shift vector $\omega^\iB$, and in the case of
the Newtonian gauge, the parameter velocity which characterizes this boost
reduces to $\hat \Phi^\iB$. This change in the tetrad used to decompose the momentum
induces a transformation of the distribution function. Up to second order the
two gauge invariant distribution functions (with obvious notation) are related by
\bea
\hat I^{(1)}_{(\partial_\eta)}&=&\hat I^{(1)}_{(\dd
  \eta)}\\*
\hat I^{(2)}_{(\partial_\eta)}&=&\hat I^{(2)}_{(\dd
  \eta)}+\frac{\partial \bar I_{(\dd \eta)}}{\partial p^\zT}p^\zT \hat \Phi^{(2)}_\iB n^\iT\,.\nonumber
\eea
It can be checked that the equation satisfied by $\hat
I^{(2)}_{(\partial_\eta)}$ is the same as $\hat I^{(2)}_{(\dd \eta)}$
[equation~(\ref{EqLiouville2I})] with $\partial_\iB \hat \Phi_{\jB}^{(2)}n^\iT
n^\jT$ replaced by $-\hat \Phi^{'(2)}_\jB n^\jT$, and thus our result is consistent with~\cite{Hu1997,Bartolo2006} where  $\hat
I^{(2)}_{(\partial_\eta)}$ is used.\\
What would be the best choice then? In our case (the same choice is performed
in~\cite{Durrer1994}), $e^{\zT}_{\,\iB}=0$. We can check that if there is no scalar
perturbation, then the coordinates of the acceleration defined by $a_\nu\equiv {(e^\zT)}^\mu\nabla_\mu
e^\zT_\nu$ read
\be
a_\iB=(e^{\zT}_{\,\iB})'+\HH e^{\zT}_{\,\iB}\,.
\ee
We thus conclude that the choice that we made corresponds to picking observers which
accelerate the less, that is which  accelerate only because of gradients in the gravitational
potential, and thus this class of observers is the closest to freely falling observers.

\section{The gauge invariant Liouville equation for baryons}\label{SecLiouvilleBar}

\subsection{The background and first order Liouville operator}

We follow the same method as for radiation. The main difference lies in the
fact that $n^\iT$ is not a unit vector anymore. Consequently we cannot drop
terms like $n_\iT n^\iT=\beta^2$. Furthermore it is not interesting to split the
dependence in the momentum into a dependence in its energy and in the
direction of its spatial part, since the magnitude of the spatial momentum ($\lambda$) cannot be
identified with the energy $q^\zT$. At the background level, the evolution of
$q^\zT$, $q^\iT$ and $\lambda$ are dictated by
\be
\flamoi \left(\frd{q^\zT}{\eta}\right)^{(0)}=-\HH q^\zT \beta^2\,,\qquad\left(\frd{\lambda}{\eta}\right)^{(0)}=-\HH \lambda\,,\qquad\left(\frd{q^\iT}{\eta}\right)^{(0)}=-\HH q^\iT\,,
\ee
and we thus obtain for the background Liouville operator
\be
\bar L^{\hashamoi}[\bar g]=\dsd{\bar{g}}{\eta}-\HH q^\iT \dsd{\bar{g}}{q^\iT}=\dsd{\bar{g}}{\eta}-\HH \lambda \dsd{\bar{g}}{\lambda}\,.
\ee 

In order to compute the Liouville operator at first order, we need the
evolution of the particle spatial momentum magnitude $\lambda$ at first order,
which from equation~(\ref{Geodesiqueentetrade}) is given in the Newtonian gauge by
\be\label{Eq_evolution_pi0_ordre_1_baryons}
\left(\frd{\lambda}{\eta}\right)^{(1)}=\frac{1}{\beta}\left(\frd{q^\zT}{\eta}\right)^{(1)}=\lambda\left[-\frac{1}{\beta^2}n^\iT \partial_\iB \Phi^{(1)}
  +\Psi^{(1)'}\right]\,.
\ee
and this leads to the following first order gauge invariant Liouville operator
\be
\flamoi L^{\hashamoi(1)}[\bar g,\hat g^{(1)}]=\frac{\partial \hat g^{(1)}}{\partial \eta}+ n^\jT \partial_\jB \hat g^{(1)} -\HH q^\iT \frac{\partial \hat g^{(1)}}{\partial
  q^\iT}+\left[- \frac{1}{\beta^2}n^\jT \partial_\jB \hat \Phi^{(1)} + \hat
  \Psi^{(1)'} \right]q^\kT \frac{\partial \bar g}{\partial q^\kT}\,.
\ee

\subsection{The second order Liouville operator}
 
At second order, the evolution of $\lambda$ is also obtained from equation~(\ref{Geodesiqueentetrade}), and in the Newtonian gauge we obtain
\bea
\flamoi \left(\frd{\lambda}{\eta}\right)^{(2)}=\frac{1}{\lambda}\left(\frd{q^\zT}{\eta}\right)^{(2)} &=&\frac{\lambda}{\beta^2}\left\{ -n^i \partial_\iB \Phi^{(2)}+
  \left[\Psi^{(2)'} n_\iT n^\iT+\left(\partial_\iB B_{\jB}^{(2)}- H_{\iB \jB}^{(2)'}\right)n^\iT
    n^\jT\right]\right.\nonumber\\*
&& \qquad \left.+2 (\Phi-\Psi)n^\iT \partial_\iB \Phi + 4 \beta^2 \Psi \Psi'\right\}.
\eea
We also need to use the first order evolution of $q^\iT$ which in the
Newtonian gauge reads
\be
\left(\frd{q^\iT}{\eta}\right)^{(1)}=q^\iT \Psi'-q^\zT \partial^\iB \Phi-
q^\zT \left(\beta^2 \gamma^{\iT \jT} -n^\iT n^\jT\right)\partial_\jB \Psi\,.
\ee
Finally we need the first order photon trajectory given in the Newtonian gauge by
\be
\left(\frd{x^\iB}{\eta}\right)^{(1)}=\left(\frac{q^\iT}{q^\zT}\right)^{(1)}=n^\iT (\Phi+\Psi)\,.
\ee 
With this we obtain the second order gauge invariant Liouville operator for
massive particles
\bea
\flamoi L^{\hashamoi(2)}[\bar g,\hat g^{(2)}] &=&\frac{\partial \hat g^{(2)}}{\partial \eta}+ n^\jT \partial_\jB \hat g^{(2)} -\HH q^\iT \frac{\partial \hat g^{(2)}}{\partial
  q^\iT}\\*
\flamoi &&+\left[- \frac{1}{\beta^2}n^\jT \partial_\jB \hat \Phi^{(2)} + \hat
  \Psi^{(2)'}+\frac{1}{\beta^2}\left(\partial_\iB \hat \Phi_{\jB}^{(2)}- \hat H_{\iB \jB}^{(2)'}\right)n^\iT
    n^\jT \right]q^\kT \frac{\partial \bar g}{\partial q^\kT}\,,\nonumber
\eea
\bea
\flamoi L^{\hashamoi(1)(1)}[\bar g,\hat g^{(1)}] &=&2 n^\iT (\hat \Phi+\hat \Psi)\partial_\iB \hat{g}^{(1)} - 2 \left[(\beta^2 \delta^{\iT \jT}-n^\iT n^\jT)\partial_\jB \hat
  \Psi+  \partial^\iB \hat \Phi \right]\frac{\partial \hat{g}^{(1)}}{\partial q^\iT}q^\zT\\*
\flamoi &+& \left[\frac{2}{\beta^2} (\hat \Phi-\hat \Psi) n^\jT \partial_\jB \hat \Phi
  + 4 \hat \Psi \hat \Psi' \right]q^\iT \frac{\partial \bar
  g}{\partial q^\iT}+ 2 \frac{\partial
  \hat{g}^{(1)}}{\partial q^\iT}q^\iT \hat \Psi'- 2 \hat \Phi L^{\hashamoi(1)}[\bar g,\hat g^{(1)}]\,.\nonumber
\eea

\subsection{The fluid limit}

If we choose a tetrad, not necessarily adapted to this bulk velocity,
then the components of the stress-energy tensor are given in function of the
energy density and pressure by
\bea
T^{\zT
  \zT}_{}&=&\rho_{}+(\rho_{}+P_{})\left[(u_{}^\zT)^2-1\right]=\int\frac{\dd^3
q^\kT}{(2 \pi)^3}g_{}(q^\kT) q^\zT\,,\\*
T^{\zT \iT}_{}&=&\rho_{} u_{}^\zT u_{}^\iT=\int\frac{\dd^3
q^\kT}{(2 \pi)^3}g_{}(q^\kT) q^\iT\,,\\*
T^{\iT \jT}_{}&=&\rho_{} u_{}^\iT u_{}^\iT +P_{} \delta^{\iT \jT}=\int\frac{\dd^3
q^\kT}{(2 \pi)^3}g_{}(q^\kT) \frac{q^\iT q^\jT}{q^\zT}\,.
\eea 
In general for a distribution function $g(q^\bT)$, its associated
stress-energy tensor conservation tensor $\nabla_\mu T^{\mu \nu}$ is obtained by~\cite{Ehlers1971}
\be\label{ConsTmunufromLiouville}
a \nabla_\mu T^{\mu \bT}=\int L^\hashamoi[g] q^\bT \frac{\dd^3 q^\iT}{(2 \pi)^3}\,,
\ee
where we recall that $a$ is the scale factor. We can then perform a perturbative expansion of this expression according to
the expansion~(\ref{defdeltamatter}) and obtain for baryons the perturbation of the continuity equation
\be\label{Consordre0}
\flamoi a \overline{\left(\nabla_\mu T^{\mu \zT} \right)}=\int\frac{\dd^3 q^\kT}{(2 \pi)^3} q^\zT \bar{L}^\hashamoi[\bar g]=\left(\bar
  T^{\zT \zT}\right)'+3\HH \bar T^{\zT \zT}+\HH \bar T^{\iT}_{\,\iT}\,,
\ee

\bea
\flamoi a\left(\nabla_\mu T^{\mu \zT} \right)^{(1)}&=&\int\frac{\dd^3 q^\kT}{(2 \pi)^3} q^\zT L^{\hashamoi(1)}[\bar g,\hat g^{(1)}]\\*
\flamoi &=&\left( \hat T^{\zT \zT(1)}\right)'+3\HH \hat T^{\zT \zT(1)}+\HH
\hat T^{\iT(1)}_{\,\iT}+\partial_\iB \hat T_{}^{\zT \iT(1)}-\hat \Psi'\left(3
  \bar T^{\zT\zT} +\bar T^\iT_{\,\iT}\right)\,,\nonumber
\eea

\bea
\flamoi a\left(\nabla_\mu T^{\mu \zT} \right)^{(2)}&=&\int\frac{\dd^3 q^\kT}{(2 \pi)^3} q^\zT L^{\hashamoi(2)}[\bar g,\hat g^{(1)},\hat g^{(2)}]\\*
\flamoi &=&\left( \hat T^{\zT \zT(2)}\right)'+3\HH \hat T^{\zT \zT(2)}+\HH
\hat T^{\iT(2)}_{\,\iT}+\partial_{\iB} \hat T^{\zT \iT(2)}-\hat \Psi^{'(2)}\left(3
  \bar T^{\zT\zT} +\bar T^\iT_{\,\iT}\right)\nonumber\\*
\flamoi &&+2 (\hat \Psi+\hat \Phi)\partial_{\iB} \hat T^{\zT\iT(1)}-2\hat \Psi' \left(3\hat
  T_{}^{\zT\zT(1)}+\hat T^{\iT(1)}_{{},\iT} \right)+4\hat
T^{\zT\iT(1)}\partial_{\iB}\left(\hat \Phi-\hat \Psi \right)\nonumber\\*
\flamoi &&-4 \left(3  \bar T^{\zT\zT} +\bar T^\iT_{\ib,\iT}\right)\hat \Psi' \hat \Psi\nonumber\\*
\flamoi &&-2 \hat \Phi\left[\left( \hat T^{\zT \zT(1)}\right)'+3\HH \hat T^{\zT \zT(1)}+\HH
\hat T^{\iT(1)}_{\,\iT}+\partial_\iB \hat T^{\zT \iT(1)}-\hat \Psi'\left(3
  \bar T^{\zT\zT} +\bar T^\iT_{\,\iT}\right) \right]\nonumber,
\eea
and the perturbation of the Euler equation
\be
\flamoi a\left(\nabla_\mu T^{\mu \iT} \right)^{(0)}=0\,,
\ee
\bea
\flamoi a\left(\nabla_\mu T^{\mu \iT} \right)^{(1)} &=&\int\frac{\dd^3 q^\kT}{(2 \pi)^3} q^\iT L^{\hashamoi(1)}[\bar g,\hat g^{(1)}]\\*
\flamoi &=&\left( \hat T^{\zT \iT(1)}\right)'+4\HH \hat T^{\zT \iT(1)}+\partial_\jB
\hat T^{\jT \iT(1)}+\partial_\jB\hat \Phi\left(\bar T^{\zT\zT} \delta^{\iT\jT}+\bar T^{\iT\jT} \right)\,,\nonumber
\eea

\bea\label{Eulerordre2}
\flamoi a\left(\nabla_\mu T^{\mu \iT} \right)^{(2)} &=& \int\frac{\dd^3 q^\kT}{(2 \pi)^3} q^\iT L^{\hashamoi(2)}[\bar g,\hat g^{(1)},\hat g^{(2)}]\\*
\flamoi &=&\left( \hat T^{\zT \iT(2)}\right)'+4\HH \hat T^{\zT \iT(2)}+\partial_\jB
\hat T^{\jT \iT(2)}+\partial_\jB\hat \Phi^{(2)}\left(\bar T^{\zT\zT}
  \delta^{\iT\jT}+\bar T^{\iT\jT} \right)\nonumber\\*
\flamoi &&-2\hat\Phi \left[\left( \hat T^{\zT \iT(1)}\right)'+4\HH
  \hat T^{\zT \iT(1)}\right]+(2 \hat \Psi-4\hat\Phi) \partial_\jB\hat
\Phi^{(1)}\left(\bar T^{\zT\zT} \delta^{\iT\jT}+\bar T^{\iT\jT}
\right)\nonumber\\*
\flamoi &&+2\partial_\jB\hat \Phi\left(\hat T^{\zT \zT(1)}
  \delta^{\iT\jT}+\hat T^{\iT\jT(1)} \right)-8\hat \Psi' \hat T^{\zT
  \iT(1)}+2 \hat \Psi \partial_\jB \hat T^{\iT\jT(1)}\nonumber\\*
\flamoi &&+2\left(\partial^\iB \hat \Psi \hat T^{\kT(1)}_{\,\kT}-3 \partial_\jB \hat \Psi \hat T^{\iT\jT(1)}
\right)\,.\nonumber
\eea

In order to further specify the continuity and Euler equations, we need to
express the perturbations of the stress-energy tensor components in function
of the perturbations of the fluid quantities. Due to the strong coupling between electrons and protons through Coulomb
interactions when matter is ionized, the baryons cannot develop significant
anisotropic stress before recombination and thus the corresponding
stress-energy tensor is of the form
\be
T^{\mu \nu}=(P+\rho)u^\mu u^\nu + P g^{\mu \nu}.
\ee
Indeed, the strong interactions between electrons and protons ensure that they
follow a Fermi-Dirac distribution of energies in their rest frame (that is where they have no bulk
velocity).  We also note that since $\bar u^\iT=0$, then the normalization condition of $u_\aT
 u^\aT=-1$ implies
\be 
u^{\zT(1)}=0\,,\qquad  u^{\zT(2)}=u^{\iT(1)}u_{\iT(1)}\,.
\ee
Furthermore we can identify the velocity $v^\iT$ and the spatial momentum $u^\iT$ up to second
order in perturbation since
\be
u^{\iT}=\frac{1}{\sqrt{1-v_\jT v^\jT}} v^\iT\,.
\ee
We then deduce the following perturbative expansion of the stress-energy tensor in
function of the fluid quantities 
\bea
\bar T^{\zT  \zT}&=&\bar \rho\,,\\*
T^{\zT\zT(1)}&=&\rho^{(1)}\,,\\*
T^{\zT\zT(2)}&=&\rho^{(2)}+2(\bar \rho +\bar P)u^{\iT(1)}u_{\,\iT}^{(1)}\,,
\eea

\bea
\bar T^{\zT  \iT}&=&0\,,\\*
T^{\zT\iT(1)}&=&(\bar \rho+\bar P) u^{\iT(1)}\,,\\*
T^{\zT\iT(2)}&=&(\bar \rho+\bar P)\left( u^{\iT(2)}\right)+2(\rho^{(1)}+P^{(1)})u^{\iT(1)}\,,\label{TOiO2}
\eea

\bea
\bar T^{\iT  \jT}&=&\bar P \delta^{\iT\jT}\,,\\*
T^{\iT\jT(1)}&=&P^{(1)}\delta^{\iT\jT}\,,\\*
T^{\iT\jT(2)}&=&P^{(2)}\delta^{\iT\jT}+2(\bar \rho+\bar P)u^{\iT(1)}u^{\jT(1)}\,.\label{TijO2}
\eea
By using these expressions, we obtain the following
gauge invariant conservation and Euler equations from equations~(\ref{Consordre0}-\ref{Eulerordre2})

\be
\flamoi a \overline{\left(\nabla_\mu T^{\mu \zT} \right)}=\bar \rho'+3
\HH(\bar \rho+\bar P)\,,
\ee

\bea\label{Eqcontbar1}
\flamoi a\left(\nabla_\mu T^{\mu \zT} \right)^{(1)}&=&\hat \rho^{(1)'}+3
\HH\left(\hat \rho^{(1)}+\hat P^{(1)}\right)\nonumber\\*
\flamoi &&+(\bar \rho+\bar
P)\partial_\iB \hat u^{\iT(1)}-3(\bar \rho+\bar P)\hat \Psi'\,,
\eea

\bea\label{Eqcontbar2}
\flamoi a\left(\nabla_\mu T^{\mu \zT} \right)^{(2)} &=& \hat \rho^{(2)'}+3
\HH\left(\hat \rho^{(2)}+\hat P^{(2)}\right)+6 \HH(\bar \rho +\bar
P)\hat u^{\iT(1)}\hat u_{\,\iT}^{(1)}\nonumber\\*
\flamoi&&+(\bar \rho+\bar
P)\partial_\iB \hat u^{\iT(2)}-3(\bar \rho+\bar P)\hat
\Psi^{(2)'}+2(\bar \rho' +\bar
P')\hat u^{\iT(1)}\hat u_{\,\iT}^{(1)}\nonumber\\*
\flamoi&&+2 \HH(\bar \rho +\bar P)\hat u^{\iT(1)}\hat
u_{\,\iT}^{(1)}+2\partial_\iB\left[(\hat \rho^{(1)}+\hat P^{(1)})
  \hat u^{\iT(1)}\right]-6\hat \Psi' \left(\hat
  \rho^{(1)}+\hat P^{(1)}\right)\nonumber\\*
\flamoi&&+ 2 (\hat \Phi +\hat \Psi)(\bar \rho+\bar
P)\partial_\iB \hat u^{\iT(1)}+4(\bar \rho+\bar P)\hat
u^{\iT(1)}\partial_\iB(\hat \Phi-\hat \Psi)\nonumber\\*
\flamoi&&-12 \hat \Psi \hat \Psi' (\bar
\rho+\bar P)+4(\bar \rho +\bar
P)\hat u^{\iT(1)}\hat u_{\,\iT}^{(1)'}- 2 \hat \Phi a\left(\nabla_\mu T^{\mu \zT} \right)^{(1)}\,,
\eea

\bea\label{EqEulerbar1}
\flamoi a\left(\nabla_\mu T^{\mu \iT} \right)^{(1)}&=&\left[(\bar \rho+\bar
P)\hat u^{\iT(1)}\right]'+4 \HH \left[(\bar \rho+\bar
P)\hat u^{\iT(1)}\right]\nonumber\\*
\flamoi &&+\partial^\iB \hat P^{(1)}+\partial^\iB\hat \Phi\left(\bar \rho +\bar P \right)\,,
\eea

\bea\label{EqEulerbar2}
\flamoi a\left(\nabla_\mu T^{\mu \iT} \right)^{(2)}&=&\left[(\bar \rho+\bar
P)\hat u^{\iT(2)}\right]'+4 \HH \left[(\bar \rho+\bar
P)\hat u^{\iT(2)}\right]\nonumber\\*
\flamoi &&+\partial^\iB \hat P^{(2)}+\left(\bar
  \rho +\bar P \right)\partial^\iB\hat \Phi^{(2)}+2(\bar \rho +\bar
P)\partial_\jB\left(\hat u^{\iT(1)} \hat u^{\jT(1)}\right)\nonumber\\*
\flamoi &&+2\left[(\hat \rho^{(1)}+\hat P^{(1)})\hat u^{\iT(1)}\right]'+8 \HH \left[(\hat \rho^{(1)}+\hat P^{(1)})\hat u^{\iT(1)}\right]\nonumber\\*
\flamoi &&-2(\hat\Phi +\hat \Psi)\left\{\left[(\bar \rho+\bar
P)\hat u^{\iT(1)}\right]'+4 \HH \left[(\bar \rho+\bar
P)\hat u^{\iT(1)}\right]\right\}\nonumber\\*
\flamoi &&-4\left(\bar \rho + \bar P \right)\hat\Phi \partial^\iB\hat \Phi+2\partial^\iB\hat \Phi\left(\hat \rho^{(1)}+\hat P^{(1)} \right)-8\hat \Psi'(\bar \rho+\bar
P)\hat u^{\iT(1)}\nonumber\\*
\flamoi &&+2 \hat \Psi a\left(\nabla_\mu T^{\mu \iT} \right)^{(1)}\,.
\eea

\subsection{Comparison with the coordinates approach}

In the literature, the fluid equation are usually derived starting directly
from the stress-energy tensor and in components associated with coordinates.
It is straightforward to compare with our results since 
\be
\nabla_\mu T^{\mu \aB}=\nabla_\mu T^{\mu \iT} e_\iT^{\aB}.
\ee
At first order we have
\be
a \left(\nabla_\mu T^{\mu \aB}\right)^{(1)}=\left(\nabla_\mu T^{\mu \aT}\right)^{(1)},
\ee
and at second order
\bea
a \left(\nabla_\mu T^{\mu \zB}\right)^{(2)}&=&\left[ \left(\nabla_\mu T^{\mu
    \zT}\right)^{(2)}+2 \hat\Phi^{(1)}\left(\nabla_\mu T^{\mu \zT}\right)^{(1)}\right]\,,\\*
a \left(\nabla_\mu T^{\mu \iB}\right)^{(2)}&=&\left[ \left(\nabla_\mu T^{\mu
    \iT}\right)^{(2)}-2 \hat\Psi^{(1)}\left(\nabla_\mu T^{\mu \iT}\right)^{(1)}\right]\,.
\eea
The presentation that we have chosen in the previous section makes the use of these
relations simple. However, in order to succeed this comparison it should be noticed that~\cite{Pitrou2007}
\bea\label{choicevectoralignement}
v^{\iT(1)}&=&V^{\iB(1)}\,,\\*
v^{\iT(2)}&=&V^{\iB(2)}-2\Psi V^{\iB(1)}\,,
\eea
since the results in the literature for the fluid approximation are so far expressed in function of $V^\iB$.

\section{Collision term}\label{Sec_Collision}

\subsection{General expression}

The general expression of the collision term is given by~\cite{Uzan1998,Portsmouth2004,Nagirner2001}
\bea\label{DefCollisionbrute1}
\flamoi {C_{\gr{u}}}_{\xS\gS}(\gr{p})&=&{S_{\gr{u}}}_{\,\xS}^{\aS}{S_{\gr{u}}}_{\,\gS}^{\bS}(2\pi)^4\int {\rm D}\gr{q}\int {\rm
  D}\gr{p}' \int {\rm D}\gr{q}'\dirac{4}\left(\gr{p}'+\gr{q}-\gr{p}-\gr{q}'\right)\\*
\flamoi &&\left\{ \,\,\,\,\gel(\gr{q}) f_{\cS\dS}(\gr{p}'){\cal  M}_{\,\,\,\,\,\aS\bS}^{\cS\dS}(\gr{p}',\gr{q};\gr{p},\gr{q}') \right.\nonumber\\*
\flamoi &&\,\, -  \gel(\gr{q}')\left[\eta^{\eS\hS}U_{\aS\bS}(\gr{p})f_{\cS\dS}(\gr{p})  \right]{\cal  M}_{\,\,\,\,\,\eS\hS}^{\cS\dS}(\gr{p},\gr{q}';\gr{p}',\gr{q})\nonumber\\*
\flamoi &&\left.\,\,+\left[\gel(\gr{q})-\gel(\gr{q}')\right] f_{\cS\dS}(\gr{p}'){\cal
  M}_{\,\,\,\,\,\eS\hS}^{\cS\dS}(\gr{p}',\gr{q};\gr{p},\gr{q}') {{\cal N}_{\gr{q}}}_{\,\,\,\,\,\,\aS\bS}^{\eS\hS}(\gr{p}) \right\}\,,\nonumber
\eea
where we used the notation $ {\rm D} \gr{p} \equiv
\dirac{1}(\gr{p}.\gr{p}-m^2)\frac{\dd^4 p^\aT }{(2 \pi)^3} $ .
This expression accounts for the incoming transitions in the second line and
the outgoing transitions in the third line. We have neglected the Pauli
blocking terms $(1-\gel/2)$ since for the physics of the CMB, the electrons are
much more diluted than the photons. However we include the stimulated
emission in the fourth line and discuss later its relevance, and this requires the use of the stimulated emission matrix
\bea
\flamoi &&{{\cal N}_{\gr{q}}}_{\,\,\,\,\,\,\aS\bS}^{\eS\hS}(\gr{p},\gr{q})\equiv\\*
\flamoi && \frac{1}{2}\left[f_{\aS\bS}(\gr{p}) {S_\gr{q}}^{\eS \hS}+\left({S_{\gr{q}}}^{\eS}_{\,\aS}{S_{\gr{q}}}^{\hS}_{\,\bS}-{S_\gr{q}}^{\eS\hS}{S_\gr{q}}_{\aS\bS}\right)I(\gr{p})+{f^\star}^{\eS\hS}(\gr{p}){S_\gr{q}}_{\aS\bS}\right]\,.\nonumber
\eea
In this definition we recall that the projectors ${S_{\gr{q}}}^{\aS\bS}$ are taken with respect to an
observer having a velocity $\gr{q}/m$, which means that
${S_{\gr{q}}}^{\aS\bS}q_\bS=0$, and ${f^\star}^{\eS\hS}$ is the complex
conjugate of ${f}^{\eS\hS}$. The stimulated emission in the case of
polarized light is discussed in \cite{Kosowsky1996,Nagirner2001,Hansen1999}.
These treatments give the stimulated emission matrix in terms of the Stokes
parameters. However, we do not want to refer to them since they imply and
choice of the angle in the polarization plane basis, but it can be checked that the
expression given here for the stimulated emission tensor is equivalent to the
expression of these references when expressed in terms of the Stokes parameters.
These parameters are an intermediary step between the polarization tensor and the multipoles which is
not compulsory, and we thus intentionally bypass them in our derivations. In the end, this method will prove more powerful because it is computationally more straightforward.\\
The Dirac function in equation~(\ref{DefCollisionbrute1}) ensures momentum conservation and ${\cal
  M}_{\,\,\,\,\,\,\,\eS\hS}^{\cS\dS}(\gr{p},\gr{q};\gr{p}',\gr{q}')$ is the
transition tensor for the process
\be
\gamma(\gr{p})+ e^{-}(\gr{q})\longleftrightarrow\gamma(\gr{p}')+ e^{-}(\gr{q}')\,.
\ee
Its expression is given by
\bea\label{DefMatrix}
\flamoi &&{\cal M}_{\,\,\,\,\,\,\aS\bS}^{\cS\dS}(\gr{p},\gr{q};\gr{p}',\gr{q}')=\\*
\flamoi &&\frac{3}{2} \pi \st
\me^2\left\{{Q_\gr{q}}_{\,\aS\bS}^{\cS\dS}+\frac{1}{4}\left[\frac{\gr{p}'.\gr{q}}{\gr{p}.\gr{q}}+\frac{\gr{p}.\gr{q}}{\gr{p}'.\gr{q}}-2\right]\left({S_\gr{q}}^{\cS\dS}{S'_\gr{q}}_{\aS\bS}+{Q_\gr{q}}^{\cS\dS}_{\,\aS\bS}-{Q_\gr{q}}^{\dS\cS}_{\,\aS\bS} \right) \right\}\nonumber
\eea
where
\be
{Q_\gr{q}}^{\mu \nu}_{\,\alpha \beta}\equiv
{S_\gr{q}}^{\mu}_{\,\,\gamma}{S_\gr{q}}^{\nu}_{\,\,\delta}{S'_\gr{q}}^{\gamma}_{\,\,\alpha}{S'_\gr{q}}^{\delta}_{\,\,\beta}\equiv
{S_\gr{q}}^{\mu}_{\,\,\gamma}(\gr{p}){S_\gr{q}}^{\nu}_{\,\,\delta}(\gr{p}){S_\gr{q}}^{\gamma}_{\,\,\alpha}(\gr{p}'){S_\gr{q}}^{\delta}_{\,\,\beta}(\gr{p}')\,.
\ee
The detail of its derivation can be found in~\cite{Portsmouth2004}.
Intuitively, it consists in transforming the incoming polarization tensor to
the rest frame of the incoming electron, and applying the Klein-Nishina formula
for the amplitude of the scattering process. Thanks to the reversibility of the scattering process, the transition tensor satisfies
\be
{\cal   M}^{\cS\dS}_{\,\,\,\,\,\,\,\aS\bS}(\gr{p}',\gr{q}' ;\gr{p},\gr{q})={S_{\gr{q}'}}^{\cS}_{\,\gS}\,{S_{\gr{q}'}}^{\dS}_{\,\hS}\,{\cal M}_{\,\,\,\,\,\,\,\eS\fS}^{\gS\hS}(\gr{p},\gr{q};\gr{p}',\gr{q}'){S_{\gr{q}'}}^{\eS}_{\,\aS}\,{S_{\gr{q}'}}^{\fS}_{\,\bS}\,,
\ee
as it can be checked directly on its expression~(\ref{DefMatrix}). It is then
possible to recast the collision term as
\bea\label{DefCollisionameliore}
\flamoi {C_{\gr{u}}}_{\xS\gS}(\gr{p})&=&{S_{\gr{u}}}_{\,\xS}^{\aS}{S_{\gr{u}}}_{\,\gS}^{\bS}
\int {\rm D}\gr{q} \int {\rm D}\gr{p}' \int {\rm D}\gr{q}' \,(2
\pi)^4\dirac{4}\left(\gr{p}'+\gr{q}-\gr{p}-\gr{q}'\right)\\*
\flamoi && \quad \times \gel(\gr{q}) f_{\cS\dS}(\gr{p}'){\cal
  M}_{\,\,\,\,\,\eS\hS}^{\cS\dS}(\gr{p}',\gr{q};\gr{p},\gr{q}')\left\{ \delta^{\eS}_{\aS} \delta^{\hS}_{\bS}+ {{\cal N}_{\gr{q}}}_{\,\,\,\,\,\,\aS\bS}^{\eS\hS}(\gr{p})\right\}\nonumber\\*
\flamoi &-&{S_{\gr{u}}}_{\,\xS}^{\aS}{S_{\gr{u}}}_{\,\gS}^{\bS}\int {\rm D}\gr{q}{\rm D}\gr{p}' {\rm D}\gr{q}' \,(2
\pi)^4\dirac{4}\left(\gr{p}+\gr{q}-\gr{p}'-\gr{q}'\right)\nonumber\\*
\flamoi && \quad \times \gel(\gr{q})\left\{\eta^{\eS\hS}U_{\aS\bS}(\gr{p})f_{\cS\dS}(\gr{p})+{{\cal
    N}_{\gr{q}}}_{\cS\dS\aS\bS}(\gr{p})f^{\eS\hS}(\gr{p}' )\right\}{\cal M}_{\,\,\,\,\,\eS\hS}^{\cS\dS}(\gr{p},\gr{q};\gr{p}',\gr{q}')\,.\nonumber
\eea
The remaining two screen projectors at the very beginning of equations~(\ref{DefCollisionbrute1},\ref{DefCollisionameliore}) ensure that the result is
read in the required frame (usually the cosmological frame). The form of the
expressions~(\ref{DefCollisionbrute1},\ref{DefCollisionameliore}) is manifestly covariant. First, the
measures ${\rm D}\gr{p}$ and ${\rm D}\gr{q}$, the Dirac function and the
electron distribution function $g_\ie(\gr{p})$ are
Lorentz invariant. Second, given the transformation rule~(\ref{Eq_Tpropertydebase}) for the
polarization tensor, the screen projectors in the transition tensor and the
stimulated emission tensor ensure that  $f_{\cS\dS}(\gr{p}'){\cal
  M}_{\,\,\,\,\,\aS\bS}^{\cS\dS}(\gr{p}',\gr{q};\gr{p},\gr{q}')$ and $f_{\cS\dS}(\gr{p}'){\cal
  M}_{\,\,\,\,\,\eS\hS}^{\cS\dS}(\gr{p}',\gr{q};\gr{p},\gr{q}'){f^\star_{\gr{q}}}^{\eS\hS}(\gr{p})$
are independent from the frame used to evaluate $f_{\cS\dS}$. Finally we note
that the screen projectors ${S_{\gr{u}}}_{\,\xS}^{\aS}{S_{\gr{u}}}_{\,\gS}^{\bS}$ are defined with respect to the observer having a velocity $\gr{u}$, which is the
velocity with respect to which the collision tensor in equations~(\ref{DefCollisionbrute1},\ref{DefCollisionameliore}) is defined (in our case $u_\mu=(\dd \eta)_\mu$). Hence they ensure that the expression of the collision tensor is independent from
the observer (that is the frame) used to define the unit polarization tensor $U_{\aS\bS}$ appearing in the third line of
equation~(\ref{DefCollisionbrute1}) and the fourth line of equation~(\ref{DefCollisionameliore}), and also independent from
the observer used to define the polarization tensor $f_{\aS\bS}$ appearing in
the stimulated emission tensor. Additionally these screen projectors makes it
straightforward to check that the collision tensor transforms as in the rule~(\ref{Eq_Tpropertydebase}).\\

Given this discussion, it is in principle possible to express this collision
term in the fundamental tetrad basis, in order to specify further the
collision tensor. However, all the screen projectors appearing in the transition matrix and in the stimulated emission
tensor are taken with respect to the observer with velocity $\gr{q}$ which varies throughout the integration $\int {\rm
  D}\gr{q}$. In order to simplify the computations, it would thus be helpful to use an intermediate tetrad basis adapted to each momentum $\gr{q}$ in this
integration. Additionally, instead of expressing the collision tensor in the
fundamental frame or tetrad, one would prefer to compute it, that is to obtain
it explicitly in function of the radiation multipoles, in the electrons rest
frame. In a second step, we would then transform the collision term in the
fundamental frame using its transformation rule which is given by
equation~(\ref{Eq_Tpropertydebase}), and the transformation rules of the
radiation multipoles and momentum components under this change of frame that we gave in section~\ref{Sec_Changeframe}. We thus need for this final
step to determine up to which order in the electrons velocity we need to compute this change of frame
to be consistent with the perturbative expansion of the Boltzmann equation.\\

\subsection{A double perturbative expansion}\label{Sec_pertscheme}

When performing a perturbative expansion, we need to expand both in the metric
perturbations and in the velocity perturbations. However these perturbations
are of different type and also have a different magnitude. The magnitude of
the metric perturbations are typically of order $10^{-5}$ for the processes
involved in the CMB, and thus the bulk velocities of the radiation and matter
distribution are also of the same amplitude. However the temperature of
radiation ($T_\ir$) and the temperature of electrons ($T_\ie$), which are
nearly identical around  recombination are approximately of order $T_\ir/\me\simeq T_\ie/\me \simeq 10^{-6}$ and thus the thermal velocity of
electrons is of order $10^{-3}$. Consequently, when going up to second order
in metric perturbations, we would need in principle to keep perturbations up
to fourth order in velocities as they can in the end lead to terms of order
$(T_\ir/\me)^2$, $T_\ir T_\ie /\me^2$ and $(T_\ie/\me)^2$, which can be comparable to the second order perturbations of
the metric if the numeric coefficients in front of these terms are quite large.
It can be shown~\cite{Challinor1997,Itoh1997,Nozawa2005} that in fact, due to Comptonization processes, these terms will be of order
$(\bar T/\me)^2 \times (T_\ir-T_\ie )/\bar T$ and will be thus completely
negligible in our context. However we would certainly need to retain terms up
to third order in the baryons velocity, that we will call accordingly to this
discussion of order $3/2$ in the metric perturbation (though there is no such
half integer perturbation for the metric).\\

We will now apply this method to perform a perturbative expansion of the
collision term
\begin{itemize}
\item in section~\ref{CollisionnobulknoT} for electrons having no thermal dispersion and choosing for this computation the rest frame of electrons which is also the rest frame of baryons,
\item in section~\ref{Collisionnobulk} for electrons having a thermal
  dispersion, choosing the same frame for the computation,
\item in section~\ref{Collisionwithbulk} for the general case, accounting
  for the bulk velocity of electrons by transforming the result obtained in the baryons rest frame toward the fundamental frame. 
\end{itemize}

\subsection{Distribution of cold electrons with no bulk velocity}\label{CollisionnobulknoT}

It will prove convenient to calculate the collision term in the Boltzmann
equation starting from a very simple case. We will first assume that the free
electrons have only a bulk velocity and no thermal dispersion in their
velocity distribution function $\gel(q^\iT)$. Furthermore, we will choose to align the first vector of the
tetrad $\gr{e}_\zT$ (an adapted tetrad) with this bulk velocity that is to
work in the baryons rest frame in order to specify the collision tensor given in
equation~(\ref{DefCollisionameliore}). In that case the electron distribution
function is explicitly given by $\gel(q^\iT)=(2
\pi)^3\dirac{3}(q^\iT)n_{\mathrm e}$ which makes the integration on electrons
momenta trivial. In the baryons rest frame, the velocity of the incoming
electron is aligned with the tetrad $\gr{e}_\zT$, and we obtain from the momentum conservation in the collision
\be\label{Devunsrum}
\frac{1}{q^{\zT}{q}^{'\zT}}=\frac{1}{\me^2}\left[1-\frac{(p^\iT-p'^\iT)(p_\iT-p'_{\iT})}{2 \me^2}\right]\,.
\ee
The second term in the brackets of this expression is of order $(T_\ir/\me)^2$, that is comparable to
second order perturbations in the metric, and according to the discussion in
section~\ref{Sec_pertscheme} they can be ignored. We then need to expand the Dirac
function. We first integrate on the spatial components of the outgoing
electron momentum since the integrand does not depend on it in the
form~(\ref{DefCollisionameliore}) of the collision tensor. The remaining one-dimensional Dirac function is handled by following the standard steps that can be found
in~\cite{Bernstein1988}, and we obtain
\be
\flamoi\dirac{1}(p'^{\zT}+q^{\zT}-p^{\zT}-q'^{\zT})=\dirac{1}(p^\zT-p'^{\zT})-\frac{(p^\iT-p'^\iT)(p_\iT-p'_\iT)}{2 \me}\frac{\partial}{\partial p'^{\zT}}\dirac{1}(p'^{\zT}-p^{\zT})\,,
\ee
\be
\flamoi\dirac{1}(p'^{\zT}+q'^{\zT}-p^{\zT}-q^{\zT})=\dirac{1}(p^\zT-p'^{\zT})+\frac{(p^\iT-p'^\iT)(p_\iT-p'_\iT)}{2 \me}\frac{\partial}{\partial p'^{\zT}}\dirac{1}( p'^{\zT}-p^{\zT})\,.
\ee
Additionally, using the fact the tetrad corresponds to the rest frame of all electrons,
\be
\frac{1}{p^\zT}=\frac{1}{p'^{\zT}}+\frac{1}{\me}\left(1-n^\iT n'_\iT \right)\,,
\ee
which is more conveniently expressed as
\be
\frac{p'^{\zT}}{p^{\zT}}=1+\frac{p'^{\zT}}{\me}\left(1-n^\iT n'_\iT\right)\,.
\ee
As a consequence, the term in square brackets in the
definition~(\ref{DefMatrix}) of the transition matrix is of order $(p^\zT/\me)^2 \sim (T_\ir/\me)^2$ and is comparable to
second order perturbations of the metric. This would not have been the case if the tetrad was not
adapted to the incoming electron velocity and in that case the term in
brackets in equation~(\ref{DefMatrix}) would be of order $T_\ir/\me$. This property
is important since, given the discussion in section~\ref{Sec_pertscheme}, they can be ignored. Finally using 
\be
n_{\mathrm e} =\int \frac{\dd^3 q^{\iT}}{(2\pi)^3}
\gel\left(q^\hT\right)\,,
\ee
the collision term is expressed as
\be
C_{\aT \bT}\left(p^\hT\right)=  n_{\mathrm e} \st\, \left[{C}_{\aT \bT,\,\mathrm{T}}\left(p^\hT\right)+{C}_{\aT \bT,\,\mathrm{R}}\left(p^\hT\right)\right]\,,
\ee
where
\be\label{Cratepere1}
\flamoi C_{\aT \bT,\,\mathrm{T}}\left(p^\hT\right)=p^\zT\left[\frac{3}{2}\int \frac{\dd^2 \Omega'}{4 \pi}
  S_{\aT}^{\,\cT} S_{\bT}^{\,\dT} f_{\cT \dT}\left(p'^\hT\right) -f_{\aT \bT}\left(p^\hT\right)\right]\,,
\ee
\bea\label{Cratepere2}
\flamoi && {C}_{\aT \bT,\,\mathrm{R}}(p^\hT)=-\frac{3}{2}\int \dd p'^\zT  p'^{\zT}\int \frac{\dd^2 \Omega'}{4 \pi}\left[\frac{( p^\zT)^2+( p'^{\zT})^2-2  p'^{\zT} p^{\zT} {n}^\iT n'_\iT}{2 \me}\right]\frac{\partial \dirac{1}( p'^{\zT}- p^{\zT})}{\partial
  p'^{\zT}}\\*
\flamoi && \times \left\{S^{\cT \dT}f_{ \cT
      \dT}\left(p'^\hT\right)\left[ f_{ \aT
    \bT}(p^\hT)-I(p^\hT)\frac{S_{\aT\bT}}{2}\right] +\left[1+I(p^\hT)\right]S_{\aT}^{\cT}S_{\bT}^{\dT}f_{\cT
    \dT}\left(p'^\hT\right)\right\}+2\frac{(p^\zT)^2}{\me}f_{ \aT
    \bT}(p^\hT).\nonumber
\eea
We recall that in these expressions, the screen projector $S^{\cT \dT}$ is
defined with respect to the photon with momentum $\gr{p}$, and thus $S^{\cT
  \dT}f_{ \cT  \dT}(p'^\hT) \neq I(p'^\hT)$. Note also that $S^{\cT
  \dT}S'_{ \cT  \dT} =1+ (\gr{n}.\gr{n}')^2$~\cite{Portsmouth2004}, and this helps recovering from
the trace of the Thomson term~(\ref{Cratepere1}) the standard form of the Thomson collision term for unpolarized radiation.
Using the integrals~(\ref{Intonn}) we find explicitly the Thomson and the
recoil terms which are given by
\bea\label{CollisionThomson}
\flamoi \frac{C_{\aT \bT,\,\mathrm{T}}(p^\hT)}{p^\zT}&=&\frac{1}{2}S_{\aT \bT}\left[- I(p^\zT,n^\iT)+I_\emptyset(p^\zT)+\frac{1}{10}I_{\cT\dT}(p^\zT)n^\cT
  n^\dT-\frac{3}{5}E_{\cT \dT}(p^\zT)n^\cT n^\dT\right]\nonumber\\*
\flamoi &&+\left[-P_{\aT \bT}(p^\zT,n^\iT)-\frac{1}{10}I_{\aT \bT}(p^\zT)
  +\frac{3}{5}E_{\aT \bT}(p^\zT) \right]^{\mathrm{TT}}\nonumber\\*
\flamoi &&+\frac{1}{2}\ii \epsilon_{\aT \bT \cT}n^\cT\left[-V(p^\zT,n^\iT)+\frac{1}{2}V_\dT(p^\zT)n^\dT \right]\,,
\eea

\bea
\flamoi\frac{C_{\aT
    \bT,\,\mathrm{R}}(p^\hT)}{p^\zT}&=&\frac{p^\zT}{\me}\left\{\frac{1}{2}S_{\aT
    \bT}\left\{2 I(p^\zT,n^\iT)+\left[1+I(p^\zT,n^\iT)\right]\left(2+p^\zT \frac{\partial}{\partial
      p^\zT}  \right)\right.\right.\nonumber\\*
\flamoi&&\qquad \left[I_\emptyset(p^\zT)-\frac{2}{5}I_\cT(p^\zT)n^\cT+\frac{1}{10}I_{\cT
      \dT}(p^\zT)n^\cT n^\dT -\frac{3}{70}I_{\cT
      \dT \hT}(p^\zT)n^\cT n^\dT n^\hT \right.\nonumber\\*
\flamoi&&\qquad\quad \left.\left.-\frac{3}{5}E_{\cT \dT}(p^\zT)n^\cT n^\dT +\frac{1}{7}E_{\cT
      \dT \hT}(p^\zT)n^\cT n^\dT n^\hT\right]\right\}\nonumber\\*
\flamoi&+&\left\{P_{\aT
    \bT}(p^\zT,n^\iT)\left[2+I'_\emptyset(p^\zT)+2 I_\emptyset(p^\zT)
  \right]+\left[1+I_\emptyset(p^\zT)\right]\left(2+p^\zT \frac{\partial}{\partial
      p^\zT}  \right)\right.\nonumber\\*
\flamoi&&\qquad\left[+\frac{3}{5}E_{\aT \bT}(p^\zT) -\frac{1}{7}E_{\aT \bT \cT}(p^\zT)n^\cT -\frac{2}{5}B_{\cT
      \aT}(p^\zT)\epsilon_{\bT}^{\,\,\cT \dT}n_{\dT}\right.\nonumber\\*
\flamoi&&\left.\left.\qquad\qquad\qquad\qquad-\frac{1}{10}I_{\aT\bT}(p^\zT) +\frac{3}{70}I_{\aT\bT
  \cT}(p^\zT)n^\cT\right]\right\}^{\mathrm{TT}}\nonumber\\*
\flamoi&+&\frac{1}{2}\ii \epsilon_{\aT \bT \cT}n^\cT\left\{V(p^\zT,n^\iT)\left[2+I'_\emptyset(p^\zT)+2 I_\emptyset(p^\zT) \right]+\left[1+I_\emptyset(p^\zT)\right]\left(2+p^\zT\frac{\partial}{\partial
      p^\zT}  \right)\right.\nonumber\\*
\flamoi&&\qquad\qquad\quad \left.\left.\left[-\frac{1}{2}V_\emptyset(p^\zT)+\frac{1}{2}V_\cT(p^\zT)n^\cT-\frac{1}{5}V_{\cT \dT}(p^\zT)n^\cT n^\dT \right]\right\}\right\}\,.
\eea
We recall that, according to the notation of section~\ref{Transfomultipoles},
$I'_{\aT_\ell}(p^\zT)\equiv p^\zT \frac{\partial}{\partial p^\zT}
I_{\aT_\ell}(p^\zT)$. We see clearly by examining the antisymmetric part of the Thomson and recoil
terms, i.e. the terms proportional to $\ii \epsilon_{\aT \bT \cT}n^\cT$, that the circular polarization is not
excited and thus remains null if it is initially so. From now on, we will
discard the antisymmetric term in the polarization tensor and in the Boltzmann
equation. Additionally, since there is a small amount of energy transferred from
the electrons to the photons in the recoil term $C_{\aT \bT,\,\mathrm{R}}$, there is also energy
transferred from the photons to the electrons, and it is incompatible with a
cold distribution of electrons. In order to have a consistent treatment of the
collision term, we can still work in the electrons rest frame but we have at
least to consider a thermal distribution of electrons. Consequently we turn to this case in
the next section, but the results obtained for a cold distribution of
electrons can be used to derive the collision tensor for this more general case. 

\subsection{Thermal distribution of electrons with no bulk velocity}\label{Collisionnobulk}

If the distribution of the velocities of free electrons is thermal then we can
apply the previous method for each of the electrons. That is for each
momentum $\gr{q}$ in the integral on distribution function $\gel(\gr{q})$, we choose a
tetrad field $\tilde{\gr{e}}_a$ whose timelike vector $\tilde{\gr{e}}_\zT$ is
aligned with this momentum ($\tilde{\gr{e}}_\zT = \gr{q}/\me$). We then
calculate the collision rate per electron defined in equations~(\ref{Cratepere1}) and (\ref{Cratepere2}) in this frame [that
is $\st\tilde{n}_\ie\tilde {C}_{\aTt \bTt,\,\mathrm{T}}(p^\cTt)$ and  $\st\tilde{n}_\ie\tilde {C}_{\aTt
  \bTt,\,\mathrm{R}}(p^\cTt)$], and since these quantities transform similarly
to equation~(\ref{Eq_Tpropertydebase}), then the collision term is expressed by
\be\label{Collisionthermalnobul}
\flamoi C_{\aT \bT}(p^\hT)= \st \int \frac{\dd^3 q^{\iT}}{(2\pi)^3q^\zT}
\me g_\ie(q^\hT) S_\aT^{\,\cTt}S_\bT^{\,\dTt}\left[\tilde{\cal C}_{\cTt \dTt,\,\mathrm{T}}\left(p^\hTt\right)+\tilde{\cal C}_{\cTt \dTt,\,\mathrm{R}}\left(p^\hTt\right)\right]\,.
\ee
In practice this means that we need to use the transformation rules of the
radiation multipoles, of the energy and of the momentum direction, in order to express everything inside this integral in
function of the multipoles taken in the fundamental frame. The interest
lies in the fact that integrals on the electron distribution function with odd powers of $\gr{v}$ (the velocity of the electron
considered) vanish. Since we need only to go up to third order in the
thermal velocities, then in that case we can restrict to second order in
thermal velocity. \\
However we have many reasons to follow a much simpler track. First the terms
coming from the second order change of frame (that is second order in
$\gr{v}$) will lead after integration on the electron momentum to terms
having a factor at least $T_{\mathrm e}/\me\equiv(\bar T +\delta T_{\mathrm
  e})/ \me$, and following the discussion of section~\ref{Sec_pertscheme},
this factor can be comparable to first order perturbations in the metric ($\bar T/\me$) or second order ($\delta T_{\mathrm
  e}/ \me$). Since we stop our expansion at second order, we can have in each
of this term either a quantity whose smallest non-vanishing perturbation is
the first order ($I_{\uline{\aT_\ell}}$ for $\ell \ge 1$) or a quantity
whose smallest non-vanishing perturbation is the background ($I_\emptyset$). If it is the
latter case, then the term contributes to the famous
Kompaneets collision term~\cite{Kompaneets1957}. If it is the former case, then the term contributes to what can considered as corrections to the Kompaneets term for anisotropic radiation, but since it is linear in first order perturbations, this type of term cannot contribute to the
bispectrum~\cite{PUB2008}. Similarly, the terms coming from the recoil term have an overall factor of
$p^\zTt/\me$ which is typically of order $T_\ir/\me$, and for these terms
the same reasoning applies. Consequently, when interested in the bispectrum, we should in
principle just keep the Kompaneets term. However the expression of this
famous Kompaneets collision term which is contained in the total collision
term~(\ref{Collisionthermalnobul}) is 
\be\label{Kompaneets}
\flamoi C_{\cT \dT,\,K}\left(p^\hT\right)=n_{\mathrm e}\st \frac{S_{\cT \dT}}{2}\frac{1}{\me
  p^\zT}\frac{\partial}{\partial p^\zT}\left\{(p^\zT)^4\left[ T_{\mathrm
      e} \frac{\partial I_\emptyset(p^\zT)}{\partial p^\zT}+I_\emptyset(p^\zT)
  \left( 1+\frac{1}{2}I_\emptyset(p^\zT)\right)\right] \right\},
\ee
and it can be checked that for $I_\emptyset(p^\zT)/2$ following a Bose-Einstein
distribution of temperature $T_\ir=T_\ie$, then the Kompaneets collision term
exactly vanishes, and if these two temperatures are close, as it is the case
when recombination occurs, then it is of order
$(T_\ir-T_{\mathrm e})/\me$. The Kompaneets contribution, which is at least a second order
quantity in the sense of the discussion in section~\ref{Sec_pertscheme}, bears
thus only a linear response in the (primordial) first order metric
perturbations through $\bar T/ \me(T^{(1)}_\ir-T^{(1)}_{\mathrm e})/ \bar T$. Consequently it cannot
contribute to the bispectrum generated by evolution, and we will ignore it in the rest of this paper. Note
also that on the form~(\ref{Kompaneets}) of the Kompaneets collision term, it
is obvious that $C^\hashamoi_{\cT \dT,\,K}\equiv a C_{\cT \dT,\,K}\,/p^\zT$
does conserve the photons number density~\cite{Bernstein1988,Hu1994} and not
the energy density (it is known to induce spectral distortions) as it is not the case on
the expression (4.38) given in \cite{Bartolo2006}. Hence, the integrated
counterpart of the Kompaneets term, ${\cal
  C}^\hashamoi_{\aT\bT}$, does not vanish and it is only because we focus on the bispectrum that
we may discard it.\\
Gathering all these remarks, we conclude that for a thermal distribution
of electrons with no bulk velocity, the collision term is given by
\be\label{CollisionApprox2}
C_{\aT \bT}(p^\cT)= n_{\mathrm e} \st S_\aT^{\,\cT}S_\bT^{\,\dT}{\cal
    C}_{\cT \dT,\,\mathrm{T}}(p^\cT)\,+\,\Or(1)\left[\Or(T_{\ie}/\me)+\Or(p^\zT/\me)\right],
\ee
and that the terms of order $\Or(T_{\mathrm e}/\me)$ and $\Or(p^\zT/\me)$ are irrelevant for the purpose of computing the
bispectrum generated by evolution. This method which consists in working in
the baryons rest frame simplifies the physical
interpretation of the collision tensor. Indeed on the
form~(\ref{CollisionThomson}) and (\ref{CollisionApprox2}) we clearly see that
in this frame in the tight coupling limit (that is when $C_{\aT
  \bT,\,\mathrm{T}}=0$) there is no polarization ($P_{\aT \bT}=0$) and the
radiation is moving with the baryons as if it was a single fluid since the
only non-vanishing multipole is ${\cal I}_\emptyset$. This terminates
rigorously the discussion in section II.C.2 of~\cite{PUB2008} (see also the
discussion at the end of section~\ref{Sec_multipoles}). 

\subsection{Thermal distribution with bulk velocity}\label{Collisionwithbulk}

If the distribution of electrons has indeed a bulk velocity in the
cosmological frame, then we just need to transform the result obtained in the
baryons rest frame to the cosmological frame, and in this
process, we keep only the terms which can contribute up to second order when
performing perturbations. For the sake of comparison with literature, we only
report the intensity part of the collision term and present here $C^\hashamoi$ rather than $C$. We obtain
\bea\label{Collisionbrutestar}
\flamoi S^{\aT \bT}C^\hashamoi_{\aT \bT}(p^\hT) &=&\tau'\left\{\left[-I(p^\zT,n^\iT)+I_\emptyset(p^\zT)+\frac{1}{10}I_{\cT\dT}(p^\zT)n^\cT
  n^\dT\right]\left(1-\gr{n}.\gr{v}\right)\right.\\*
\flamoi &&\quad -I'_\emptyset(p^\zT)\gr{n}.\gr{v}+\frac{7}{10}I_\cT(p^\zT)v^\cT
+\frac{3}{10}I'_\cT(p^\zT)v^\cT+\frac{2}{10}I_{\cT\dT}(p^\zT)n^\cT
n^\dT\gr{n}.\gr{v}\nonumber\\*
\flamoi &&\quad-\frac{1}{10}I'_{\cT\dT}(p^\zT)n^\cT
n^\dT\gr{n}.\gr{v}-\frac{1}{10}I_{\cT}(p^\zT)n^\cT\gr{n}.\gr{v}+\frac{1}{10}I'_{\cT}(p^\zT)n^\cT\gr{n}.\gr{v}\nonumber\\*
\flamoi &&\quad -\frac{1}{5}I_{\cT\dT}(p^\zT)n^\cT
v^\dT+\frac{6}{35}I_{\cT\dT\hT}(p^\zT)n^\cT n^\hT v^\dT+\frac{3}{70}I'_{\cT\dT\hT}(p^\zT)n^\cT n^\hT v^\dT \nonumber\\*
\flamoi &&\quad +I'_\emptyset(p^\zT)\gr{v}.\gr{v}+\frac{3}{20}I''_\emptyset(p^\zT)\gr{v}.\gr{v}+\frac{11}{20}I''_\emptyset(p^\zT)\left(\gr{n}.\gr{v}\right)^2+I'_\emptyset(p^\zT)\left(\gr{n}.\gr{v}\right)^2\nonumber\\*
\flamoi &&\quad-\frac{3}{5}E_{\cT \dT}(p^\zT)n^\cT
  n^\dT(1+\gr{v}.\gr{n})+\frac{3}{5}E'_{\cT \dT}(p^\zT)n^\cT
  n^\dT\gr{v}.\gr{n}+\frac{6}{5}E_{\cT \dT}(p^\zT)n^\cT v^\dT\nonumber\\*
\flamoi &&\quad-\frac{4}{7}E_{\cT \dT \hT}(p^\zT)n^\cT n^\dT  v^\hT-\frac{1}{7}E'_{\cT \dT \hT}(p^\zT)n^\cT n^\dT
  v^\hT\nonumber\\*
\flamoi &&\quad\left.+\frac{2}{5}B_{\cT\dT}(p^\zT)\epsilon^{\fT\hT\dT}v_{\hT}n_{\fT} n^\cT+\frac{2}{5}B'_{\cT\dT}(p^\zT)\epsilon^{\fT\hT\dT}v_{\hT}n_{\fT} n^\cT \right\}+\Or(T_{\mathrm e}/\me,p^\zT/\me)\,,\nonumber
\eea
where $\tau'\equiv a n_{\mathrm e} \st$\footnote{Note here that $n_{\mathrm
    e}$ is the electron number density as seen in the electrons rest frame. Though we have changed the
frame for expressing the collision term we keep this definition for the
electron number density since this is the most physical. However, a
change of frame only affects the number density by a factor $\gamma\simeq
1+\gr{v}.\gr{v}/2$, and since the background collision term vanishes, the
second order perturbation of the collision term is not affected by the precise choice of frame used to define the electron number density.}.\\
Though we have also obtained the full collision term for the polarization
part, we only report here the result of its energy integrated counterpart, in order to
simplify the expressions obtained.
\bea
\flamoi {\cal C}^\hashamoi_{\aT \bT}(n^\iT)&=&\tau'\frac{1}{2}S_{\aT \bT}\left[\left(
  -{\cal I}(n^\iT)+{\cal I}_\emptyset+\frac{1}{10}{\cal I}_{\cT\dT}n^\cT
  n^\dT\right)\left(1-\gr{n}.\gr{v}\right)\right.\\*
\flamoi &&\qquad\quad-\frac{1}{2}{\cal I}_\cT v^\cT-\frac{1}{2}{\cal I}_\cT n^\cT
\gr{n}.\gr{v}+\frac{3}{5}{\cal I}_{\cT\dT} n^\cT n^\dT \gr{n}.\gr{v}\nonumber\\*
\flamoi &&\qquad\quad-\frac{1}{5}{\cal I}_{\cT\dT} n^\cT v^\dT +4{\cal I}_\emptyset \gr{n}.\gr{v}-{\cal I}_\emptyset\gr{v}.\gr{v}+7{\cal I}_\emptyset(\gr{v}.\gr{n})^2\nonumber\\*
\flamoi &&\qquad\quad\left.-\frac{3}{5}{\cal E}_{\cT \dT}n^\cT
  n^\dT(1+5\gr{v}.\gr{n})+\frac{6}{5}{\cal E}_{\cT \dT}n^\cT
  v^\dT-\frac{6}{5}{\cal B}_{\cT\dT}\epsilon^{\fT\hT\dT}v_{\hT}n_{\fT} n^\cT\right]\nonumber\\*
\flamoi&& + \tau'\left[-{\cal P}_{\aT
      \bT}(n^\iT)(1-\gr{n}.\gr{v})+\left(-\frac{1}{10}{\cal I}_{\aT \bT}
      +\frac{3}{5}{\cal E}_{\aT \bT}\right)(1+3\gr{v}.\gr{n})\right.\nonumber\\*
\flamoi &&\qquad\left.+\frac{6}{5}{\cal E}_{\aT \cT}n^\cT
  v_\bT-\frac{1}{5}{\cal I}_{\aT \cT}n^\cT v_\bT+\frac{6}{5}{\cal
    B}_{\aT}^{\,\,\dT}\epsilon_{\bT\hT\dT}v^{\hT}+\frac{1}{2}{\cal I}_\aT
  v_\bT-{\cal I}_\emptyset v_\aT v_\bT\right]^{\mathrm{TT}}\nonumber\\*
\flamoi&&+{\cal O}(T_{\mathrm e}/\me,p^\zT/\me)\,.\nonumber
\eea
The trace part of this energy-integrated collision term as well as its
non-energy integrated counterpart which is given in equation~(\ref{Collisionbrutestar}) reproduce the results obtained
previously in \cite{Hu1994,Bartolo2006,Dodelson1993} for unpolarized radiation.

\subsection{Multipole decomposition}\label{Sec_multipoles}

This last expression can then be decomposed in multipoles, in order to
separate the intensity from the linear polarization part.
\bea\label{Collision2Ibrut}
\flamoi \frac{1}{\tau'}{\cal C}^\hashamoi[{\cal I}]_{\uline{\aT_\ell}}&=&-{\cal I}_{\uline{\aT_\ell}}+\frac{(\ell+1)}{(2\ell+3)}{\cal I}_{\uline{\aT_\ell}
\bT}v^\bT+{\cal I}_{\langle \uline{\aT_{\ell-1}}}v_{\aT_\ell\rangle}+\delta_\ell^0\left[\frac{4}{3}{\cal I}_\emptyset v_\bT v^\bT-\frac{2}{3}{\cal I}_\bT v^\bT+{\cal I}_\emptyset\right]\\*
\flamoi && +\delta_\ell^1\left[3 {\cal I}_\emptyset
  v_{\aT_1}\right]+\delta_\ell^3\left[\frac{1}{2}{\cal I}_{\langle \aT_1 \aT_2}v_{\aT_3
    \rangle}-3 {\cal E}_{\langle \aT_1 \aT_2}v_{\aT_3 \rangle} \right]\nonumber\\*
\flamoi &&+\delta_\ell^2\left[\frac{1}{10}{\cal I}_{\aT_1 \aT_2}-\frac{3}{5}{\cal E}_{\aT_1 \aT_2}-\frac{1}{2}{\cal I}_{\langle
    \aT_1}v_{\aT_2\rangle}+7{\cal I}_\emptyset v_{\langle \aT_1}v_{\aT_2\rangle}-\frac{6}{5}\epsilon_{\bT\cT
    \langle \aT_1}v^\bT
  {\cal B}^\cT_{\,\,\aT_2 \rangle}\right]\,,\nonumber
\eea
\bea
\flamoi \frac{1}{\tau'}{\cal C}^\hashamoi[{\cal E}]_{\uline{\aT_\ell}}&=&-{\cal E}_{\uline{\aT_\ell}}+\frac{(\ell-1)(\ell+3)}{(2\ell+3)(\ell+1)}{\cal E}_{\uline{\aT_\ell}
\bT}v^\bT+{\cal E}_{\langle \uline{\aT_{\ell-1}}}v_{\aT_\ell\rangle}\\*
\flamoi &&+\delta_\ell^3\left[-\frac{1}{2}{\cal I}_{\langle \aT_1 \aT_2}v_{\aT_3
    \rangle}+3 {\cal E}_{\langle \aT_1 \aT_2}v_{\aT_3 \rangle} \right]-\frac{2}{\ell+1}\epsilon_{\bT\cT
    \langle \aT_\ell}v^\bT  {\cal B}^\cT_{\,\,\uline{\aT_{\ell-1}} \rangle}\nonumber\\*
\flamoi &&+\delta_\ell^2\left[-\frac{1}{10}{\cal I}_{\aT_1 \aT_2}+\frac{3}{5}{\cal E}_{\aT_1 \aT_2}+\frac{1}{2}{\cal I}_{\langle  \aT_1}v_{\aT_2\rangle}- {\cal I}_\emptyset v_{\langle \aT_1}v_{\aT_2\rangle}+\frac{6}{5}\epsilon_{\bT\cT
    \langle \aT_1}v^\bT  {\cal B}^\cT_{\,\,\aT_2 \rangle}\right]\,,\nonumber
\eea

\bea\label{Collision2Bbrut}
\flamoi \frac{1}{\tau'}{\cal C}^\hashamoi[{\cal B}]_{\uline{\aT_\ell}}&=&-{\cal B}_{\uline{\aT_\ell}}+\frac{(\ell-1)(\ell+3)}{(2\ell+3)(\ell+1)}{\cal B}_{\uline{\aT_\ell}
\bT}v^\bT+{\cal B}_{\langle \uline{\aT_{\ell-1}}}v_{\aT_\ell\rangle}\\*
\flamoi &&+\frac{2}{\ell+1}\epsilon_{\bT\cT \langle \aT_\ell}v^\bT  {\cal
  E}^\cT_{\,\,\uline{\aT_{\ell-1}}
  \rangle}+\delta_\ell^2\left[\frac{4}{5}\epsilon_{\bT\cT \langle \aT_1}v^\bT
  {\cal E}^\cT_{\,\,\aT_2 \rangle}-\frac{2}{15}\epsilon_{\bT\cT \langle \aT_1}v^\bT  {\cal I}^\cT_{\,\,\aT_2 \rangle}\right]\,.\nonumber
\eea
We recall that the multipoles are projected so all indices $\aT,\bT,\cT\dots$
in the above expressions could in fact be replaced by $\iT,\jT,\kT\dots$.\\
In the case where there is no polarization, the
expression~(\ref{Collision2Ibrut}) for the intensity part of the collision is nearly consistent with equations~($64$-$67$) of ~\cite{Maartens1999}. Indeed
the only discrepancy is the coefficient in front of ${\cal I}_\emptyset v_{\langle
  \aT_1}v_{\aT_2\rangle}$, which in our case is $7$ and in the case
of~\cite{Maartens1999} is $3$. In their notation, this discrepancy arises from a missing factor
$6$ in their equation~($60$) in front of the factor $\left(v_B^c
  e_c \right)^2$. The terms in the first bracket of the r.h.s in this
equation can be traced directly to the expansion of $[\gamma_B(1-v_B^c
e_c)]^{-3}$, arising from the change of frame between the baryons rest frame and the fundamental frame. Indeed, in the definition~\ref{EqLcurly} applied to
the collision tensor, $C[]$ is a scalar but (noting here $\tilde{\gr{e}}_\aT$
the tetrad in the baryons rest frame) $(p^\zT)^3 \dd p^\zT=(p^\zTt)^3 \dd p^\zTt \left[\gamma(1-n^\iT
  v_\iT)\right]^{-4}$, and in the definition~\ref{Defstar} of ${\cal C}^\hashamoi$ there
  is an additional factor 1/$p^\zT= \gamma (1- n^\iT v_\iT)/p^\zTt$. As a consequence, the factor $-3$ in front of the term
$\rho_R v_B^{\langle a} v_B^{b\rangle}$ in equation~($63$) of~\cite{Maartens1999} should be $-7$. \\

In the tight coupled limit, that is when we can neglect (when compared to the
Liouville operator) the r.h.s of equations~(\ref{Collision2Ibrut}-\ref{Collision2Bbrut}), then it is obvious
that
\bea
&&{\cal E}_{\uline{\iT_\ell}}={\cal B}_{\uline{\iT_\ell}}={\cal
  I}_{\uline{\iT_\ell}}=0,\quad\mathrm{if}\quad \ell \ge 3\,,\\
&&{\cal E}_{\iT\jT}={\cal B}_{\iT\jT}=0\,,\nonumber\\
&&\frac{2}{15}{\cal  I}_{\iT\jT}=T_{\langle \iT \jT \rangle}=\frac{4}{3}{\cal
  I}_\emptyset v_{\langle\iT} v_{\jT\rangle}\,,\nonumber\\
&&{\cal  I}_{\iT}=4 {\cal   I}_\emptyset v_\iT \,.\nonumber
\eea
We recover of course that since there is no polarization in the baryons rest frame, as
noted at the end of section~\ref{Collisionnobulk}, then there is no
polarization in any frame, and we check from the comparison with a perturbed
stress-energy tensor with no anisotropic stress given in
equations~(\ref{TOiO2}-\ref{TijO2}) that the multipoles ${\cal
  I}_\aT$ and ${\cal I}_{\aT\bT}$, which vanish in the baryons rest frame, are
non-vanishing only because of the change of frame, and consequently in this
tight coupled limit there is a quadrupole but no anisotropic stress \cite{Fidler}.\\

\subsection{Perturbative expansion of the collision term}

Though this result might appear to be general since we have not performed a
perturbative expansion around a FL space-time, it is not general since we have
kept terms quadratic in the change of frame velocity $\gr{v}$ only when they
multiply ${\cal I}_\emptyset$, since this is the only quantity which will
have a background contribution. The full second order transformation (in
$\gr{v}$) can be computed using the transformation rules at second order in
$\gr{v}$ for the multipoles involved in the Thomson collision term~(\ref{CollisionThomson})
and we have reported them in~\ref{AppTrule2}. This full result at second order in the
bulk velocity $\gr{v}$ has already been derived in~\cite{Psaltis1997}, but without
polarization. By developing with the expansion~(\ref{decomposition-ordre2}) the intensity, the electric and
the magnetic multipoles along with $\tau'$ and $\gr{v}$, we can perform
a perturbative expansion of the collision term following the same notations as
for the Liouville operator expansion, that is as in equation~(\ref{DecL}). The background
collision term vanishes. Since we have decided to discard the first order
magnetic multipole of radiation (because we can neglect first order vector and
tensor perturbations), there
is no contribution to the magnetic multipole of the collision term. The first
order intensity and electric multipoles read
\be
\flamoi {\cal C}^{\hashamoi(1)}[\bar{\cal I},\hat{\cal
  I}^{(1)}]_{\uline{\aT_\ell}}=\bar{\tau '}\left\{-\hat{\cal
    I}^{(1)}_{\uline{\aT_\ell}}+\delta_\ell^0 \hat{\cal I}^{(1)}_\emptyset
  +\delta_\ell^1 4 \bar{\cal I}_\emptyset \hat v^{(1)}_{\aT_1}+\delta_\ell^2\left[\frac{1}{10}\hat{\cal I}^{(1)}_{\aT_1 \aT_2}-\frac{3}{5}\hat{\cal E}^{(1)}_{\aT_1 \aT_2}\right]\right\}\,,
\ee

\be
\flamoi {\cal C}^{\hashamoi(1)}[\hat{\cal E}^{(1)}]_{\uline{\aT_\ell}}=\bar{\tau '}\left\{-\hat{\cal E}^{(1)}_{\uline{\aT_\ell}}+\delta_\ell^2\left[-\frac{1}{10}\hat{\cal I}^{(1)}_{\aT_1 \aT_2}+\frac{3}{5}\hat{\cal E}^{(1)}_{\aT_1 \aT_2}\right]\right\}\,.
\ee
At second order, the linear contribution is given by
\be\label{CollisionI2}
\flamoi {\cal C}^{\hashamoi(2)}[\bar{\cal I},\hat{\cal
  I}^{(2)}]_{\uline{\aT_\ell}}=\bar{\tau '}\left\{-\hat{\cal
    I}^{(2)}_{\uline{\aT_\ell}}+\delta_\ell^0\hat{\cal
    I}^{(2)}_\emptyset+\delta_\ell^1 4 \bar{\cal I}_\emptyset
  \hat v^{(2)}_{\aT_1}+\delta_\ell^2\left[\frac{1}{10}\hat{\cal I}^{(2)}_{\aT_1 \aT_2}-\frac{3}{5}\hat{\cal E}^{(2)}_{\aT_1 \aT_2}\right]\right\}\,,
\ee

\be
\flamoi {\cal C}^{\hashamoi(2)}[\hat{\cal E}^{(2)}]_{\uline{\aT_\ell}}=\bar{\tau '}\left\{-\hat{\cal E}^{(2)}_{\uline{\aT_\ell}}+\delta_\ell^2\left[-\frac{1}{10}\hat{\cal I}^{(2)}_{\aT_1 \aT_2}+\frac{3}{5}\hat{\cal E}^{(2)}_{\aT_1 \aT_2}\right]\right\}\,,
\ee
\be
\flamoi {\cal C}^{\hashamoi(2)}[\hat{\cal B}^{(2)}]_{\uline{\aT_\ell}}=- \bar{\tau '}\hat{\cal B}^{(2)}_{\uline{\aT_\ell}}\,.
\ee
As for the quadratic contribution, it is given by
\bea
\flamoi {\cal C}^{\hashamoi(1)(1)}[\bar{\cal I},\hat{\cal
  I}^{(1)}]_{\uline{\aT_\ell}}&=& 2 \bar{\tau '} \left\{\frac{(\ell+1)}{(2\ell+3)}\hat{\cal I}^{(1)}_{\uline{\aT_\ell}
\bT}\hat v^{\bT(1)}+\hat{\cal I}^{(1)}_{\langle \uline{\aT_{\ell-1}}}\hat
v^{(1)}_{\aT_\ell\rangle}+\delta_\ell^1 3 \hat{\cal I}^{(1)}_\emptyset
  \hat v^{(1)}_{\aT_1}\right.\\*
\flamoi && +\delta_\ell^0\left[\frac{4}{3}\bar{\cal I}_\emptyset \hat v^{(1)}_\bT \hat
  v^{\bT(1)} -\frac{2}{3}\hat{\cal I}^{(1)}_\bT \hat
  v^{\bT(1)}\right]+\delta_\ell^2\left[-\frac{1}{2}\hat{\cal I}^{(1)}_{\langle \aT_1}\hat
  v^{(1)}_{\aT_2\rangle}+7\bar{\cal I}_\emptyset \hat v^{(1)}_{\langle
    \aT_1}\hat v^{(1)}_{\aT_2\rangle} \right]\nonumber\\*
\flamoi && \left.+\delta_\ell^3\left[\frac{1}{2}\hat{\cal
    I}^{(1)}_{\langle \aT_1 \aT_2}\hat v^{(1)}_{\aT_3
    \rangle}-3 {\cal E}^{(1)}_{\langle \aT_1 \aT_2}\hat v^{(1)}_{\aT_3
    \rangle} \right]\right\}+2 \hat{\tau'}^{(1)}{\cal C}^{\hashamoi(1)}[\bar{\cal I},\hat{\cal I}^{(1)}]_{\uline{\aT_\ell}}\,,\nonumber
\eea

\bea
\flamoi {\cal C}^{\hashamoi(1)(1)}[\hat{\cal E}^{(1)}]_{\uline{\aT_\ell}}&=& 2
\bar{\tau '}\left\{\frac{(\ell-1)(\ell+3)}{(2\ell+3)(\ell+1)}\hat{\cal E}^{(1)}_{\uline{\aT_\ell}
\bT}\hat v^{\bT(1)}+\hat{\cal E}^{(1)}_{\langle \uline{\aT_{\ell-1}}}\hat v^{(1)}_{\aT_\ell\rangle}\right.\nonumber\\*
\flamoi &&+\delta_\ell^3\left[-\frac{1}{2}\hat{\cal I}^{(1)}_{\langle \aT_1 \aT_2}\hat
  v^{(1)}_{\aT_3 \rangle}+3 \hat{\cal E}^{(1)}_{\langle \aT_1 \aT_2}\hat v^{(1)}_{\aT_3 \rangle} \right]-\frac{2}{\ell+1}\epsilon_{\bT\cT
    \langle \aT_\ell}\hat v^{\bT(1)}  \hat{\cal B}^{\cT(1)}_{\,\,\,\uline{\aT_{\ell-1}} \rangle}\nonumber\\*
\flamoi &&\left.+\delta_\ell^2\left[\frac{1}{2}\hat{\cal I}^{(1)}_{\langle  \aT_1}\hat
    v^{(1)}_{\aT_2\rangle}-  \bar{\cal I}_\emptyset \hat v^{(1)}_{\langle
      \aT_1}\hat v^{(1)}_{\aT_2\rangle} \right]\right\}+2 \hat{\tau'}^{(1)}{\cal C}^{\hashamoi(1)}[\hat{\cal E}^{(1)}]_{\uline{\aT_\ell}}\,,
\eea

\bea\label{CollisionB11}
\flamoi {\cal C}^{\hashamoi(1)(1)}[\hat{\cal
  B}^{(1)}]_{\uline{\aT_\ell}}&=& 2 \bar{\tau '}\left\{\frac{2}{\ell+1}\epsilon_{\bT\cT \langle
  \aT_\ell}\hat v^{\bT(1)}  \hat{\cal E}^{\cT(1)}_{\,\,\,\uline{\aT_{\ell-1}}
  \rangle}\right.\\*
\flamoi &&\qquad \left.+\delta_\ell^2\left[\frac{4}{5}\epsilon_{\bT\cT \langle \aT_1}\hat
  v^{\bT(1)}  {\cal E}^{\cT(1)}_{\,\,\,\aT_2
    \rangle}-\frac{2}{15}\epsilon_{\bT\cT \langle \aT_1}\hat v^{\bT(1)}  \hat{\cal I}^{\cT(1)}_{\,\,\,\aT_2 \rangle}\right]\right\}\,.\nonumber
\eea
\subsection{The fluid equations for baryons}

The total stress-energy tensor is conserved 
\be
\nabla_\mu T_{\ib+\ir}^{\mu  \nu}=\nabla_\mu T_\ib^{\mu  \nu}+\nabla_\mu T_\ir^{\mu  \nu}=0\,.
\ee
This arises from the Bianchi identities, and thus the action of baryons on
photons is opposite to the action of photons on baryons. This means that we
can define the force resulting from the action of photons on baryons and the
force resulting from the action of baryons on photons by
\be\label{Force}
\nabla_\mu T_\ib^{\mu  \nu}\equiv F_{\ir \rightarrow \ib}^{\nu}=-\nabla_\mu
T_\ir^{\mu  \nu}\equiv -F_{\ib \rightarrow \ir}^{\nu}\,.
\ee
The expression of the force in the tetrad basis is further given by
\be
F^{\aT}_{\ir \rightarrow \ib}=\frac{1}{a}\int q^\aT
C^\hashamoi[g](q^\iT)\frac{\dd^3 q^\iT}{(2 \pi)^3}=-\frac{1}{a}\int  N^\aT {\cal C}^\hashamoi[I](n^\iT)\frac{\dd^2 \Omega}{4 \pi}\,\,,
\ee
and from equations~(\ref{ConsTmunufromLiouville}) and~(\ref{EqBoltzmann}), its components are easily related to the moments of the radiation collision term by
\bea\label{Forces}
a F^{\zT}_{\ir \rightarrow \ib}&=&-{{\cal C}^\hashamoi[I]}^\emptyset\,,\\*
a F^{\iT}_{\ir \rightarrow \ib}&=&-\frac{1}{3}{{\cal C}^\hashamoi[I]}^\iT\,.
\eea
 It is then straightforward to deduce the perturbative expansion of the force,
 which is inherited directly from the perturbative expansion of the radiation
 collision term. Note that in particular $F^{\zT(1)}_{\ir \rightarrow \ib}=0$. 

\section{From PSTF multipoles to normal modes components}\label{SecSTFGlm}

\subsection{The normal modes in Fourier space}

The equations~(\ref{LiouvilleI2}-\ref{curlL11I}), (\ref{LiouvilleE2}-\ref{LiouvilleB11}), (\ref{Eqcontbar2}-\ref{EqEulerbar2}) 
with equations~(\ref{CollisionI2}-\ref{CollisionB11}) and (\ref{Forces}) are a key result of this paper,
since they provide the full second order Boltzmann hierarchy for the coupled
system of baryons and photons. However in order to make contact with the standard approach of CMB numerical integration, it is more
convenient to describe the angular dependence of the radiation functions using
the normal modes components~\cite{Hu1997}. And in order to perform a numerical
integration in the most simple way, we also turn to Fourier space in order to
solve only for a time integral. It has been shown~\cite{Challinor2000b} that
we can convert the PSTF multipoles into normal modes components by decomposing
the quantities ${\cal I}$, ${\cal E}$ and ${\cal B}$  according to
\be
\flamoi {\cal X}(x^\aB,n^\iT)=\sum_{\ell m}\int \frac{\dd^3 \gr{k}}{(2\pi)^{3/2}}{\cal X}_{\ell}^{m}(\gr{k},\eta) G^{\cal X}_{\ell
  m}(\gr{k},x^\iB,n^\jT)=\sum_{\ell=0}^{\infty}{\cal X}_{\uline{\aT_\ell}}(x^\aB)n^{\langle
  \uline{\aT_\ell} \rangle}\,,
\ee
where ${\cal X}$ stands for ${\cal I}$, ${\cal E}$ or ${\cal B}$. Here we have
used the notation $\gr{k}$, that is a boldface font, for a mode in the Fourier transform though it is
only a three dimensional vector. The modes $G^{\cal X}_{\ell
  m}(n^\iT)$ according to which we decompose are defined by
\be
G^{\cal I}_{\ell m}(\gr{k},x^\iB,\gr{n})=\bar{\cal I}\frac{1}{N_\ell} e^{\ii
k_\iB x^\iB} Y^{\ell m}(\gr{n}) \,,
\ee

\be
G^{\cal E}_{\ell m}(\gr{k},x^\iB,\gr{n})=-G^{\cal B}_{\ell
  m}(\gr{k},x^\iB,\gr{n})= \bar{\cal I}\frac{M_\ell}{\sqrt{2}N_\ell}e^{\ii
k_\iB x^\iB} Y^{\ell m}(\gr{n})\,,
\ee
with
\be
\N{\ell} \equiv \ii^\ell \sqrt{\frac{(2\ell+1)}{4 \pi}}\,.
\ee
The components ${\cal X}_{\ell m}(\gr{k},\eta)$ are the time-dependent components in Fourier space of the spherical harmonics
decomposition, also called normal modes. Note that we have also factorized the background intensity so
these quantities are now dimensionless, but we did not divide by
$4$ as is usually done when we want to have a quantity which can be interpreted as a temperature.
Instead we decompose ${\cal I}/\bar {\cal I}$ which is the fractional
variation of the energy density of radiation as it is what is really measured.
The ${\cal X}_{\ell m}(\gr{k},\eta)$ can be obtained from the PSTF multipoles using the formulas
\be
\hat{\cal I}_{\ell}^{m}(\gr{k},\eta)=\frac{1}{\bar{\cal I}}N_\ell \Delta_\ell\int
\frac{\dd^3 x^\iB}{(2 \pi)^{3/2}}e^{-\ii k_\iB x^\iB}
\Ysu{\ell}{m}{\uline{\aT_\ell}}\hat{\cal I}_{\uline{\aT_\ell}}(x^\aB)\,,
\ee
\be
\hat{\cal E}_{\ell}^{m}(\gr{k},\eta)=\frac{1}{\bar{\cal
    I}}\frac{N_\ell\sqrt{2}}{M_\ell} \Delta_\ell\int \frac{\dd^3 x^\iB}{(2
  \pi)^{3/2}}e^{-\ii k_\iB x^\iB} \Ysu{\ell}{m}{\uline{\aT_\ell}}\hat{\cal E}_{\uline{\aT_\ell}}(x^\aB)\,,
\ee
\be
\hat{\cal B}_{\ell}^{m}(\gr{k},\eta)=-\frac{1}{\bar{\cal
    I}}\frac{N_\ell\sqrt{2}}{M_\ell} \Delta_\ell\int \frac{\dd^3 x^\iB}{(2
  \pi)^{3/2}}e^{-\ii k_\iB x^\iB} \Ysu{\ell}{m}{\uline{\aT_\ell}}\hat{\cal B}_{\uline{\aT_\ell}}(x^\aB)\,.
\ee
The functions $\Yu{\ell}{m}{\uline{\aT_\ell}}$ are null if one of the indices in $\uline{\aT_\ell}$ is $\zT$, since it is already the case for the
indices in the intensity, the electric and the magnetic multipoles. 
These quantities are built in details in~\cite{Thorne1980}, and in the
case where $m\ge0$ they are defined by
\bea
\flamoi \Yu{\ell}{m}{\uline{\iT_\ell}}&=&(-1)^m\left[\frac{2 \ell+1}{4
    \pi}\frac{(\ell-m)!}{(\ell+m)!}\right]^{1/2}\sum_{j=0}^{[(\ell-m)/2]}\frac{(-1)^j(2
  \ell-2 j)!}{2^\ell j! (\ell-j)!(\ell-m-2j)!}\\*
\flamoi &&\times\left(\delta_1^{(\iT_1}+\ii\delta_2^{(\iT_1}\right)\dots\left(\delta_1^{\iT_m}+\ii\delta_2^{\iT_m}\right)\delta_3^{\iT_{m+1}}\dots\delta_3^{\iT_{\ell-2
  j}}\delta^{\iT_{\ell-2
  j+1} \iT_{\ell-2j+2} }\dots\delta^{\iT_{\ell-1} \iT_{\ell)} }\,.\nonumber
\eea
If $m<0$, then it is defined by $\Yu{\ell}{m}{\uline{\iT_\ell}}=(-1)^m\Ysu{\ell}{-m}{\uline{\iT_\ell}}$.
The PSTF multipoles can be conversely obtained from the normal modes components
by the inverse formulae
\be
\hat{\cal E}_{\uline{\aT_\ell}}(x^\aB)= \bar{\cal
  I}\frac{M_\ell}{\sqrt{2}N_\ell}\int \frac{\dd^3 \gr{k}}{(2 \pi)^{3/2}}e^{\ii
  k_\iB x^\iB}\sum_{m=-\ell}^{\ell}\Yd{\ell}{m}{\uline{\aT_\ell}}\hat{\cal E}_{\ell}^{m}(\gr{k},\eta)\,,
\ee

\be
\hat{\cal I}_{\uline{\aT_\ell}}(x^\aB)= \bar{\cal I} \frac{1}{N_\ell}\int
\frac{\dd^3 \gr{k}}{(2 \pi)^{3/2}}e^{\ii
  k_\iB x^\iB}\sum_{m=-\ell}^{\ell}\Yd{\ell}{m}{\uline{\aT_\ell}}\hat{\cal I}_{\ell}^{m}(\gr{k},\eta)\,,
\ee

\be
\hat{\cal B}_{\uline{\aT_\ell}}(x^\aB)=- \bar{\cal
  I}\frac{M_\ell}{\sqrt{2}N_\ell}\int \frac{\dd^3 \gr{k}}{(2 \pi)^{3/2}}e^{\ii
  k_\iB x^\iB}\sum_{m=-\ell}^{\ell}\Yd{\ell}{m}{\uline{\aT_\ell}}\hat{\cal B}_{\ell}^{m}(\gr{k},\eta)\,.
\ee

\subsection{The basis for the decomposition of tensors}

In order to decompose the metric perturbations and the fluid perturbations
according to this spherical harmonics scheme, we
start from the vectors
\be
\flamoi \bar{\gr{e}}_{(0)} \equiv -\bar{\gr{e}}_3,\qquad \bar{\gr{e}}_{(1)}\equiv
\frac{1}{\sqrt{2}}\left(\bar{\gr{e}}_1+ \ii\bar{\gr{e}}_2 \right),\qquad
\bar{\gr{e}}_{(-1)}\equiv -\frac{1}{\sqrt{2}}\left(\bar{\gr{e}}_1 - \ii\bar{\gr{e}}_2  \right)\,.
\ee
These are not true vectors since they live on the background space-time and
their indices are raised and lowered by the euclidian metric $\delta_{\iB\jB}$
and its inverse $\delta^{\iB\jB}$. Their associated forms are thus
\be
\flamoi \bar{\gr{e}}^{(0)} \equiv -\bar{\gr{e}}^3,\qquad \bar{\gr{e}}^{(1)}\equiv
\frac{1}{\sqrt{2}}\left(\bar{\gr{e}}^1+ \ii\bar{\gr{e}}^2 \right),\qquad
\bar{\gr{e}}^{(-1)}\equiv -\frac{1}{\sqrt{2}}\left(\bar{\gr{e}}^1 - \ii\bar{\gr{e}}^2  \right)\,.
\ee
We then build a tensor basis in Fourier space out of this basis. We start from the
scalar basis which in Fourier space is given by
\be
Q^{(0)}\equiv \exp\left(e^{\ii k_\iB x^\iB}\right)\,,
\ee
and we align $\gr{k}$ with $\bar{\gr{e}}_3$.
It is then used to build the scalar basis of higher rank tensors according to
\bea
Q_{\iB}^{(0)} &\equiv& -\partial_\iB Q^{(0)}/k =\ii \bar  e^{(0)}_{\,\,\iB} Q^{(0)}\,,\\*
Q_{\iB \jB}^{(0)}&\equiv& \partial_{\langle \iB}\partial_{\jB \rangle} Q^{(0)}/k^2=\left(-\bar e^{(0)}_{\,\,\iB} \bar e^{(0)}_{\,\,\jB}+\delta_{\iB\jB}/3\right)Q^{(0)}\,,
\eea
since perturbation variables "live" on the background space-time. We then remark that
\be
Q^{(0)}=G^{{\cal I}}_{00},\qquad n^\iT Q_{\iB}^{(0)}=G^{{\cal I}}_{10},\qquad n^\iT
n^\jT Q_{\iB\jB}^{(0)}=\frac{2}{3}G^{{\cal I}}_{20}\,,
\ee
and this enables us to make contact with the normal modes decomposition first performed for the directional dependence of radiation.
For vectors, we follow the same method and we start from the basis
\be\label{Qi}
Q_{\iB}^{(\pm 1)}\equiv \ii\bar  e_{\,\,\,\iB}^{(\pm 1)}Q^{(0)}\,,
\ee
from which we can build a basis for vector type perturbations of higher order rank
tensors (we restrict to rank two tensors, since in the problem at hand we
have no rank higher than $2$)
\be
Q_{\iB \jB}^{(\pm 1)}\equiv -\partial_{(\iB} Q_{\jB)}^{(\pm 1)} /k=\ii\bar  e^{(0)}_{\,\,(\iB}Q_{\jB)}^{(\pm 1)}\,.
\ee
We then make contact with the normal modes decomposition by
noting that
\be\label{PropQi}
n^\iT Q_{\iB}^{(\pm 1)} = G^{{\cal I}}_{1 \pm 1},\qquad n^\iT n^\jT Q_{\iB\jB}^{(\pm
  1)}=\frac{1}{\sqrt{3}}G^{{\cal I}}_{2 \pm 1}\,.
\ee
Note the difference in the definition of $Q_{\iB}^{\pm1}$ in equation~(\ref{Qi}) with
respect to the expression given in~\cite{Hu1997}. However it agrees with the
expression given in~\cite{Hu2000a} up to factors $(-1)^m$ which arise because
of a different convention in the spherical harmonics. \\  
Finally for tensors we use the basis
\be
Q_{\iB\jB}^{(\pm 2)}\equiv -\sqrt{\frac{3}{2}} \bar e^{\pm 1}_\iB \bar  e^{\pm 1}_\jB Q^{(0)}\,,
\ee
which conveniently satisfies
\be
n^\iT n^\jT Q_{\iB \jB}^{\pm 2}=G^{{\cal I}}_{2 \pm 2}\,.
\ee
The (second order) vector perturbation of the metric is then decomposed as
\be\label{VectorQIJ}
\hat \Phi^{(2)}_\iB(x^\aB) =\int \frac{\dd^3 \gr{k}}{(2 \pi)^{3/2}} \sum_{m=\pm 1} Q_\iB^{(m)}(\gr{k})\hat \Phi_m^{(2)}(\gr{k},\eta)\,.
\ee
Similarly the (second order) tensor perturbations are decomposed according to
\be
\hat H_{\iB \jB}^{(2)}(x^\aB)=\int \frac{\dd^3 \gr{k}}{(2 \pi)^{3/2}} \sum_{m=\pm1}Q_{\iB\jB}^{(2 m)}(\gr{k}) \hat H_{m}^{(2)}(\gr{k},\eta)\,.
\ee
As for gradients of the first order and second order scalar perturbations ($\Phi$
and $\Psi$) they are decomposed similarly to the vector perturbations in
\be
\partial_\iB \hat \Phi (x^\aB) =\int \frac{\dd^3 \gr{k}}{(2 \pi)^{3/2}} \sum_{m=-1}^{1} Q_\iB^{(m)}(\gr{k}) k^{(m)} \Phi(\gr{k},\eta)\,.
\ee
This is equivalent to decomposing the Fourier mode $\gr{k}$ into a scalar part
(in the direction $\bar{\gr{e}}_{(0)}$) and a vector part (in the plane orthogonal to $\bar{\gr{e}}_{(0)}$) according to
\be\label{Decompk}
k^\iB=k^{(0)}\bar e_{(0)}^{\,\,\iB}+k^{(1)}\bar e_{(1)}^{\,\,\iB}+k^{(-1)}\bar e_{(-1)}^{\,\,\iB}=\sum_{m=-1}^1 k^{(m)}\bar e_{(m)}^{\,\,\iB}\,.
\ee
The relation of these components to the Cartesian components of $\gr{k}$ are
\be
k^{(0)}=-k^3\,,\quad k^{(1)}=\frac{1}{\sqrt{2}}(k^1-\ii k^2)\,,\quad
k^{(-1)}=-\frac{1}{\sqrt{2}}(k^1+\ii k^2)\,.
\ee
For the second order perturbations (or for first order perturbations when we
solve the first order equations) we choose to align the mode $\gr{k}$ with
$\bar{\gr{e}}_{(3)}$, which implies in the previous decomposition that $k^{(1)}=k^{(-1)}=0$,
and $k^{(0)}=-k$. However for first order quantities appearing in quadratic terms in the second
order equations, this is not possible because of the convolution generated by the Fourier transform.  
In the following we will also use the notation\footnote{Note here that the
  hat denotes a unit vector and has nothing to do with a gauge invariant
  variable.} $\hat{\gr{k}}=\gr{k}/k$ with the same decomposition as in equation~(\ref{Decompk}).\\

As for the second order velocity of the electrons $\gr{v}$, it has a scalar degree of
freedom and a vector degree of freedom according to the
decomposition~(\ref{decv}), and the Fourier components of this vector degree
of freedom could be decomposed similarly to $\hat{\Phi}^\iB$, that is
according to equation~(\ref{VectorQIJ}) with $m$ running from $-1$ to $1$. However since
the velocity of the electrons always appear with tetrad basis components, it
is not very convenient to go in the coordinate basis in order to split into
scalar and vector modes and then transform back to tetrad components. We use instead
for $v_\aT$ the same type of decomposition as we did for ${\cal I}_\aT$.
Namely, this decomposition reads
\be
\hat v_\aT (x^\aB)= \frac{1}{N_1}\int \frac{\dd^3 \gr{k}}{(2 \pi)^{3/2}}e^{\ii
  k_\iB x^\iB}\sum_{m=-1}^{1}\Yd{1}{m}{\aT} \hat v_m (\gr{k},\eta)\,.
\ee

\subsection{Comment on the SVT decomposition}

We have arbitrarily chosen to perform the multipole decomposition on $\hat v^\aT$ and not $\hat
V^\aB$. 
If we use $\hat V^\aB$ for this multipole decomposition, then $\hat V_0$ matches
the scalar component and $\hat V_{\pm1}$ match the vector components in
the usual SVT decomposition of equation~(\ref{decv}). We did not follow this
approach for the baryons velocity, but we did it for the vector degrees of freedom in the metric $\hat \Phi^\iB$,
and also for the tensor degrees of freedom $\hat H^{\iB \jB} $. We see from
equation~(\ref{choicevectoralignement}) that $\hat v^\aT \neq \hat V^\aB$ and
consequently,  when performing the multipole decomposition using $\hat v^\aT$ (which is
what we have done in this paper) the normal modes components components are
different ($\hat v_m \neq \hat V_m$). It would
thus be misleading to name $\hat v_0$ a scalar component and $\hat
v_{\pm1}$ vector components. For the sake of the discussion here we will
nonetheless use the terms \emph{coordinate SVT} and \emph{tetrad SVT} to name these
different decompositions. Essentially the difference comes from the fact that when
we perform the coordinate SVT decomposition of a vector, the decomposition is made on a
Euclidian space-time which is the background space-time. Indeed the indices on the mode $k^\iB$ are lowered and raised by a simple
$k_\iB=\delta_{\iB \jB}k^\iB$. The coordinate SVT decomposition of tensors is not motivated by physical reason but just by the Fourier transformation. On
the other hand, in the tetrad SVT decomposition we use the tetrad indices
$\iT$, that is we work in the locally Euclidian frame of the physical
space-time, and indices are lowered and raised by $\hat v_\iT=\delta_{\iT
  \jT}\hat v^\jT$. The
fictitious background Euclidian frame is different from the physical locally
Euclidian frame. As it can be seen from equation~(\ref{choicevectoralignement}), a
local volume expansion higher than the average ($\Psi$) leads to a difference in what
is considered as a length, and the SVT components found in one decomposition are
a rescaled version of those obtained in the other decomposition. If we had chosen not to neglect
the first order gravitational waves, then the r.h.s of the second of the
equations~(\ref{choicevectoralignement}) would be supplemented by a term
$2E^{\iT(1)}_\jT V^{\jB(1)}$. Again, this would lead to a difference in
the direction used to perform the decomposition and the scalar part of a
vector in one approach would mix the scalar and the vector components of the
other approach. Note that since we have neglected the first order vector and
tensor perturbations of the metric, the decomposition of the second order
vector and tensor perturbations of the metric in either approach leads to the same result. A more detailed discussion on the
issue of SVT decomposition can be found in~\cite{Osano2007}.

\subsection{Transforming the Boltzmann equation in normal modes components}

Since the second order equations are quadratic, we need to know how to compose
the $\Yd{\ell}{m}{\uline{\aT_{\ell}}}$. This is deduced from the composition rules of
the spherical harmonics, and the key result is

\bea\label{EqFaye}
\flamoi \Yd{\ell_1}{m_1}{(  \uline{\iT_{\ell_1}}}\Yd{\ell_2}{m_2}{\uline{\iT_{\ell_2}})}&=&\sum_{\underset{\,(\ell_1+\ell_2-\ell_3)\,\,\mathrm{even}}{\ell_3=0\,}}^{\ell_1+\ell_2}\sqrt{\frac{(2
  \ell_1 +1)(2 \ell_2 +1)}{4 \pi (2 \ell_3 +1)}}C_{\ell_1 m_1 \ell_2
m_2}^{\ell_3 (m_1+m_2)}C_{\ell_1 0 \ell_2 0}^{\ell_3 0}\nonumber\\*
\flamoi &&\qquad \qquad\times \Yd{\ell_3}{(m_1 +m_2)}{( \uline{\iT_{\ell_3}}}\delta_{\iT_{\ell_3+1}\iT_{\ell_3+2}}\dots\delta_{\iT_{\ell_1+\ell_2-1}\iT_{\ell_1+\ell_2})}\,,
\eea
where the $C_{\ell_1 m_1 \ell_2 m_2}^{\ell_3 m_3}$ are the Clebsch-Gordan
coefficients. A set of useful relations can be obtained by considering the
cases where $\ell_1=\ell_2$, $\ell_1=\ell_2-1$ or $\ell_1=\ell_2+1$. These are
reported in~\ref{UsefulForms}.

\subsection{The Boltzmann hierarchy in normal modes components at first order}

Throughout this section we will use the following definitions that will
simplify the notation
\bea
\kPlmsn{\ell}{m}{s}{0}&\equiv& \kMlmsn{\ell}{m}{s}{0}\equiv
\sqrt{\frac{(\ell^2-m^2)(\ell^2-s^2)}{\ell^2}}\,,\\
\kPlmsn{\ell}{m}{s}{\pm1}&\equiv& -\sqrt{\frac{(\ell \pm m)(\ell\pm  m
    +1)(\ell^2-s^2)}{2 \ell^2}}\,,\\*
\kMlmsn{\ell}{m}{s}{\pm1}&\equiv& \sqrt{\frac{(\ell\pm m)(\ell\pm m
    -1)(\ell^2-s^2)}{2 \ell^2}}\,,
\eea
\be
\llmn{\ell}{m}{0}\equiv-\frac{m}{\ell}\,,\qquad\llmn{\ell}{m}{\pm 1}\equiv\pm
\frac{1}{\ell}\sqrt{\frac{(\ell\mp m+1)(\ell\pm m)}{2}}\,.
\ee

At first order we choose to align the mode considered $\gr{k}$ with the
direction $\bar{\gr{e}}_3$, and the dependence in $\gr{k}$ becomes only a
dependence in its magnitude $k$. The set of equations obtained at first order is (dropping the obvious dependence of
all quantities in $\eta$ in the rest of this paper)
\bea\label{LI1}
{\cal L}^{\hashamoi(1)}[\hat{\cal I}]_\ell^0(k)&=&\hat{\cal
  I}_\ell^{'0}(k)+k\left[\frac{\kPlmsn{\ell+1}{0}{0}{0}}{2\ell+3}\hat{\cal
  I}_{\ell+1}^{0}(k)-\frac{\kMlmsn{\ell}{0}{0}{0}}{2\ell-1}\hat{\cal
  I}_{\ell-1}^{0}(k) \right]\nonumber\\*
&&-\delta_\ell^0 4 \Psi'(k)-\delta_\ell^1 4 k \hat\Phi(k)\,,
\eea
\be\label{LE1}
{\cal L}^{\hashamoi(1)}[\hat{\cal E}]_\ell^0(k)=\hat{\cal
  E}_\ell^{'0}(k)+k\left[\frac{\kPlmsn{\ell+1}{0}{2}{0}}{2\ell+3}\hat{\cal
  E}_{\ell+1}^{0}(k)-\frac{\kMlmsn{\ell}{0}{2}{0}}{2\ell-1}\hat{\cal
  E}_{\ell-1}^{0}(k) \right]\,,
\ee

\bea\label{CI1}
{\cal C}^{\hashamoi(1)}[\hat{\cal I}]_\ell^0(k)&=&\bar{\tau'}\left\{-\hat{\cal
  I}_\ell^0(k)+\delta_\ell^0 \hat{\cal
  I}_0^0(k)+4 \delta_\ell^1  \hat{v}_0(k)\right.\nonumber\\*
&&\qquad\left.+\delta_\ell^2\frac{1}{10}\left[\hat{\cal I}_2^0(k)-\sqrt{6}\hat{\cal E}_2^0(k)\right] \right\}\,,
\eea

\be\label{CE1}
{\cal C}^{\hashamoi(1)}[\hat{\cal E}]_\ell^m(k)=\bar{\tau'}\left\{-\hat{\cal E}_\ell^m(k)-\delta_\ell^2\frac{\sqrt{6}}{10}\left[\hat{\cal I}_2^0(k)-\sqrt{6}\hat{\cal E}_2^0(k)\right]\right\}\,.
\ee
As already mentionned, the first order magnetic modes are not excited since
the first order vector and tensor modes are negligible, so we did not report them in
the above equations. Additionally, the intensity and electric multipoles are
only excited for $m=0$, that is the reason why we also only reported this case. 

\subsection{The Boltzmann hierarchy in normal modes components at second order}

At second order, a Fourier mode $\gr{k}$ on a quadratic term will appear as a
convolution on modes $\gr{k}_1$ and $\gr{k}_2$ whose sum is $\gr{k}$. It implies an integral of the form
\be
\mathcal{K}(\gr{k}_1,\gr{k}_2,\gr{k} )\equiv \int \frac{\dd^3\gr{k}_1
 \dd^3\gr{k}_2}{(2 \pi)^{3/2}}\, \dirac{3}(\gr{k}_1+\gr{k}_2-\gr{k})\,, 
\ee
that we abbreviate in ${\cal K}$. 
We can only align $\gr{k}$ with the direction $\bar{\gr{e}}_3$, but not $\gr{k}_1$
or $\gr{k}_2$ at the same time. However for a first order quantity the components $X_{\ell}^m$ for a mode in a given direction
$\hat{\gr{k}}_1$ can be obtained by rotating the components $X_{\ell}^0$ obtained when
we had decided to align this direction with $\bar{\gr{e}}_3$. Namely this rotation
leads for a mode $\gr{k}$ to
\be\label{Rotatemode}
X_{\ell}^m(\gr{k})=\sqrt{\frac{4 \pi}{2 \ell +1}}Y^{\star \ell
  m}(\hat{\gr{k}})X_{\ell}^0\left(k \bar{\gr{e}}_3\right)\,,
\ee
and in particular for the first order velocity
\be
\hat v_n(\gr{k})=-\hat k^{(n)} \hat v_0(k\bar{\gr{e}}_3)\,.
\ee
Note that the scalar product of two mode vectors is given by
\be
\gr{k}_1
.\gr{k}_2=k_1^{(0)}k_2^{(0)}-k_1^{(1)}k_2^{(-1)}-k_1^{(-1)}k_2^{(1)}=\sum_{n=-1}^1
(-1)^n k_1^{(n)} k_2^{(-n)}\,.
\ee

\subsubsection{The Liouville operator}

We finally obtain for the linear terms in the second order Liouville operator
\bea\label{LiouvilleNMI2}
\flamoi {\cal L}^{\hashamoi(2)}[\hat{\cal I}^{(2)}]_\ell^m(k)&=&\hat{\cal
  I}_\ell^{'m(2)}(k)+k\left[\frac{\kPlmsn{\ell+1}{m}{0}{0}}{2\ell+3}\hat{\cal
  I}_{\ell+1}^{m(2)}(k)-\frac{\kMlmsn{\ell}{m}{0}{0}}{2\ell-1}\hat{\cal
  I}_{\ell-1}^{m(2)}(k) \right]\nonumber\\*
\flamoi &&-\delta_\ell^0\delta_m^04 \Psi'^{(2)}(k)-\delta_\ell^1\delta_m^0 4 k
\hat\Phi^{(2)}(k)\nonumber\\*
\flamoi && +\delta_\ell^2\delta_m^1\frac{4
  k}{\sqrt{3}}\hat{\Phi}_1^{(2)}(k)+\delta_\ell^2\delta_m^{-1}\frac{4 k}{\sqrt{3}}\hat{\Phi}_{-1}^{(2)}(k)\nonumber\\*
\flamoi &&+\delta_\ell^2\delta_m^2 4 \hat{H}_1^{'(2)}+\delta_\ell^2\delta_m^{-2}4 \hat{H}_{-1}^{'(2)}\,,
\eea

\bea
\flamoi {\cal L}^{\hashamoi(2)}[\hat{\cal E}^{(2)}]_\ell^m(k)&=&\hat{\cal
  E}_\ell^{'m(2)}(k)\\*
\flamoi &&+k\left[\frac{\kPlmsn{\ell+1}{m}{2}{0}}{2\ell+3}\hat{\cal
  E}_{\ell+1}^{m(2)}(k)+\frac{2 m}{\ell(\ell+1)}\hat{\cal
  B}_{\ell}^{m(2)}(k)-\frac{\kMlmsn{\ell}{m}{2}{0}}{2\ell-1}\hat{\cal
  E}_{\ell-1}^{m(2)}(k) \right]\,,\nonumber
\eea

\bea
\flamoi {\cal L}^{\hashamoi(2)}[\hat{\cal B}^{(2)}]_\ell^m(k)&=&\hat{\cal
  B}_\ell^{'m(2)}(k)\\*
\flamoi &&+k\left[\frac{\kPlmsn{\ell+1}{m}{2}{0}}{2\ell+3}\hat{\cal
  B}_{\ell+1}^{m(2)}(k)-\frac{2 m}{\ell(\ell+1)}\hat{\cal
  E}_{\ell}^{m(2)}(k)-\frac{\kMlmsn{\ell}{m}{2}{0}}{2\ell-1}\hat{\cal
  B}_{\ell-1}^{m(2)}(k) \right]\,.\nonumber
\eea
As for the quadratic terms, we can use the composition rules given in~\ref{UsefulForms} to obtain
\bea
\flamoi {\cal L}^{\hashamoi(1)(1)}[\hat{\cal I}]_\ell^m(k)&=&2{\mathcal K}\left\{-\sum_{n=-1}^1\frac{\kPlmsn{\ell+1}{m}{0}{n}}{2\ell+3}\hat{\cal
    I}_{\ell+1}^{m+n}(\gr{k}_2)\right.\\*
\flamoi &&\quad\quad\times \left[-(\ell-2)k_1^{(-n)}\hat{\Phi}(k_1)+\left(k_2^{(-n)}-(\ell+2)k_1^{(-n)}\right)\hat{\Psi}(k_1)\right]\nonumber\\*
\flamoi &&\quad+\sum_{n=-1}^1
\frac{\kMlmsn{\ell}{m}{0}{n}}{2\ell-1}\hat{\cal
  I}_{\ell-1}^{m-n}(\gr{k}_2)\nonumber\\*
\flamoi&&\quad \quad \times \left[(\ell+3)k_1^{(n)}\hat{\Phi}(k_1)+\left(k_2^{(n)}+(\ell-1)k_1^{(n)}\right)\hat{\Psi}(k_1)\right]\nonumber\\*
\flamoi &&\quad-4 \hat{\Psi}'(k_1)\hat{\cal I}_\ell^m(\gr{k}_2)-
\hat{\Phi}(k_1)\hat{\cal I}_\ell^{'m}(\gr{k}_2)\nonumber\\*
\flamoi &&\quad +4 \delta_{\ell}^1\left(\hat{\Psi}(k_1)-2\hat{\Phi}(k_1)\right)\hat{\Phi}(k_2)k_2^{(m)}\nonumber\\*
\flamoi &&\quad \left. +4\delta_{\ell}^0\,\left[\hat{\Phi}(k_1)-2 \hat{\Psi}(k_1)\right]\hat{\Psi}'(k_2)\right\}\,,\nonumber
\eea
\bea
\flamoi{\cal L}^{\hashamoi(1)(1)}[\hat{\cal E}]_\ell^m(k)&=&2{\mathcal
  K}\left\{-\sum_{n=-1}^1\frac{\kPlmsn{\ell+1}{m}{2}{n}}{2\ell+3}\hat{\cal
    E}_{\ell+1}^{m+n}(\gr{k}_2)\right.\\*
\flamoi &&\quad\quad \times \left[-(\ell-2)k_1^{(-n)}\hat{\Phi}(k_1)+\left(k_2^{(-n)}-(\ell+2)k_1^{(-n)}\right)\hat{\Psi}(k_1)\right]\nonumber\\*
\flamoi && \quad+\sum_{n=-1}^1
\frac{\kMlmsn{\ell}{m}{2}{n}}{2\ell-1}\hat{\cal
  E}_{\ell-1}^{m-n}(\gr{k}_2)\nonumber\\*
\flamoi && \quad\quad\times \left[(\ell+3)k_1^{(n)}\hat{\Phi}(k_1)+\left(k_2^{(n)}+(\ell-1)k_1^{(n)}\right)\hat{\Psi}(k_1)\right]\nonumber\\*
\flamoi &&\left. \quad-4 \hat{\Psi}'(k_1)\hat{\cal E}_\ell^m(\gr{k}_2)- \hat{\Phi}(k_1)\hat{\cal E}_\ell^{'m}(\gr{k}_2)\right\}\,,\nonumber
\eea
\bea\label{L11B}
 \flamoi {\cal L}^{\hashamoi(1)(1)}[\hat{\cal B}]_\ell^m(k)&=& 2{\mathcal K}\frac{2}{(\ell+1)}\left\{\sum_{n=-1}^1
  \llmn{\ell}{m}{n}\hat{\cal
    E}_{\ell}^{m-n}(\gr{k}_2)\times\right.\\*
\flamoi&&\left.\qquad\qquad\times\left[(k_2^{(n)}-k_1^{(n)})\left(\hat{\Phi}(k_1)+\hat{\Psi}(k_1)\right)+4 k_1^{(n)}\hat{\Phi}(k_1)\right]\right\}\nonumber
\eea
This last expression and its version in PSTF multipoles~(\ref{LiouvilleB11}),
can be traced directly to the lensing term in the expression~(\ref{EqPab11})
and we recover the well known result that at second order, gravitational
lensing generates magnetic type multipoles out of electric type multipoles.

\subsubsection{The collision tensor}

Using a similar method applied to the collision term leads to
\be\label{C2a}
\flamoi{\cal C}^{\hashamoi(2)}[\hat{\cal I}]_\ell^m(k)=\bar{\tau'}\left[-\hat{\cal
  I}_\ell^{m(2)}(k)+\delta_\ell^0 \delta_m^0 \hat{\cal
  I}_0^{0(2)}(k)+4 \delta_\ell^1 \hat{v}_m^{(2)}(k)+\delta_\ell^2 \hat P^{m(2)}(k) \right]\,,
\ee
where $\hat P^{m(2)}(k)$ is not null only if $-2 \le m\le 2 $ and is defined in
that case by
\be
\hat P^{m(2)}(k)=\frac{1}{10}\left[\hat{\cal
      I}_2^{m(2)}(k)-\sqrt{6}\hat{\cal E}_2^{m(2)}(k)\right]\,.
\ee
For the electric and magnetic type collision terms, their second order linear components read
\be
{\cal C}^{\hashamoi(2)}[\hat{\cal E}]_\ell^m(k)=\bar{\tau'}\left[-\hat{\cal
    E}_\ell^m(k)-\delta_\ell^2 \sqrt{6}\hat P^{m(2)}(k)\right]\,,
\ee
\be
{\cal C}^{\hashamoi(2)}[\hat{\cal B}]_\ell^m(k)=-\bar{\tau'}\hat{\cal B}_\ell^m(k)\,.
\ee
The quadratic terms are then given by
\bea
\flamoi{\cal C}^{\hashamoi(1)(1)}[\hat{\cal I}^{(1)}]_\ell^m(k)&=&2 \bar{\tau'}{\cal
  K}\left\{-\sum_{n=-1}^1\frac{\kPlmsn{\ell+1}{m}{0}{n}}{2\ell+3}\hat{\cal
      I}_{\ell+1}^{m+n}(\gr{k}_2)\hat{v}_{-n}(\gr{k}_1)\right.\nonumber\\*
\flamoi &&\quad\quad+\sum_{n=-1}^1
\frac{\kMlmsn{\ell}{m}{0}{n}}{2\ell-1}\hat{\cal
    I}_{\ell-1}^{m-n}(\gr{k}_2)\hat{v}_n(\gr{k}_1)\nonumber\\*
\flamoi
&&\quad\quad+\delta_{\ell}^0\left[-\frac{4}{3}\sum_{n=-1}^{1}(-1)^n\hat{v}_n(\gr{k}_1)\hat{v}_{-n}(\gr{k}_2)\right.\nonumber\\*
&&\left.\;\quad\quad\qquad+\sum_{n=-1}^12 \frac{\kPlmsn{1}{0}{0}{n}}{3} \,\hat{\cal I}_{1}^{n}(\gr{k}_2)\hat{v}_{-n}(\gr{k}_1)\right]\nonumber\\*
\flamoi &&\quad\quad+\delta_{\ell}^13\sum_{n=-1}^1
\kMlmsn{1}{m}{0}{n}\hat{\cal I}_{0}^{m-n}(\gr{k}_2)\hat{v}_n(\gr{k}_1)\nonumber\\*
\flamoi &&\quad\quad+\delta_{\ell}^2\sum_{n=-1}^1
\frac{\kMlmsn{2}{m}{0}{n}}{3}\left[-\frac{1}{2}\hat{\cal
    I}_{1}^{m-n}(\gr{k}_2)+7 \hat{v}_{m-n}(\gr{k}_2)\right]\hat{v}_n(\gr{k}_1)\nonumber\\*
\flamoi &&\quad\quad\left.+\delta_{\ell}^3\frac{1}{2}\sum_{n=-1}^1\frac{\kMlmsn{3}{m}{0}{n}}{5}\left[\hat{\cal
      I}_{2}^{m-n}(\gr{k}_2)-\sqrt{6}\hat{\cal
      E}_{2}^{m-n}(\gr{k}_2)\right]\hat{v}_n(\gr{k}_1) \right\}\nonumber\\*
\flamoi &&+2 {\cal K}\hat{\tau'}^{(1)}(\gr{k}_1){\cal C}^{\hashamoi(1)}[\hat{\cal I}^{(1)}]_\ell^m(\gr{k}_2)\,,
\eea

\bea
\flamoi{\cal C}^{\hashamoi(1)(1)}[\hat{\cal E}^{(1)}]_\ell^m(k)&=& 2 \bar{\tau'}{\cal
  K}\left\{-\sum_{n=-1}^1\frac{\kPlmsn{\ell+1}{m}{2}{n}}{2\ell+3}\hat{\cal
      E}_{\ell+1}^{m+n}(\gr{k}_2)\hat{v}_{-n}(\gr{k}_1)\right.\nonumber\\*
\flamoi &&\quad +\sum_{n=-1}^1
\frac{\kMlmsn{\ell}{m}{2}{n}}{2\ell-1}\hat{\cal
    E}_{\ell-1}^{m-n}(\gr{k}_2)\hat{v}_n(\gr{k}_1)\nonumber\\*
\flamoi &&\quad +\delta_{\ell}^2\sum_{n=-1}^1
\frac{\kMlmsn{2}{m}{0}{n}}{3}\left[\frac{\sqrt{6}}{2}\hat{\cal
    I}_{1}^{m-n}(\gr{k}_2)-\sqrt{6} \hat{v}_{m-n}(\gr{k}_2)\right]\hat{v}_n(\gr{k}_1)\nonumber\\*
\flamoi &&\quad \left.+\delta_{\ell}^3\frac{1}{2}\sum_{n=-1}^1\frac{\kMlmsn{3}{m}{2}{n}}{5}\left[-\sqrt{6}\hat{\cal I}_{2}^{m-n}(\gr{k}_2)+6
    \hat{\cal E}_{2}^{m-n}(\gr{k}_2)\right]\hat{v}_n(\gr{k}_1)\right\}\nonumber\\*
\flamoi &&+2 {\cal K}\hat{\tau'}^{(1)}(\gr{k}_1){\cal C}^{\hashamoi(1)}[\hat{\cal E}^{(1)}]_\ell^m(\gr{k}_2)\,,
\eea

\bea\label{CollisionNMB11}
\flamoi{\cal C}^{\hashamoi(1)(1)}[\hat{\cal B}^{(1)}]_\ell^m(k)&=& -2 \bar{\tau'}{\cal K}\left\{\frac{-2 }{(\ell+1)}\sum_{n=-1}^1
  \llmn{\ell}{m}{n} \hat{v}_n(\gr{k}_1) \hat{\cal E}_{\ell}^{m-n}(\gr{k}_2)\right.\\*
\flamoi&&\qquad\left.-\delta_{\ell}^2\sum_{n=-1}^1 \llmn{2}{m}{n} \hat{v}_n(\gr{k}_1) \left[\frac{4}{5}\hat{\cal E}_{2}^{m-n}(\gr{k}_2)-\frac{2}{15}\sqrt{6}\hat{\cal I}_{2}^{m-n}(\gr{k}_2)\right] \right\}\,.\nonumber
\eea

\subsection{Continuity and Euler equations for baryons}

As discussed in section~\ref{Fluidlimit} we describe the baryons by cold matter that is
by a fluid with no pressure. From equations~(\ref{Eqcontbar1}) (\ref{Eqcontbar2}) and (\ref{Force}) we obtain
the continuity equations 
\be
\flamoi \left[\frac{\hat \rho^{(1)}(k)}{\bar \rho}\right]'+k
\hat{v}_{0}^{(1)}(k)-3 \hat \Psi^{'(1)}(k)=0\,,\label{ContinuiteBaryons1}
\ee

\bea\label{ContinuiteBaryons2}
\flamoi \left[\frac{\hat \rho^{(2)}(k)}{\bar \rho}\right]'+k
\hat{v}_{0}^{(2)}(k)-3 \hat \Psi^{'(2)}(k)\nonumber\\*
\flamoi +2{\cal  K}\left\{-\sum_{n=-1}^1(-1)^n\left[\HH\hat{v}_n(\gr{k}_1)\hat{v}_{-n}(\gr{k}_2)+\frac{\hat
      \rho(\gr{k}_1)}{\bar \rho}k_1^{(n)} \hat{v}_{-n}(\gr{k}_2)+ 2 \hat{v}_n(\gr{k}_1)\hat{v}'_{-n}(\gr{k}_2)\right]\right.\nonumber\\*
\flamoi \qquad\quad+\left[\frac{\hat \rho(k_1)}{\bar \rho}+\hat \Phi(k_1)+\hat
  \Psi(k_1)\right] k_2 \hat v_0(k_2)-3 \hat \Psi'(k_1)\frac{\hat \rho(k_2)}{\bar
  \rho}-6\hat \Psi(k_1)\hat \Psi'(k_2)\nonumber\\*
\flamoi \qquad\quad\left.-2 \sum_{n=-1}^1(-1)^n \left[\hat
    \Phi(k_2)-\hat \Psi(k_2)\right]k_2^{(n)}\hat
  v_{-n}(\gr{k}_1)\right\}\nonumber\\*
\flamoi =-\frac{2 \bar{\tau'}}{R}{\cal  K}\sum_{n=-1}^1(-1)^{n}\left[\frac{1}{3}\hat{\cal
      I}_{1}^{n}(\gr{k}_2)\hat{v}_{-n}(\gr{k}_1)-\frac{4}{3}\hat{v}_n(\gr{k}_1)\hat{v}_{-n}(\gr{k}_2)\right]\,\nonumber
\eea
where $R\equiv \bar \rho / \bar {\cal I}$.\\
From equations~(\ref{Force}) (\ref{EqEulerbar1}) and (\ref{EqEulerbar2}) we obtain the
Euler equation for baryons
\be\label{EulerBaryons1}
\flamoi \hat v^{'(1)}_m(k)+\HH \hat v^{(1)}_m(k)-\delta_m^0 k \hat
\Phi^{(1)}(k)=-\frac{\bar{\tau'}}{3 R}\left[-\hat{\cal
  I}_1^{m(1)}(k) + 4 \hat{v}_m^{(1)}(k)\right]\,,
\ee

\bea\label{EulerNM2}
\flamoi&& \hat v^{'(2)}_m(k)+\HH \hat v^{(2)}_m(k)- \delta_m^0 k \hat
\Phi^{(2)}(k)\\*
\flamoi&&+2{\cal K}\left\{\left[\frac{\hat \rho(k_1)}{\bar \rho}-\hat \Phi(k_1)
  \right] \left[\hat v'_m(\gr{k}_2)+\HH\hat v_m(\gr{k}_2) \right]+\left[\frac{\hat \rho(k_1)}{\bar \rho}+\hat \Psi(k_1)
  \right]k_2^{(m)}\hat \Phi(k_2)\right.\nonumber\\*
\flamoi&&\left.+\hat v_m(\gr{k}_2)\left[\left(\frac{\hat \rho(k_1)}{\bar \rho}\right)'-4\hat \Psi'(k_1)\right]+\delta_m^0 k\hat \Phi(k_1)\hat
  \Phi(k_2) -\hat v_m(\gr{k}_2)\sum_{n=-1}^{1}(-1)^n v_n(\gr{k}_1) k^{(-n)}
\right\}\nonumber\\*
\flamoi&&=-\frac{\bar{\tau'}}{3 R}\left[-\hat{\cal
  I}_1^{m(2)}(k) + 4 \hat{v}_m^{(2)}(k)\right]-\frac{2}{3 R}{\cal K}\hat{\tau'}^{(1)}(\gr{k}_1)\left[-\hat{\cal  I}_1^{m(1)}(\gr{k_2}) + 4 \hat{v}_m^{(1)}(\gr{k}_2)\right]\nonumber\\*
\flamoi&&\quad -\frac{2}{3 R} \bar{\tau'} {\cal K}\left[4 \hat{\cal
      I}_{0}^{0}(\gr{k}_2)\hat{v}_{m}(\gr{k}_1)-\sum_{n=-1}^1\frac{\kPlmsn{2}{m}{0}{n}}{5}\hat{\cal
      I}_{2}^{m+n}(\gr{k}_2)\hat{v}_{-n}(\gr{k}_1)\right]\,.\nonumber
\eea

\subsection{The road to numerical integration}

The set of equations which need to be integrated is very intricate. 
First the scalar ($m=0$) first order equations need to be integrated for each Fourier
mode magnitude $k$ (see~\cite{Seljak1997,Zaldarriaga1997,Hu1998}). This encompasses
the first order Einstein equations to determine the first order scalar potentials, the baryons continuity equation~(\ref{ContinuiteBaryons1}) and Euler
equation~(\ref{EulerBaryons1}), together with the Boltzmann infinite hierarchy of coupled equations
for the intensity and electric type multipoles given by the first order
Liouville operator multipoles~(\ref{LI1}) and (\ref{LE1}) equated with the first order collision multipoles~(\ref{CI1})
 and (\ref{CE1}). The first order Boltzmann hierarchy couples the modes $\ell$
 with modes $\ell\pm1$. The results obtained depend only on the magnitude of a given
 mode since we chose to align the mode with the azimuthal direction of the
 spherical harmonics. The multipoles for any direction of the Fourier mode is
 obtained by the rotation~(\ref{Rotatemode}). Finally, we obtain from
 this first order numerical integration a first order transfer function ${\cal
   T}^{m(1)}_\ell(\gr{k},\eta)$ defined by
\be
\hat{\cal I}_{\ell}^{m(1)}(\gr{k},\eta)= {\cal T}^{m(1)}_\ell(\gr{k},\eta) \bar{\cal I}_{\ell}^{m}(\eta)\zeta(k)\,,
\ee
where $\zeta(k)$ is the primordial curvature perturbation in comoving gauge.\\

At second order the numerical integration has to be performed in the same
manner. We have to integrate simultaneously the second order Einstein equations in order to obtain
the metric perturbations $\hat \Phi^{(2)}$, $\hat \Psi^{(2)}$, $\hat \Phi^{(2)}_\iB$ and $\hat
H^{(2)}_{\iB \jB}$ (the expressions of the second order Einstein tensor can be found
for instance in~\cite{Nakamura2007,Acquaviva2003,Nakamura2008,PitrouThese}),
the second order continuity equation for baryons~(\ref{ContinuiteBaryons2})
and the second order Euler equation~(\ref{EulerNM2}), together with the second order Boltzmann
infinite hierarchy for intensity, electric type, and magnetic type multipoles
which is obtained from the second order Liouville operator multipoles~(\ref{LiouvilleNMI2}-\ref{L11B}) equated
with the second order collision multipoles~(\ref{C2a}-\ref{CollisionNMB11}). After
numerical integration we will obtain a second order transfer function ${\cal
  T}^{m(2)}_\ell(\gr{k}_1,\gr{k}_2)$ defined by
\be
\frac12 \hat{\cal I}_{\ell}^{m(2)}(k,\eta)= {\cal K} {\cal T}^{m(2)}_\ell(\gr{k}_1,\gr{k}_2,\eta)\bar{\cal I}_{\ell}^{m}(\eta)\zeta(k_1)\zeta(k_2)\,.
\ee
However, we already see that the Boltzmann hierarchy at second order is much
more complex than its first order counterpart. First it depends on the two
modes $\gr{k}_1$ and $\gr{k}_2$ (though their sum $\gr{k}$ is chosen to be
aligned with the azimuthal direction of the spherical harmonics) inside the
convolution in quadratic terms. Second, it couples a mode $(\ell,m)$ with the
modes $(\ell\pm1,m\pm1)$. And finally, one problem which needs to be solved is the determination of $\tau'^{(1)}$. Indeed, photons can only scatter off free electrons and its correct expression is $\tau'=a n_\ie
x_\ie \st$, where $x_\ie$ is the fraction of ionized electrons. Both $\hat
n_\ie^{(1)}$ and $\hat x_\ie^{(1)}$ contribute to the first order perturbation since $\hat \tau'^{(1)}=a \st \hat x_\ie^{(1)} \bar n_\ie+a \st \bar
x_\ie \hat n_\ie^{(1)}$. Though $\hat n_\ie^{(1)}$ is easily obtained from the
first order equations using $\hat n_\ie^{(1)}/\bar n_\ie=\hat
\rho_\ie^{(1)}/\bar \rho_\ie$, the perturbation of $x_\ie$ requires to determine
and integrate the equations which rule the physics of recombination on the first order perturbed space-time.


\section*{Conclusion}
In this paper we have extended our previous investigation~\cite{Pitrou2007} to
the case of polarized radiation, also including the Compton scattering of photons off free
electrons, in order to obtain a fully consistent second order treatment of the
radiation transfer. We also studied the case of massive particles in the
kinetic theory and checked that in the case $T/m \ll 1$ it was consistent with a fluid approximation up
to second order. Our analysis is based on the careful definition of a locally
Minkowski frame which thanks to the equivalence principle simplifies the
treatment of local interactions. Though this method was already implicitly followed in
the existing literature, we have performed a deep analysis of its implications
for the gauge transformation, the SVT decomposition and the evolution
equations. At first order it provides a more satisfactory understanding of the
formalism used in the literature, whereas at second order it appears that it is completely
necessary to master the formal aspects that we have presented in this paper in order to understand the physical meaning of the dynamical equations.
For this purpose we have introduced a font and color based notation which
should clarify the formalism. The results that we have presented should now be
the starting point to a numerical integration of the second order radiative
transfer, but it appears already to be a huge and long term task if we want it
to be computationally efficient. Similarly to the first order, approximate schemes
should be first developed in order to obtain the main features, and this first
step has already been taken in~\cite{PUB2008} for small angular scales.

\ifcqg
\ack
\else
{\tt acknowledgements}
\fi
I thank J. Martin-Garcia for his help in using the tensorial calculus packages
xPert~\cite{Brizuela2008} and xAct~\cite{xAct} that were used to derive the second order
expressions of this paper and to calculate the integrals in
equation~(\ref{Eqinttochangeframe}). I thank G. Faye for his help on the
equation~(\ref{EqFaye}). I thank Jean-Philippe Uzan for having guided me into
the realm of CMB, and IAP where most of this work was carried out.

\appendix
\section{Transformation rules for multipoles at second order in $\gr{v}$}\label{AppTrule2}

Following the method of section~\ref{Sec_Changeframe} to compute the
transformation rules of the radiation multipoles under a change of frame, we
find the following rules up to second order in $\gr{v}$ for the multipoles
used in section~\ref{Collisionwithbulk} (note that contrary to the choice we made in section~\ref{Transfomultipoles},
we express the result in function of the $I_{\uline{\aT_\ell}}^{\{n\}}(p^\zT)$
and not the $I_{\uline{\aT_\ell}}^{\{n\}}(p^\zTt)$ since this is what was required to transform
the collision tensor in section~\ref{Collisionwithbulk})
\bea
\flamoi \tilde
I_{\emptyset}(p^{\zTt})&=&I_{\emptyset}(p^{\zT})-I'_{\emptyset}(p^{\zT})\gr{n}.\gr{v}+I'_{\emptyset}(p^{\zT})\gr{v}.\gr{v}+\frac{1}{6}I''_{\emptyset}(p^{\zT})\gr{v}.\gr{v}+\frac{1}{2}I''_{\emptyset}(p^{\zT})\left(\gr{n}.\gr{v}\right)^2 \nonumber\\*
\flamoi &&+ \frac{2}{3}v^\iT I_{\iT}(p^{\zT})+\frac{1}{3}v^\iT
I'_{\iT}(p^{\zT})-\frac{1}{3}I''_{\iT}(p^{\zT})v^\iT \gr{v}.\gr{n}-I'_{\iT}(p^{\zT})v^\iT \gr{v}.\gr{n}\nonumber\\*
\flamoi &&+\frac{2}{5}I_{\iT \jT}(p^{\zT})v^\iT v^\jT+\frac{2}{5}I'_{\iT \jT}(p^{\zT})v^\iT v^\jT+\frac{1}{15}I''_{\iT \jT}(p^{\zT})v^\iT v^\jT\,\,,
\eea

\bea
\flamoi \tilde I_{\iTt \jTt}(p^{\zTt})&=&I_{\iT
  \jT}(p^{\zT})+\frac{1}{2}I''_{\emptyset}(p^{\zT})v_{\langle \iT}v_{\jT
  \rangle}-I_{\langle \iT}(p^{\zT})v_{\jT \rangle}+I'_{\langle
  \iT}(p^{\zT})v_{\jT \rangle}-I''_{\langle \iT}(p^{\zT})v_{\jT
  \rangle}\gr{n}.\gr{v}\nonumber\\*
\flamoi &&-I'_{\iT \jT}(p^{\zT})\gr{n}.\gr{v}-\frac{4}{7}I_{\iT
  \jT}(p^{\zT})\gr{v}.\gr{v}+\frac{5}{7}I'_{\iT
  \jT}(p^{\zT})\gr{v}.\gr{v}+\frac{1}{14}I''_{\iT
  \jT}(p^{\zT})\gr{v}.\gr{v}+\frac{1}{2}I''_{\iT
  \jT}(p^{\zT})\left(\gr{n}.\gr{v}\right)^2\nonumber\\*
\flamoi &&-\frac{9}{7}I_{\kT \langle
  \iT}(p^{\zT})v_{\jT\rangle}v^\kT+\frac{6}{7}I'_{\kT \langle
  \iT}(p^{\zT})v_{\jT\rangle}v^\kT+\frac{2}{7}I''_{\kT \langle
  \iT}(p^{\zT})v_{\jT\rangle}v^\kT\nonumber\\*
\flamoi &&+\frac{12}{7}I_{\iT \jT\kT}(p^{\zT})v^{\kT}+\frac{3}{7}I'_{\iT \jT
  \kT}(p^{\zT})v^{\kT}-\frac{15}{7}I'_{\iT \jT
  \kT}(p^{\zT})v^{\kT}\gr{n}.\gr{v}-\frac{3}{7}I''_{\iT \jT
  \kT}(p^{\zT})v^{\kT}\gr{n}.\gr{v}\nonumber\\*
\flamoi &&+\frac{40}{21}I_{\iT \jT \kT \lT}(p^{\zT})v^\kT v^\lT+\frac{20}{21}I'_{\iT \jT \kT \lT}(p^{\zT})v^\kT v^\lT+\frac{2}{21}I''_{\iT \jT \kT \lT}(p^{\zT})v^\kT v^\lT\,,
\eea
\bea
\flamoi \tilde E_{\iTt \jTt}(p^{\zTt})&=&E_{\iT
  \jT}(p^{\zT})-\frac{2}{21}E_{\iT \jT}(p^{\zT})\gr{v}.\gr{v}+\frac{11}{14}E'_{\iT
  \jT}(p^{\zT})\gr{v}.\gr{v}+\frac{11}{42}E''_{\iT
  \jT}(p^{\zT})\gr{v}.\gr{v}\nonumber\\*
\flamoi&&+\frac{5}{7}E_{\kT \langle  \iT}(p^{\zT})v_{\jT\rangle}v^\kT-\frac{6}{7}E'_{\kT \langle
  \iT}(p^{\zT})v_{\jT\rangle}v^\kT-\frac{2}{7}E''_{\kT \langle
  \iT}(p^{\zT})v_{\jT\rangle}v^\kT\nonumber\\*
\flamoi&&+\frac{20}{21}E_{\iT \jT \kT}(p^{\zT})v_{\kT}+\frac{5}{21}E'_{\iT \jT
  \kT}(p^{\zT})v_{\kT}\nonumber\\*
\flamoi&&+\frac{50}{63}E_{\iT \jT \kT \lT}(p^{\zT})v^\kT v^\lT+\frac{25}{63}E'_{\iT
  \jT \kT \lT}(p^{\zT})v^\kT v^\lT+\frac{5}{126}E''_{\iT \jT \kT
  \lT}(p^{\zT})v^\kT v^\lT\nonumber\\*
\flamoi&& +\frac{1}{3}v_\kT \epsilon^{\kT \lT}_{\,\,\,\,\,\langle \iT}B_{\jT\rangle
  \lT}(p^{\zT})+\frac{1}{3}v_\kT \epsilon^{\kT \lT}_{\,\,\,\,\,\langle \iT}B'_{\jT\rangle
  \lT}(p^{\zT})\\*
\flamoi&&-\frac{10}{21}\epsilon^{\lT \mT}_{\,\,\,\,\,\langle \iT}B_{\jT\rangle \kT
  \mT}(p^{\zT})v^\kT v_\lT-\frac{5}{7}\epsilon^{\lT \mT}_{\,\,\,\,\,\langle \iT}B'_{\jT\rangle \kT
  \mT}(p^{\zT})v^\kT v_\lT-\frac{5}{42}\epsilon^{\lT \mT}_{\,\,\,\,\,\langle \iT}B''_{\jT\rangle \kT
  \mT}(p^{\zT})v^\kT v_\lT\nonumber\,\,.
\eea
As for their energy integrated counterparts, they transform as
\bea
\tilde
{\cal I}_{\emptyset}&=&{\cal I}_{\emptyset}+\frac{4}{3}{\cal I}_{\emptyset}\gr{v}.\gr{v}
- \frac{2}{3} {\cal I}_{\iT}v^\iT +\frac{2}{15}{\cal I}_{\iT \jT}v^\iT v^\jT\,\,,
\eea
\bea
\tilde {\cal I}_{\iT \jT}={\cal I}_{\iT
  \jT}+10\,{\cal I}_{\emptyset}v_{\langle \iT}v_{\jT
  \rangle}-5\,{\cal I}_{\langle  \iT}v_{\jT \rangle}+{\cal I}_{\kT \langle
  \iT}v_{\jT\rangle}v^\kT\,,
\eea
\be
\tilde {\cal E}_{\iT \jT}={\cal E}_{\iT
  \jT}-3{\cal E}_{\kT \langle \iT}v_{\jT\rangle}v^\kT+2{\cal E}_{\iT
  \jT}\gr{v}.\gr{v}+2v_\kT \epsilon^{\kT \lT}_{\,\,\,\,\,\langle \iT}{\cal B}_{\jT\rangle
  \lT}\,.
\ee

\section{Sources terms in second order transformations}\label{app_sources}

The perturbation variables in the decomposition (\ref{metric}) are extracted
as follows
\begin{eqnarray}\label{extraction_metric}
\Phi^{(n)} &=& -\frac{1}{2 a^2} g^{(n)}_{\zB\zB},\\*
 \Psi^{(n)} &=& - \frac{1}{4 a^2} {\mathrm P_{v}}^{\iB\jB} g^{(n)}_{\iB\jB}, \nonumber\\*
 B^{(n)} &=& \frac{1}{a^2} {\mathrm P_{s}}^\iB g^{(n)}_{\zB\iB},\nonumber\\*
 B^{(n)}_{\iB} &=&\frac{1}{a^2} {\mathrm P_{v}}_{\iB}^{\jB}g^{(n)}_{\zB\jB},\nonumber\\*
 E^{(n)} &=& \frac{1}{4 a^2} \left(\Delta \Delta \right)^{-1} \left(3\partial^\iB \partial^\jB- \Delta
  \delta^{\iB \jB}\right) g^{(n)}_{\iB \jB} ,\nonumber\\*
 E^{(n)}_{\nB}&=& \frac{1}{a^2}{\mathrm P_{v}}_{\nB}^{\lB}{\mathrm P_{s}}^\kB\left(\delta_{\kB}^{\iB} \delta_{\lB}^{\jB}-\frac{1}{3}\delta_{\kB \lB}
  \delta^{\iB \jB} \right) g^{(n)}_{\iB \jB}\nonumber\\*
 H^{(n)}_{\mB\nB} &=& \frac{1}{2 a^2}{\mathrm P_{v}}_{\mB}^{\kB}{\mathrm P_{v}}_{\nB}^{\lB}\left(\delta_{\kB}^{\iB} \delta_{\lB}^{\jB}-\frac{1}{3}\delta_{\kB \lB}
  \delta^{\iB \jB} \right) g^{(n)}_{\iB \jB}\,,\nonumber
\end{eqnarray}
where $n=1,2$ is the order and where we have used the definitions
\be
{\mathrm P_{s}}^\iB\equiv\Delta^{-1} \partial^\iB \,,\qquad {\mathrm P_{v}}^{\iB\jB}\equiv\delta^{\iB\jB}- \Delta^{-1}\partial^\iB \partial^\jB\,.
\ee
Using this method we can read the source terms defined in
equation~(\ref{transfo_order_2}), which are quadratic in the gauge change variables
$T,L$ and the perturbation variables $\Phi,\Psi,B,E,E_{\iB \jB}$
\begin{eqnarray}
\flamoi S_{\Phi} &=& T\left(T''+ 5 \HH T' + (\HH'+2\HH^2)T +4 \HH \Phi+ 2\Phi'\right)+ \partial_\iB L' \partial^\iB\left(T- 2 B - L'\right) \\*
\flamoi &&+ T'\left(2T'+4 \Phi\right) + \partial_\iB L \partial^\iB\left( T' + \HH T + 2 \Phi\right) \nonumber\\*
\flamoi S_{\Psi} &=& - T \left(\HH T' + (\HH' + 2 \HH^2) T - 2 \Psi' -4 \HH \Psi\right) - \partial_\iB\left(\HH T-2 \Psi\right)\partial^\iB L\nonumber\\*
\flamoi &&-\frac{1}{2}\left(\delta^{\iB \jB}-
  \Delta^{-1}\partial^\iB \partial^\jB\right){\cal X}_{\iB \jB}
\end{eqnarray}
\begin{eqnarray}
\flamoi S_{B} &=&{\mathrm P_{s}}^\iB {\cal X}_\iB\\*
\flamoi S_{B_\iB}&=&{\mathrm P_{v}}^{\jB}_\iB {\cal X}_\jB\\*
\flamoi S_{E}&=&(\Delta \Delta)^{-1} \left(\frac{3}{2}\partial^\iB \partial^\jB
  -\frac{1}{2}\Delta \delta^{\iB\jB}\right){\cal X}_{\iB \jB}\\*
\flamoi {S_{E_{\nB}}}&=&2{\mathrm P_{v}}^{\lB}_\nB{\mathrm P_{s}}^\kB\left(\delta_{\kB}^{\iB} \delta_{\lB}^{\jB}-\frac{1}{3}\delta_{\kB \lB}
  \delta^{\iB \jB} \right){\cal X}_{\iB \jB}\\*
\flamoi {S_{H_{\mB \nB}}}&=&{\mathrm P_{v}}^{\lB}_\nB{\mathrm P_{v}}^{\kB}_\mB \left(\delta_{\kB}^{\iB} \delta_{\lB}^{\jB}-\frac{1}{3}\delta_{\kB \lB}
  \delta^{\iB \jB} \right) {\cal X}_{\iB \jB}\,,
\end{eqnarray}
with
\bea
\flamoi {\cal X}_\iB &\equiv& \left\{ T' \partial_\iB(2 B + L' -T )+ 2 \HH T \partial_\iB(2B+L'-T) \right. \nonumber\\*
\flamoi &&\quad+ \partial^\jB L' \left[2 \partial_\iB \partial_\jB L + 2 \left(\HH T-2\Psi\right) \delta_{\iB\jB} + 4 \left(H_{\iB\jB}+\partial_\iB \partial_\jB E \right)\right]  \nonumber\\*
\flamoi &&\quad+ \partial^\jB \partial_\iB L \partial_\jB\left(2 B + L'-T \right) + \partial^\jB L \partial_\jB \partial_\iB \left(2B+L'-T\right) \nonumber\\*
\flamoi &&\quad\left.+ \partial_\iB T (-4 \Phi -2 T' -2\HH T) + T \partial_\iB(2 B' + L'' - T')\right\}\,,
\eea
\bea
\flamoi {\cal X}_{\iB\jB}&\equiv&\left\{\partial_\jB\left(2 B+ L'-T\right)\partial_\iB T + T \partial_\iB \partial_\jB (L'+2\HH L)\right.\nonumber\\*
\flamoi &&\quad + \partial_\iB \partial^\kB L \left[2 \partial_\kB \partial_\jB L + 4\partial_\kB \partial_\jB E+4 H_{\kB\jB}+ (2 \HH T -4
\Psi) \delta_{\kB\jB}\right] \nonumber\\*
\flamoi && \quad+ T \left(2 H'_{\iB\jB}+2 \partial_\iB \partial_\jB E' +4\HH H_{\iB\jB}+4 \HH \partial_\iB \partial_\jB E\right) \nonumber\\*
\flamoi &&\quad\left.+ \partial^\kB L \partial_\kB\left(\partial_\iB \partial_\jB L +2 H_{\iB\jB}+ 2 \partial_\iB \partial_\jB E\right)\right\}.
\eea
As for the matter perturbation variables, the source terms in the
transformation rules are
\begin{eqnarray}
\flamoi S_{\rho}&=&T(\bar{\rho}''T + \bar{\rho}'T' + 2 \delta \rho ') + \partial^\iB L \partial_\iB (2 \delta \rho + \bar{\rho}' T)\\*
\flamoi S_{P}&=&T(\bar{P}''T + \bar{P}'T' + 2 \delta P ') + \partial^\iB L \partial_\iB (2 \delta P + \bar{P}' T)\\*
\flamoi S_{V}&=&{\mathrm P_{s}}_\iB \left[\HH T\partial^\iB(L' - 2 V) +
  T\partial^\iB(2 V' -L'')+ \partial^\jB (L' -2 V)\partial_\jB \partial^\iB L
\right.\nonumber\\*
\flamoi &&\qquad \left.\quad  + L^\jB \partial_\jB \partial^\iB(2 V -L') + \partial^\iB L'\left(\HH T + T' + 2\Phi\right) \right] \\*
\flamoi S_{\tilde{V}^\kB}&=&{\mathrm P_{v}}^{\kB}_{\,\iB} \left[\HH T\partial^\iB(L' - 2 V) +  T\partial^\iB(2 V' -L'')+ \partial^\jB (L' -2 V)\partial_\jB \partial^\iB L \right.\nonumber\\*
\flamoi &&\qquad \left.\quad  + L^\jB \partial_\jB \partial^\iB(2 V -L') + \partial^\iB L'\left(\HH T + T' + 2\Phi\right) \right] \\*
\flamoi S_{\pi^{\iB \jB}}&=&2 T\left(\pi^{\iB  \jB}\right)' +2 \partial^\kB L \partial_\kB \pi^{\iB \jB} - 2 \pi^{\iB \kB} \partial_\kB \partial^\jB L - 2 \pi^{\jB \kB} \partial_\kB \partial^\iB L\,.
\end{eqnarray}

\section{Perturbation of tetrads and Ricci rotation coefficients}\label{app_TRS}

\subsection{The perturbation of tetrads}

At first order the coefficients of $R_{\aT \bT}$ and $S_{\aT \bT}$ are
(remembering that we discard the first order vector modes)
\begin{eqnarray}\label{B1}
_X R^{(1)}_{\zT\zT} &=& -\,_XS^{(1)}_{\zT\zT} =\Phi^{(1)}\\*
_X R^{(1)}_{\zT \iT} &=& -\,_XS^{(1)}_{\zT \iT} =- \partial_\iB B^{(1)}\nonumber\\*
_X R^{(1)}_{\iT \zT}&=& -\,_XS^{(1)}_{\iT \zT} = 0\nonumber\\*
_XR^{(1)}_{\iT \kT}&=& -\,S^{(1)}_{\iT \kT}=\Psi^{(1)}\delta_{\iB \kB}
- \partial_{\kB}\partial_{\iB}E^{(1)} -H^{(1)}_{\iB \kB}\nonumber
\end{eqnarray}
We can read directly from these expressions the transformation rules for the
tetrad when going from a gauge $X$ to a gauge $Y$
\begin{eqnarray}\label{Ttetrads1}
_Y \gr{e}^{(1)}_\zT=\Tr\left(\,_X\gr{e}^{(1)}_\zT\right) &=&-\Tr(\Phi^{(1)})\gr{\eb}_\zT -\gr{\eb}_{\iT} \partial^{\iB}\Tr(B^{(1)})  \\*
_Y \gr{e}^{(1)}_{\iT}=\Tr\left(\,_X \gr{e}^{(1)}_{\iT}\right) &=& \Tr(\Psi^{(1)})\gr{\eb}_{\iT} - \gr{\eb}_{\kT}\partial^{\kB}\partial_{\iB}\Tr(E^{(1)}).  \nonumber
\end{eqnarray}
At second order the coefficients of $R_{\aT \bT}$ and $S_{\aT \bT}$ are
\begin{eqnarray}\label{B3}
\flamoi _X R^{(2)}_{\zT \zT} &=& \Phi^{(2)} - 3 \Phi^2 + \partial_\iB B \partial^\iB B\\*
\flamoi _X R^{(2)}_{\zT \iT} &=& - \partial_{\iB}B^{(2)}-B^{(2)}_\iB + (2 \Phi -4
\Psi)\partial_{\iB}B  + 4 \partial^{\jB}B \left(\partial_{\iB}\partial_{\jB}E
  + H_{\iB \jB}\right)\nonumber\\*
\flamoi _X R^{(2)}_{\iT \zT}&=& -\,_XS^{(2)}_{\iT \zT} = 0\nonumber\\*
\flamoi _XR^{(2)}_{\iT \kT}&=& -\,_XS^{(2)}_{\iT \kT} \nonumber\\*
\flamoi &=& \Psi^{(2)}\delta_{\iB \kB} - \left(\partial_{\kB}\partial_{\iB}E^{(2)}+\partial_{(\kB}E^{(2)}_{\iB)} +
  H^{(2)}_{\kB \iB}\right)  + 3 \Psi^2 \delta_{\iB \kB}\nonumber\\*
\flamoi && + 3 \left(\partial_{\iB}\partial^{\lB}E + H^{\lB}_{\iB}\right)
\left(\partial_{\lB}\partial_{\kB}E+H_{\lB \kB}\right) -6 \Psi \left(\partial_{\iB}\partial_{\kB}E+H_{\iB \kB}\right)\nonumber\\*
\flamoi -\,_XS^{(2)}_{\zT\zT} &=& \Phi^{(2)} - \Phi^2 + \partial_\iB B \partial^\iB B\nonumber\\*
\flamoi -\,_XS^{(2)}_{\zT \iT} &=& - \partial_{\iB}B^{(2)} -2 \Psi\partial_{\iB}B  +
2 \partial^{\jB}B \left(\partial_{\iB}\partial_{\jB}E+H_{\iB \jB}\right)\nonumber
\end{eqnarray}
The transformations rules for the tetrads can then be read
\begin{eqnarray}\label{Ttetrads2}
\flamoi \Tr\left(\,_X \gr{e}^{(2)}_\zT\right) &=& -\left[\Tr(\Phi^{(2)})- 3 \Tr(\Phi)^2 + \partial_\iB \Tr(B) \partial^\iB \Tr(B) \right]\gr{\eb}_\zT \\*
\flamoi && +\left\{ -\partial^{\iB}\Tr(B^{(2)})-\Tr(B^{\iB(2)}) + \left[2 \Tr(\Phi) -4 \Tr(\Psi)\right]\partial^{\iB} \Tr(B)  \right.\nonumber\\*
\flamoi && \qquad \left.+ 4 \partial^{\jB}\Tr(B) \left[\partial^{\iB}\partial_{\jB}\Tr(E)+H^{\iB}_{\,\,\jB}\right]\right\}\gr{\eb}_{\iT}   \nonumber\\*
\flamoi \Tr\left(\,_X \gr{e}^{(2)}_{\iT}\right) &=&\left[\Tr(\Psi^{(2)}) + 3 \Tr(\Psi)^2 \right]\gr{\eb}_{\iT} \nonumber\\*
\flamoi
&&+\left\{-\partial^{\kB}\partial_{\iB}\Tr(E^{(2)})-\partial^{(\kB}\Tr(E^{(2)}_{\iB)})
- 6 \Tr(\Psi)\left[\partial^{\kB}\partial_{\iB}\Tr(E)+H^{\kB}_{\,\iB}\right]\right.\nonumber\\*
\flamoi &&\qquad\left. +3 \left[\partial_{\iB}\partial^{\jB}\Tr(E) +H^{\jB}_{\,\iB}\right] \left[\partial^{\kB}\partial_{\jB}\Tr(E)+ H^{\kB}_{\,\jB}\right]\right\}\gr{\eb}_{\kT}.
\end{eqnarray}

\subsection{The perturbation of Ricci rotation coefficients}\label{App_connections}

If we use a tetrad basis, then the covariant derivative is characterized by
the Ricci rotation coefficients~\cite{Wald1984} defined by
\be
\omega_{\aT \bT \cT} \equiv \eta_{\bT \dT }e^\dT_{\,\,\nu}e_\aT^{\,\,\mu}\nabla_{\mu}e_\cT^{\,\,\nu}=-\omega_{\aT\cT\bT}\,. 
\ee
They are related to the Christoffel symbols according to
\be
\omega_{\aT\,\,\,\,\cT}^{\,\,\,\,\bT}=\Christoffel{\bB}{\aB}{\cB}e_{\aT}^{\,\cB}e_{\cT}^{\,\aB}e^{\bT}_{\,\bB}+e_{\aT}^{\,\cB}e^{\bT}_{\,\bB}\partial_{\cB}e_{\cT}^{\,\bB}\,.
\ee
We choose a background tetrad field which is adapted to the Cartesian coordinate
system (see section~\ref{Sec_GToftetrads}). We then expand the Ricci rotation coefficients by expanding both the
Christoffel symbols and the tetrads. Since the Ricci rotation coefficients are
antisymmetric in their two last indices,
$\omega_{\zT\zT\zT}=\omega_{\iT\zT\zT}=0$ up to any order. 
For the background metric the only non-vanishing components are
\be
\bar \omega_{\iT\zT\jT}=-\bar \omega_{\iT\jT\zT}=-\frac{\HH}{a}\delta_{\iT\jT}\, .
\ee 
We can check that at this order the vectors in the tetrad commute (since they arise from a
coordinate system) as
\be
0=\bar\omega_{\aT\bT\cT}-\bar \omega_{\cT\bT\aT}=\eta_{\dT\bT}\bar e^{\dT}_{\,\nu}[\bar
e_{\aT},\bar e_{\cT}]^{\nu}\,.
\ee 
At first order, restricting to the Newtonian gauge and neglecting the vector
perturbations, the non-vanishing components are
\bea
\omega^{(1)}_{\zT\zT\iT}&=&-\omega^{(1)}_{\zT\iT\zT}=-\frac{1}{a}\partial_\iB \Phi\nonumber\\*
\omega^{(1)}_{\iT\zT\jT}&=&-\omega^{(1)}_{\iT\jT\zT}=\frac{1}{a}\left[-H_{\iB\jB}'+(\HH
  \Phi+\Psi')\delta_{\iB \jB} \right]\nonumber\\*
\omega^{(1)}_{\jT\iT\kT}&=&-\omega^{(1)}_{\jT\kT\iT}=\frac{2}{a}\left(\partial_{[\kB} H_{\iB]\jB}-\partial_{[\kB} \Psi  \delta_{\iB]\jB} \right)\nonumber\\*
\omega^{(1)}_{\zT\iT\jT}&=&0.
\eea
As for the second order, the components that we have used in this paper are in
the Newtonian gauge
\bea
\flamoi \omega^{(2)}_{\zT\zT\iT}&=&-\omega^{(2)}_{\zT\iT\zT}=\frac{1}{a}\left[-\partial_\iB
  \Phi^{(2)}+4\Phi\partial_\iB\Phi+2H _{\iB \kB}\partial^\kB \Phi-2\Psi\partial_\iB\Phi\right]\\*
\flamoi \omega^{(2)}_{\iT\zT\jT}&=&-\omega^{(2)}_{\iT\jT\zT}=\frac{1}{a}\left\{\partial_{(\iB}B_{\jB)}^{(2)}- H^{(2)'}_{\iB\jB}+\left[\HH  \Phi^{(2)}+\Psi^{(2)'}\right]\delta_{\iB\jB} -3\HH \Phi^2\delta_{\iB\jB}\right. \nonumber\\*
\flamoi &&\qquad\qquad +2(H_{\iB\jB}'-\Psi'\delta_{\iB\jB})\Phi+2[H_{\iB\kB}'-\Psi'\delta_{\iB\kB}][H^{\kB}_{\,\jB}-\Psi\delta^{\kB}_{\,\jB}]\nonumber\\*
\flamoi &&\qquad\qquad\left.+2[H_{\jB\kB}'-\Psi'\delta_{\jB\kB}][H^{\kB}_{\,\iB}-\Psi\delta^{\kB}_{\,\iB}] \right\}\,.
\eea
It can be checked that the vectors in the tetrad do not commute at the
perturbed level as  $\omega^{(1)}_{\aT\bT\cT}\neq\omega^{(1)}_{\cT\bT\aT}$ and
$\omega^{(2)}_{\aT\bT\cT}\neq\omega^{(2)}_{\cT\bT\aT}$. 

\section{Useful formulas for extracting the normal modes}\label{UsefulForms}

From the general composition rule~(\ref{EqFaye}), we obtain the following useful
particular cases which can be used to extract the normal modes on quadratic terms
%
\be
\flamoi \frac{\N{\ell}}{\N{\ell+1}}\Delta_\ell \Ysu{\ell}{m}{\uline{\aT_\ell}}\Yd{(\ell+1)}{m}{\uline{\aT_\ell}\bT} =-\frac{\sqrt{(\ell+1)^2-m^2}}{\ell+1}\N{1}\Delta_1 \,\Ysd{1}{0}{\bT}\,,
\ee

\be
\flamoi \frac{\N{\ell}}{\N{\ell+1}} \Delta_\ell
\Ysu{\ell}{m}{\uline{\aT_\ell}}\Yd{(\ell+1)}{(m\pm1)}{\uline{\aT_\ell}\bT}=\frac{\sqrt{(\ell+1\pm
    m)(\ell+2\pm m)}}{\sqrt{2}(\ell+1)}\N{1}\Delta_1\,\Ysd{1}{(\mp1)}{\bT}\,,
\ee

\be
\flamoi \frac{\N{\ell}}{\N{\ell-1}}\Delta_\ell \Ysu{\ell}{m}{ \uline{\aT_{\ell-1}}\bT}\Yd{(\ell-1)}{m}{\uline{\aT_{\ell-1}}}= \frac{\sqrt{\ell^2-m^2}}{(2\ell-1)}\,\N{1}\Delta_1\Ysd{1}{0}{\bT}\,,
\ee

\be
\flamoi \frac{\N{\ell}}{\N{\ell-1}}\Delta_\ell \Ysu{\ell}{m}{
  \uline{\aT_{\ell-1}}\bT}\Yd{(\ell-1)}{(m\pm1)}{\uline{\aT_{\ell-1}}}=
\frac{\sqrt{(\ell\mp m)(\ell\mp m-1)}}{\sqrt{2}(2\ell-1)}\,\N{1}\Delta_1\Ysd{1}{(\mp1)}{\bT}\,,
\ee
In order to extract normal modes in quadratic terms involving a ${\mathrm{
  curl}}$, we also need the following useful formulas
\be
\flamoi \ii  e^{(0)}_\bT \epsilon^{\bT \cT}_{\,\,\,\,\langle \aT_1}\,\Yd{\ell}{m}{\uline{\aT_{(\ell-1)}}\rangle \cT} =-\frac{m}{\ell}\Yd{\ell}{m}{\uline{\aT_\ell}}\,,
\ee
\be
\flamoi \ii  e^{(\pm 1)}_\bT \epsilon^{\bT \cT}_{\,\,\,\,\langle
  \aT_1}\,\Yd{\ell}{(m\mp 1)}{\uline{\aT_{(\ell-1)}}\rangle \cT}
=\pm \frac{1}{\ell}\sqrt{\frac{(\ell\pm m)(\ell+1\mp m)}{2}}\Yd{\ell}{m}{\uline{\aT_\ell}}\,.
\ee
\section*{References}

\ifcqg
\bibliographystyle{h-physrev}
\bibliography{comptonsanstitres}
\else
\bibliographystyle{h-physrev}
\bibliography{comptonprlsanstitres}
\fi


\end{document}